\documentclass[12pt]{cernprep}
\usepackage{graphicx}
\usepackage{amssymb}
\usepackage{epsfig,color,rotating}
\begin{document}
\newcommand{\dedx}{\mbox{${\rm d}E/{\rm d}x$}}
\newcommand{\EcB}{$E \! \times \! B$}
\newcommand{\omt}{$\omega \tau$}
\newcommand{\omtsq}{$(\omega \tau )^2$}
\newcommand{\rphi}{\mbox{$r \! \cdot \! \phi$}}
\newcommand{\srphi}{\mbox{$\sigma_{r \! \cdot \! \phi}$}}
\newcommand{\dg}{\mbox{`durchgriff'}}
\newcommand{\mg}{\mbox{`margaritka'}}
\newcommand{\pT}{\mbox{$p_{\rm T}$}}
\newcommand{\GeVc}{\mbox{GeV/{\it c}}}
\newcommand{\MeVc}{\mbox{MeV/{\it c}}}
\def\kr{$^{83{\rm m}}$Kr\ }
\begin{titlepage}
\docnum{CERN--PH--EP/2009--012}
\date{6 June 2009} 
%
%
%
%
\vspace{1cm}
\title{CROSS-SECTIONS OF LARGE-ANGLE HADRON PRODUCTION \\
IN PROTON-- AND PION--NUCLEUS INTERACTIONS IV: \\
COPPER NUCLEI AND BEAM MOMENTA 
FROM \mbox{\boldmath $\pm3$}~GeV/\mbox{\boldmath $c$} 
TO \mbox{\boldmath $\pm15$}~GeV/\mbox{\boldmath $c$}}

\begin{abstract}
We report on double-differential inclusive cross-sections of 
the production of secondary protons, charged pions, and deuterons,
in the interactions with a 5\% $\lambda_{\rm abs}$ 
thick stationary copper target, of proton and pion beams with
momentum from $\pm3$~GeV/{\it c} to $\pm15$~GeV/{\it c}. Results are 
given for secondary particles with production 
angles $20^\circ < \theta < 125^\circ$. 
\end{abstract}

\vfill  \normalsize
\begin{center}
The HARP--CDP group  \\  

\vspace*{2mm} 

A.~Bolshakova$^1$, 
I.~Boyko$^1$, 
G.~Chelkov$^{1a}$, 
D.~Dedovitch$^1$, 
A.~Elagin$^{1b}$, 
M.~Gostkin$^1$,
A.~Guskov$^1$, 
Z.~Kroumchtein$^1$, 
Yu.~Nefedov$^1$, 
K.~Nikolaev$^1$, 
A.~Zhemchugov$^1$, 
F.~Dydak$^2$, 
J.~Wotschack$^{2*}$, 
A.~De~Min$^{3c}$,
V.~Ammosov$^4$, 
V.~Gapienko$^4$, 
V.~Koreshev$^4$, 
A.~Semak$^4$, 
Yu.~Sviridov$^4$, 
E.~Usenko$^{4d}$, 
V.~Zaets$^4$ 
\\
 
\vspace*{5mm} 

$^1$~{\bf Joint Institute for Nuclear Research, Dubna, Russia} \\
$^2$~{\bf CERN, Geneva, Switzerland} \\ 
$^3$~{\bf Politecnico di Milano and INFN, 
Sezione di Milano-Bicocca, Milan, Italy} \\
$^4$~{\bf Institute of High Energy Physics, Protvino, Russia} \\

\vspace*{5mm}

\submitted{(To be submitted to Eur. Phys. J. C)}
\end{center}

\vspace*{5mm}
\rule{0.9\textwidth}{0.2mm}

\begin{footnotesize}

$^a$~Also at the Moscow Institute of Physics and Technology, Moscow, Russia 

$^b$~Now at Texas A\&M University, College Station, USA 

$^c$~On leave of absence at 
Ecole Polytechnique F\'{e}d\'{e}rale, Lausanne, Switzerland 

$^d$~Now at Institute for Nuclear Research RAS, Moscow, Russia

$^*$~Corresponding author; e-mail: joerg.wotschack@cern.ch
\end{footnotesize}

\end{titlepage}


\newpage 

\section{Introduction}

The HARP experiment arose from the realization that the 
inclusive differential cross-sections of hadron production 
in the interactions of few GeV/{\it c} protons with nuclei were 
known only within a factor of two to three, while 
more precise cross-sections are in demand for several 
reasons.
Pion production data on a variety of nuclei are required for (i) the understanding of the underlying physics and the modelling of Monte Carlo generators of hadron--nucleus collisions, (ii) the optimization of the design parameters of the proton driver of a neutrino factory, (iii) flux predictions for conventional neutrino beams, and (iv) the calculation of the atmospheric neutrino flux.

Consequently, the HARP detector was designed to carry 
out a programme of systematic and precise 
(i.e., at the few per cent level) measurements of 
hadron production by protons and pions with momenta from 
1.5 to 15~GeV/{\it c}. 

The detector combined a forward spectrometer with a 
large-angle spectrometer. The latter comprised a 
cylindrical Time Projection 
Chamber (TPC) around the target and an array of 
Resistive Plate Chambers (RPCs) that surrounded the 
TPC. The purpose of the TPC was track 
reconstruction and particle identification by \dedx . The 
purpose of the RPCs was to complement the 
particle identification by time of flight.

The HARP experiment was performed at the CERN Proton Synchrotron 
in 2001 and 2002 with a set of stationary targets ranging from hydrogen to lead.

Here, we report on the large-angle production (polar angle $\theta$ in the 
range $20^\circ < \theta < 125^\circ$) of secondary protons and charged pions, and of deuterons, in 
the interactions with a 5\% $\lambda_{\rm abs}$ Cu target of protons and pions with beam momenta of $\pm3.0$, 
$\pm5.0$, $\pm8.0$, $\pm12.0$, and $\pm15.0$~GeV/{\it c}. 

This is the fourth of a series of cross-section papers with results from the HARP experiment. In the first paper,
Ref.~\cite{Beryllium1}, we described the detector 
characteristics and our analysis algorithms, on the example of 
$+8.9$~GeV/{\it c} and $-8.0$~GeV/{\it c} beams impinging on a 
5\% $\lambda_{\rm abs}$ Be target. The second paper~\cite{Beryllium2} 
presented results for all beam momenta from this Be target, and the third 
paper~\cite{Tantalum} results
from the interactions with a 5\% $\lambda_{\rm abs}$ Ta target.  

Our work involves only the HARP large-angle spectrometer, the characteristics of which are described in detail in Refs.~\cite{TPCpub} and \cite{RPCpub}.

\section{The T9 proton and pion beams, and the target}

The protons and pions were delivered by
the T9 beam line in the East Hall of CERN's Proton Synchrotron.
This beam line supports beam momenta between 1.5 and 15~GeV/{\it c},
with a momentum bite $\Delta p/p \sim 1$\%.

Beam particle identification was provided for by two threshold  
Cherenkov counters, BCA and BCB, filled with nitrogen, and by time of 
flight over a flight path of 24.3~m. Table~\ref{beampartid} 
lists the beam instrumentation that was used at different
beam momenta for p/$\pi^+$
and for $\pi$/e separation. 

\begin{table}[h]
\caption{Beam instrumentation for p/$\pi^+$
and $\pi$/e separation}
\label{beampartid}
\begin{center}
\begin{tabular}{|c|c|c|c|}
\hline
Beam momentum [GeV/{\it c}] & p/$\pi^+$ separation & $\pi$/e separation \\
\hline
\hline
$\pm3.0$  & TOF                 & BCB (1.05 bar) \\
\hline
$\pm5.0$  & TOF                 & BCA (0.60 bar) \\
\hline
$\pm8.0$  & BCA (1.25 bar)      &                \\
   & BCB (1.50 bar)      &                \\ 
\hline
$\pm12.0$ and $\pm15.0$ &  BCA (3.50 bar) &             \\
          &  BCB (3.50 bar) &             \\ 
\hline          
\end{tabular}
\end{center}
\end{table}

The pion beam had a contamination by muons from pion decays. 
It also had a contamination by electrons from converted
photons from $\pi^0$ decays. Only for the beam momenta of 3 and 
5~GeV/{\it c} were electrons identified by a beam Cherenkov
counter and rejected.

The fractions of muon and electron contaminations of the pion beam
were experimentally determined~\cite{T9beammuons,T9beamelectrons} 
and are listed in Table~\ref{pioncontaminations} for all beam 
momenta. For the determination of interaction cross-sections of pions, 
the muon and 
electron contaminations must be subtracted from 
the incoming flux of pion-like particles (except electrons at the beam 
momenta of 3 and 5~GeV/{\it c}).     
\begin{table}[h]
\caption{Contaminations of the pion beams by muons and electrons}
\label{pioncontaminations}
\begin{center}
\begin{tabular}{|c|c|c|c|}
\hline
Beam momentum [GeV/{\it c}] & Muon fraction & Electron fraction \\
\hline
$\pm3.0$   & $(4.1 \pm 0.4)$\% & rejected \\
$\pm5.0$   & $(5.1 \pm 0.4)$\% & rejected  \\
$\pm8.0$   & $(1.9 \pm 0.5)$\% & $(1.2 \pm 0.5)$\% \\
$\pm12$  & $(0.6 \pm 0.6)$\% & $(0.5 \pm 0.5)$\% \\
$\pm15$  & $(0.0 \pm 0.5)$\% & $(0.0 \pm 0.5)$\% \\
\hline
\end{tabular}
\end{center}
\end{table}

There is also a kaon contamination of a few per cent in the proton 
and pion beams. Kaons are suppressed by the beam instrumentation,
except at 5 GeV/{\it c} beam momentum where they are 
indistinguishable from pions. Because the kaon interaction 
cross-sections are close to the pion interaction 
cross-sections, this kaon contamination is ignored. 

The beam trajectory was determined by a set of three multiwire 
proportional chambers (MWPCs), located upstream of the target,
several metres apart. The transverse error of the 
impact point on the target was 0.5~mm from the 
resolution of the MWPCs, plus a
contribution from multiple scattering of the beam particles
in various materials in the beam line. Excluding the target itself, the 
latter contribution is 0.2~mm for a 8~GeV/{\it c} beam 
particle.

We select  
`good' beam particles by requiring the unambiguous reconstruction
of the particle trajectory with good $\chi^2$. In addition we 
require that the particle type is unambiguously identified. 
We select `good' accelerator spills by requiring a minimal beam intensity and
a `smooth' variation of beam intensity across the 400~ms long 
spill\footnote{A smooth variation of beam intensity eases 
corrections for dynamic TPC track distortions.}.

The target was a disc made of 
high-purity (99.99\%) copper, with a density of 8.91~g/cm$^3$,
a radius of 15.1~mm, and a thickness of $7.52 \pm 0.05$~mm 
(5\% $\lambda_{\rm abs}$).

The finite thickness of the target leads to a
small attenuation of the number of incident beam particles. The
attenuation factor is $f_{\rm att} = 0.975$.

The size of the beam spot at the position of the target was several
millimetres in diameter, determined by the setting of the beam
optics and by multiple scattering. The nominal 
beam position\footnote{A 
right-handed Cartesian and/or spherical polar coordinate 
system is employed; the $z$ axis coincides with the beam line, with
$+z$ pointing downstream; the coordinate origin is at the 
upstream end of the copper target, 500~mm
downstream of the TPC's pad plane; 
looking downstream, the $+x$ coordinate points to
the left and the $+y$ coordinate points up; the polar angle
$\theta$ is the angle with respect to the $+z$ axis.} 
was at $x_{\rm beam} = y_{\rm beam} = 0$, however, excursions 
by several millimetres
could occur\footnote{The only relevant issue is that the trajectory
of each individual beam particle is known, whether shifted or not, 
and therefore the amount of matter to be traversed by the 
secondary hadrons.}. 
A loose fiducial cut 
$\sqrt{x^2_{\rm beam} + y^2_{\rm beam}} < 12$~mm
ensured full beam acceptance. The muon and electron 
contaminations of the 
pion beam, stated above, refer to this acceptance cut.

\section{Performance of the HARP large-angle detectors}

Our calibration work on the HARP TPC and RPCs
is described in detail in Refs.~\cite{TPCpub} and \cite{RPCpub},
and in references cited therein. In particular, we recall that 
static and dynamic TPC track distortions up to 10~mm have been 
corrected to better than 300~$\mu$m. Therefore, TPC track 
distortions do not affect the precision of our cross-section
measurements.  

The resolution $\sigma (1/p_{\rm T})$
is typically 0.2~(GeV/{\it c})$^{-1}$ 
and worsens towards small relative particle
velocity $\beta$ and small polar angle $\theta$.

The absolute momentum scale is determined to be correct to 
better than 2\%, both for positively and negatively
charged particles.
 
The polar angle $\theta$ is measured in the TPC with a 
resolution of $\sim$9~mrad, for a representative 
angle of $\theta = 60^\circ$. To this a multiple scattering
error has to be added which is on the average 
$\sim$7~mrad for a proton with 
$p_{\rm T} = 500$~MeV/{\it c} in the TPC gas and $\theta = 60^\circ$, 
and $\sim$4~mrad for a pion with the same characteristics.
The polar-angle scale is correct to better than 2~mrad.     

The TPC measures \dedx\ with a resolution of 16\% for a 
track length of 300~mm.

The intrinsic efficiency of the RPCs that surround 
the TPC is better than 98\%.

The intrinsic time resolution of the RPCs is 127~ps and
the system time-of-flight resolution (that includes the
jitter of the arrival time of the beam particle at the target)
is 175~ps. 

To separate measured particles into species, we
assign on the basis of \dedx\ and $\beta$ to each particle a 
probability of being a proton,
a pion (muon), or an electron, respectively. The probabilities
add up to unity, so that the number of particles is conserved.
These probabilities are used for weighting when entering 
tracks into plots or tables.

\section{Monte Carlo simulation}

We used the Geant4 tool kit~\cite{Geant4} for the simulation 
of the HARP large-angle spectrometer.

Geant4's QGSP\_BIC physics list provided us with 
reasonably realistic spectra of secondaries from incoming beam 
protons with momentum
less than 12~GeV/{\it c}.
For the secondaries from beam protons at 12 and 15~GeV/{\it c}
momentum, and from beam pions at all momenta, we found the standard 
physics lists of Geant4 unsuitable~\cite{GEANTpub}. 

To overcome this problem,
we built our own HARP\_CDP physics list
for the production of secondaries from incoming beam pions. 
It starts from Geant4's standard QBBC physics list, 
but the Quark--Gluon String Model is replaced by the 
FRITIOF string fragmentation model for
kinetic energy $E>6$~GeV; for $E<6$~GeV, the Bertini 
Cascade is used for pions, and the Binary Cascade for protons; 
elastic and quasi-elastic scattering is disabled.
Examples of the good performance of the HARP\_CDP physics list
are given in Ref.~\cite{GEANTpub}.

\section{Systematic errors}

The systematic uncertainty of our inclusive cross-sections 
is at the few-per-cent level, from errors
in the normalization, in the momentum measurement, in
particle identification, and in the corrections applied
to the data.

The systematic error of the absolute flux normalization is 
taken as 2\%. This error arises from uncertainties in the
target thickness, in the contribution of large-angle 
scattering of beam particles, in the attenuation of beam 
particles in the target, and in the subtraction of
the muon and electron contaminations of the beam. Another contribution 
comes from the removal of events with an abnormally large 
number of TPC hits\footnote{In less than 0.5\% of the
number of good events, because of apparatus malfunction,
the number of TPC hits was much larger than possible for
a physics event. Such events were considered unphysical and 
eliminated.}.

The systematic error of the track finding  
efficiency is taken as 1\% which reflects differences 
between results from different persons who conducted
eyeball scans. We also take the statistical errors of
the parameters of a fit to scan results  
as systematic error into account~\cite{Beryllium1}.
The systematic error of the correction 
for losses from the requirement of at least 10 TPC clusters 
per track is taken as 20\% of the correction which 
itself is in the range of 5 to 30\%. This estimate arose
from differences between the four TPC sectors that
were used in our analysis, and from the observed 
variations with time. 

The systematic error of the $p_{\rm T}$ scale is taken as
2\% as discussed in Ref.~\cite{TPCpub}. For the data from
the $+15$~GeV/{\it c} beams, this error
was doubled to account for a larger than usual uncertainty of 
the correction for dynamic TPC track distortions.

The systematic errors of the proton, pion, and electron
abundances are taken as 10\%. We stress that errors on 
abundances only lead to cross-section errors in case of a strong overlap of the resolution functions
of both identification variables, \dedx\ and $\beta$. 
The systematic error of the correction for migration, absorption
of secondary protons and pions in materials, and for pion
decay into muons, is taken as 20\% of the correction, or 1\% of the cross-section, whichever is larger. These estimates reflect our experience 
with remanent differences between data and Monte Carlo 
simulations after weighting Monte Carlo events with smooth functions 
with a view to reproducing the data simultaneously in 
several variables in the best possible way.

All systematic errors are propagated into the momentum 
spectra of secondaries and then added in quadrature.

\section{Cross-section results}

In Tables~\ref{pro.procu3}--\ref{pim.pimcu15}, collated
in the Appendix of this paper, we give
the double-differential inclusive cross-sections 
${\rm d}^2 \sigma / {\rm d} p {\rm d} \Omega$
for various combinations of
incoming beam particle and secondary particle, including
statistical and systematic errors. In each bin,  
the average momentum at the vertex and the average polar angle are also given.

The data of Tables~\ref{pro.procu3}--\ref{pim.pimcu15} are available 
in ASCII format in Ref.~\cite{ASCIItables}.

Some bins in the tables are empty. Cross-sections are 
only given if the total error is not larger 
than the cross-section itself.
Since our track reconstruction algorithm is optimized for
tracks with $p_{\rm T}$ above $\sim$70~MeV/{\it c} in the
TPC volume, we do not give cross-sections from tracks 
with $p_{\rm T}$ below this value.
Because of the absorption of slow protons in the material between the
vertex and the TPC gas, 
and with a view to keeping the correction
for absorption losses below 30\%, cross-sections from protons are 
limited to $p > 450$~MeV/{\it c} at the interaction vertex. 
Proton cross-sections are also not given if a 
10\% error on the proton energy loss in materials between the 
interaction vertex and the TPC volume leads to a momentum 
change larger than 2\%. Since the proton energy loss is large
in the copper target, particularly at polar angles close 
to 90~degrees, the latter condition imposes significant restrictions.
Pion cross-sections are not given if pions are separated from 
protons by less than twice the time-of-flight resolution.

The large errors and/or absence of results from the 
$+15$~GeV/{\it c} pion beams
are caused by scarce statistics because the beam
composition was dominated by protons.

\begin{figure*}[h]
\begin{center}
\begin{tabular}{cc}
\includegraphics[height=0.30\textheight]{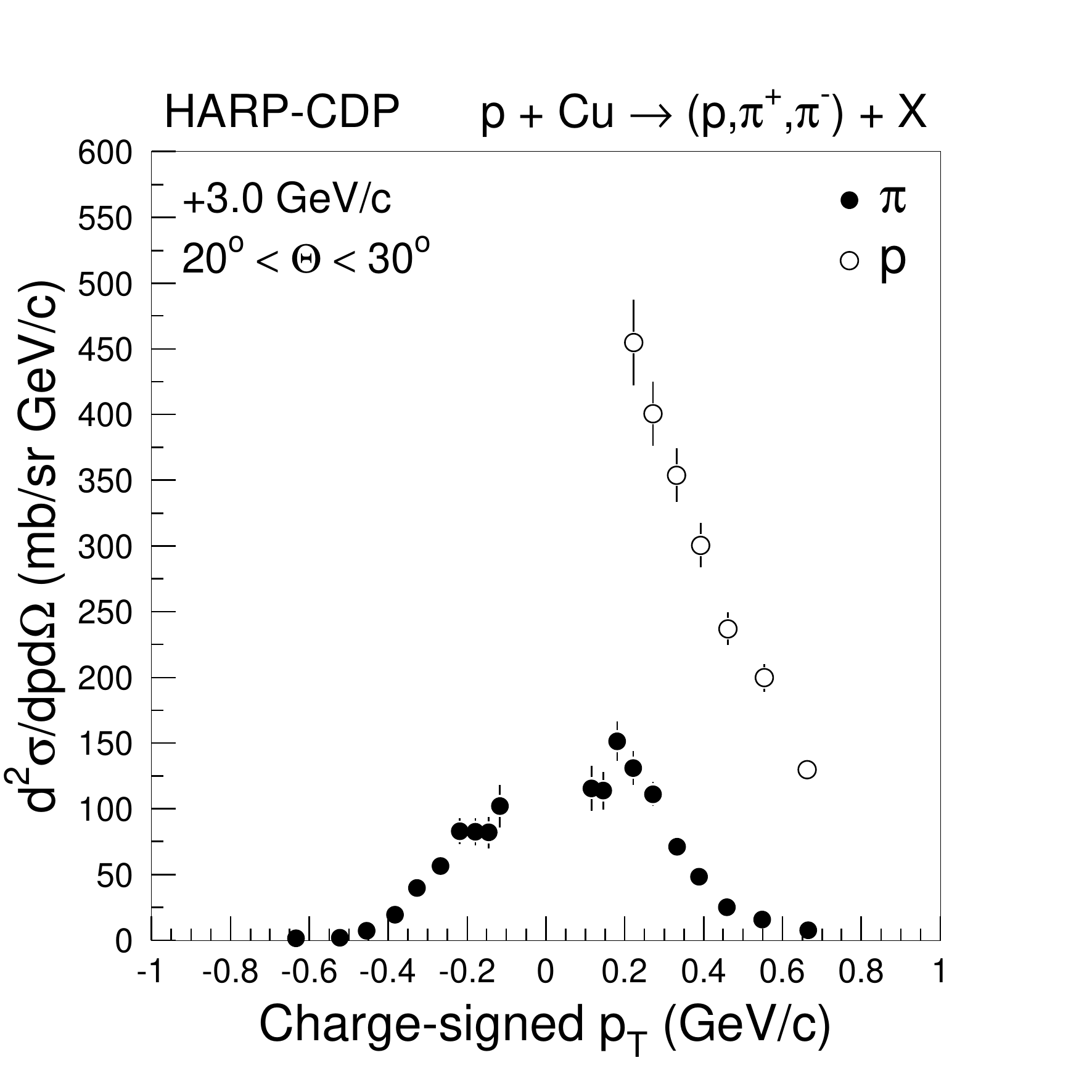} &
\includegraphics[height=0.30\textheight]{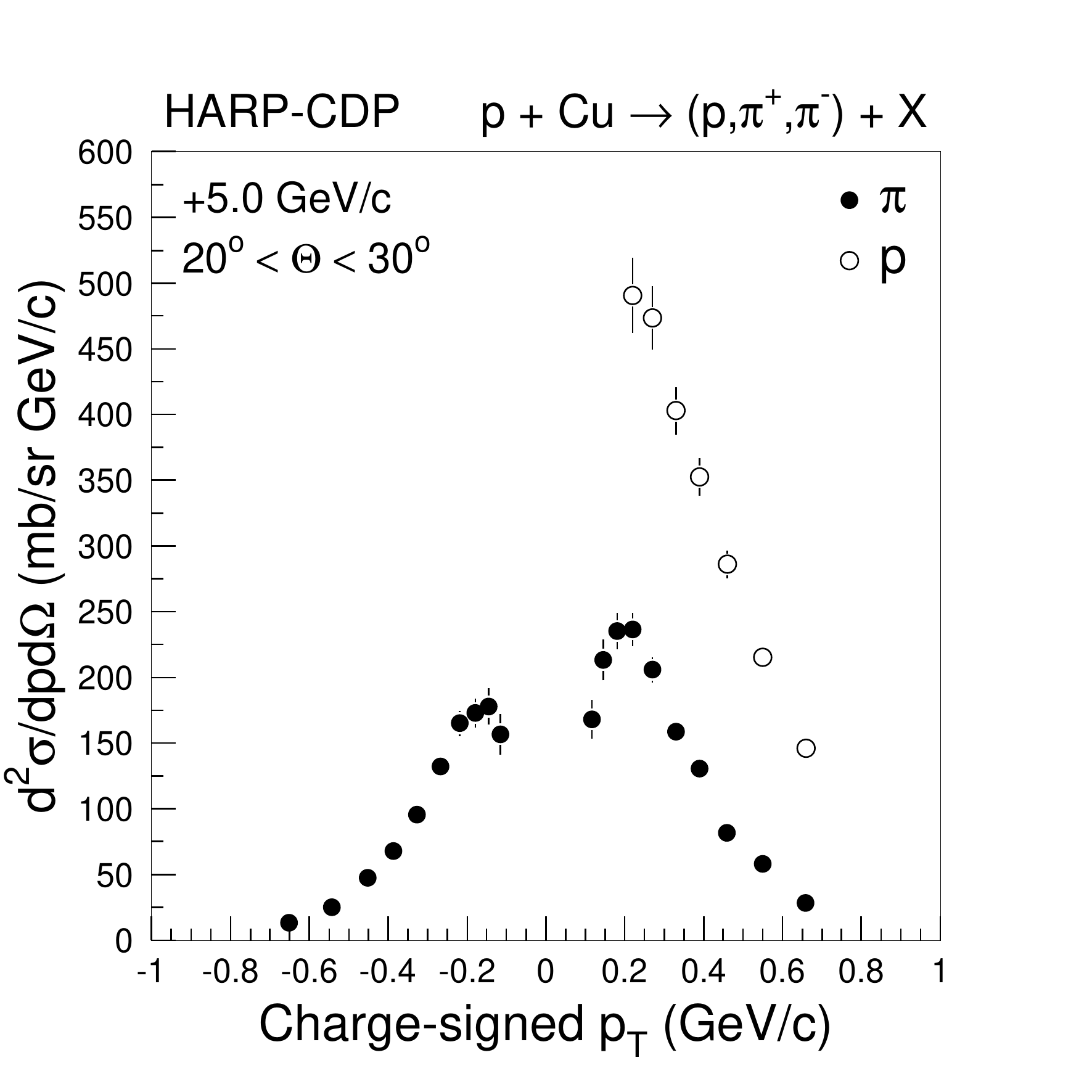} \\
\includegraphics[height=0.30\textheight]{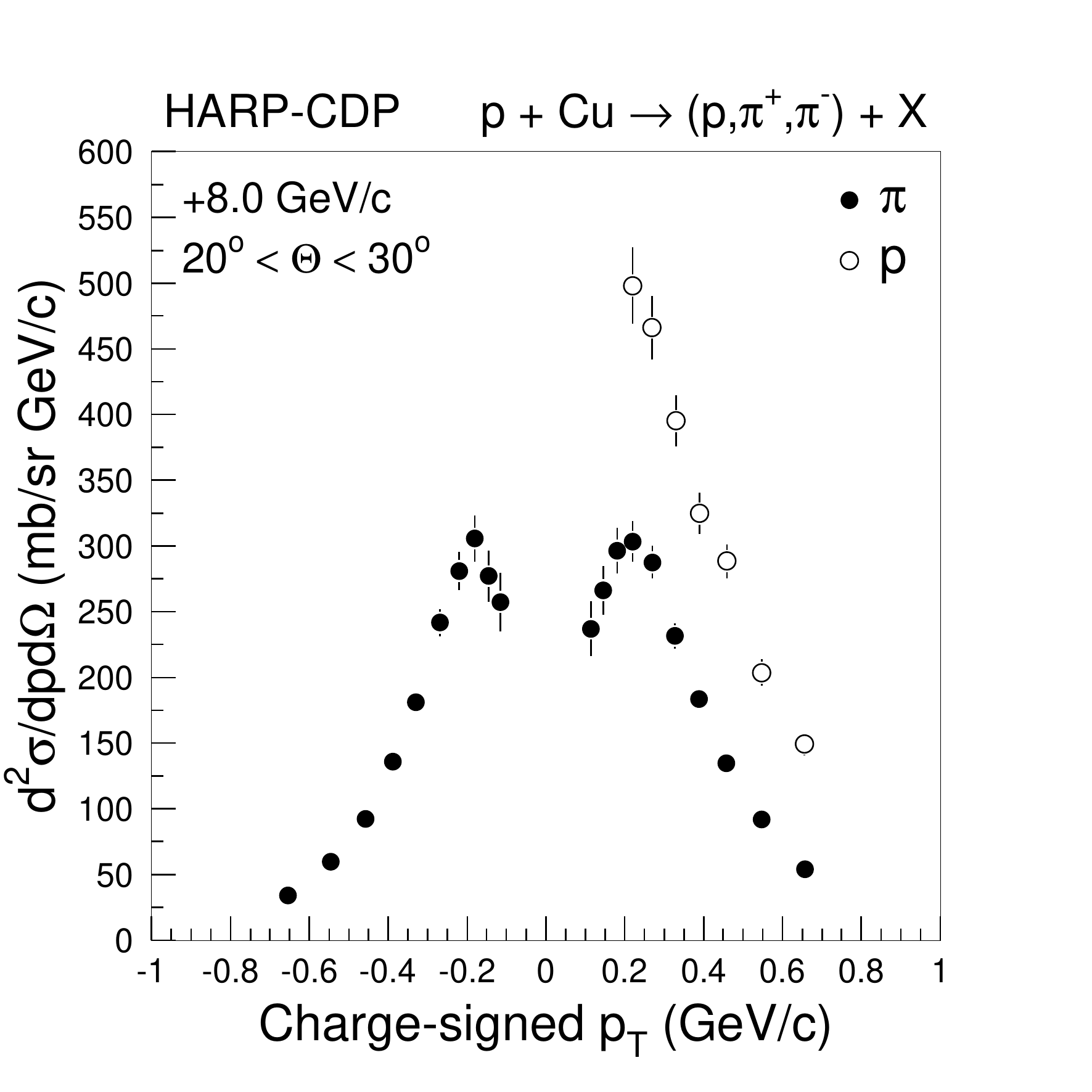} &
\includegraphics[height=0.30\textheight]{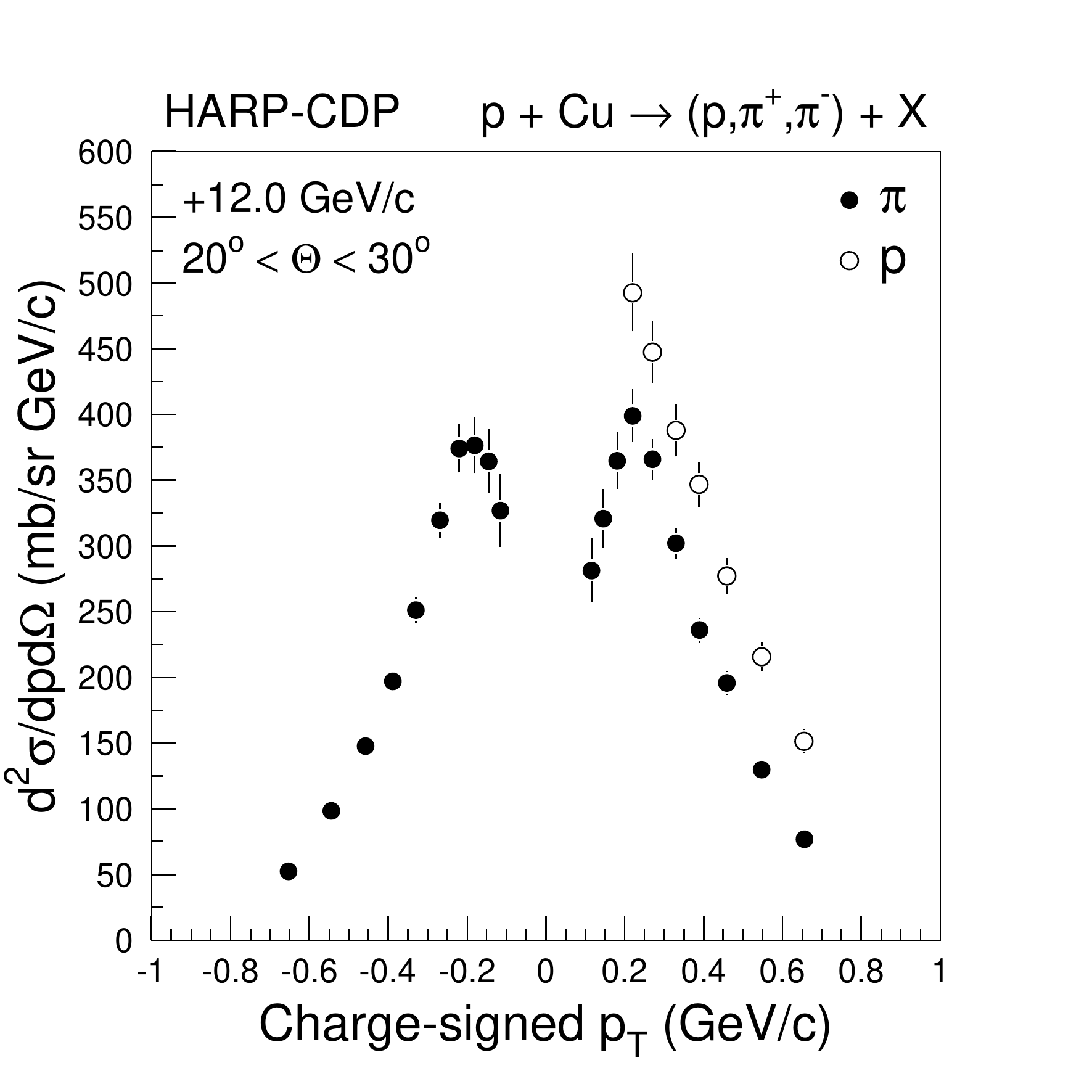} \\
\includegraphics[height=0.30\textheight]{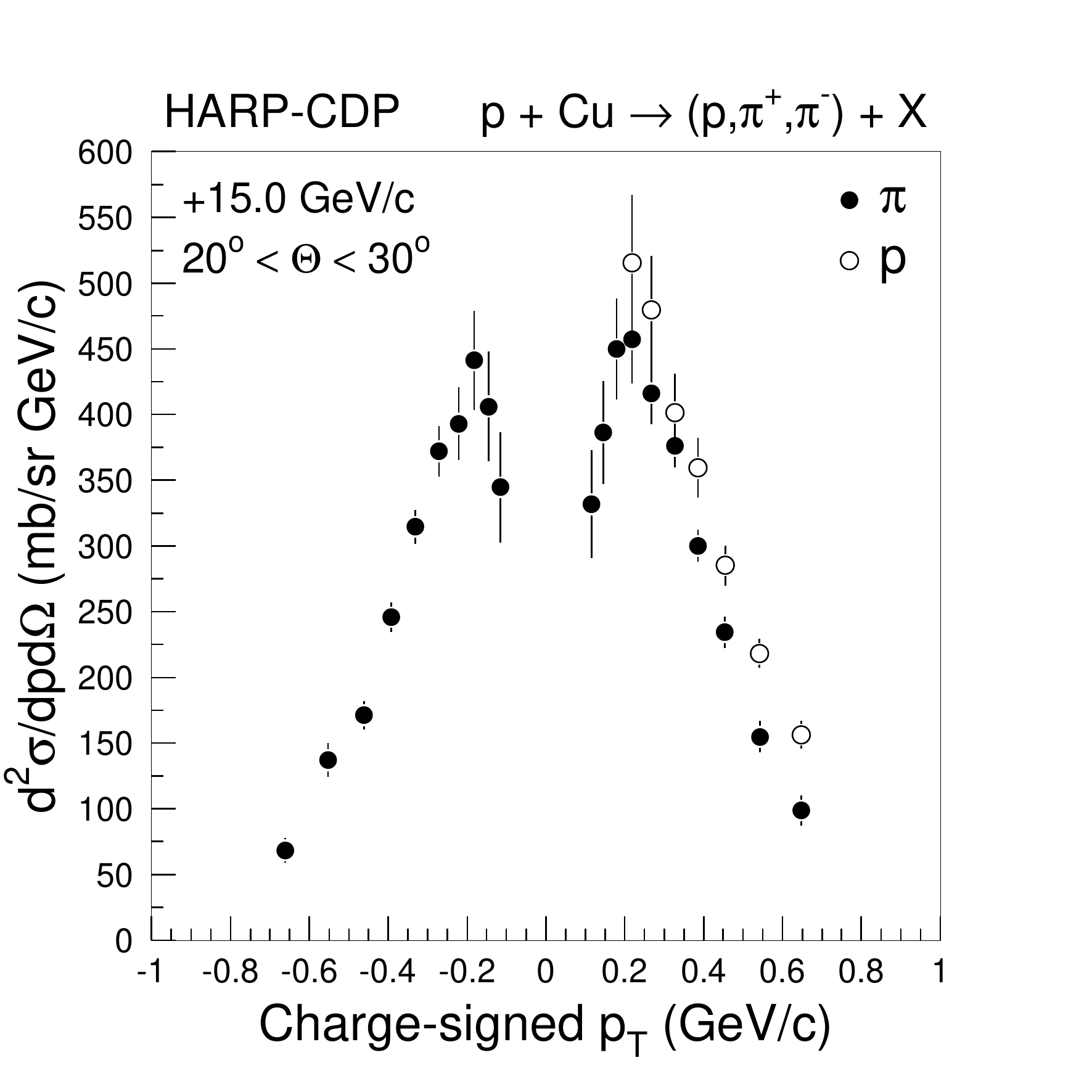} &  \\
\end{tabular}
\caption{Inclusive cross-sections of the production of secondary
protons, $\pi^+$'s, and $\pi^-$'s, by protons on copper nuclei, 
in the polar-angle range $20^\circ < \theta < 30^\circ$, for
different proton beam momenta, as a function of the charge-signed 
$p_{\rm T}$ of the secondaries; the shown errors are total errors.} 
\label{xsvsmompro}
\end{center}
\end{figure*}

\begin{figure*}[h]
\begin{center}
\begin{tabular}{cc}
\includegraphics[height=0.30\textheight]{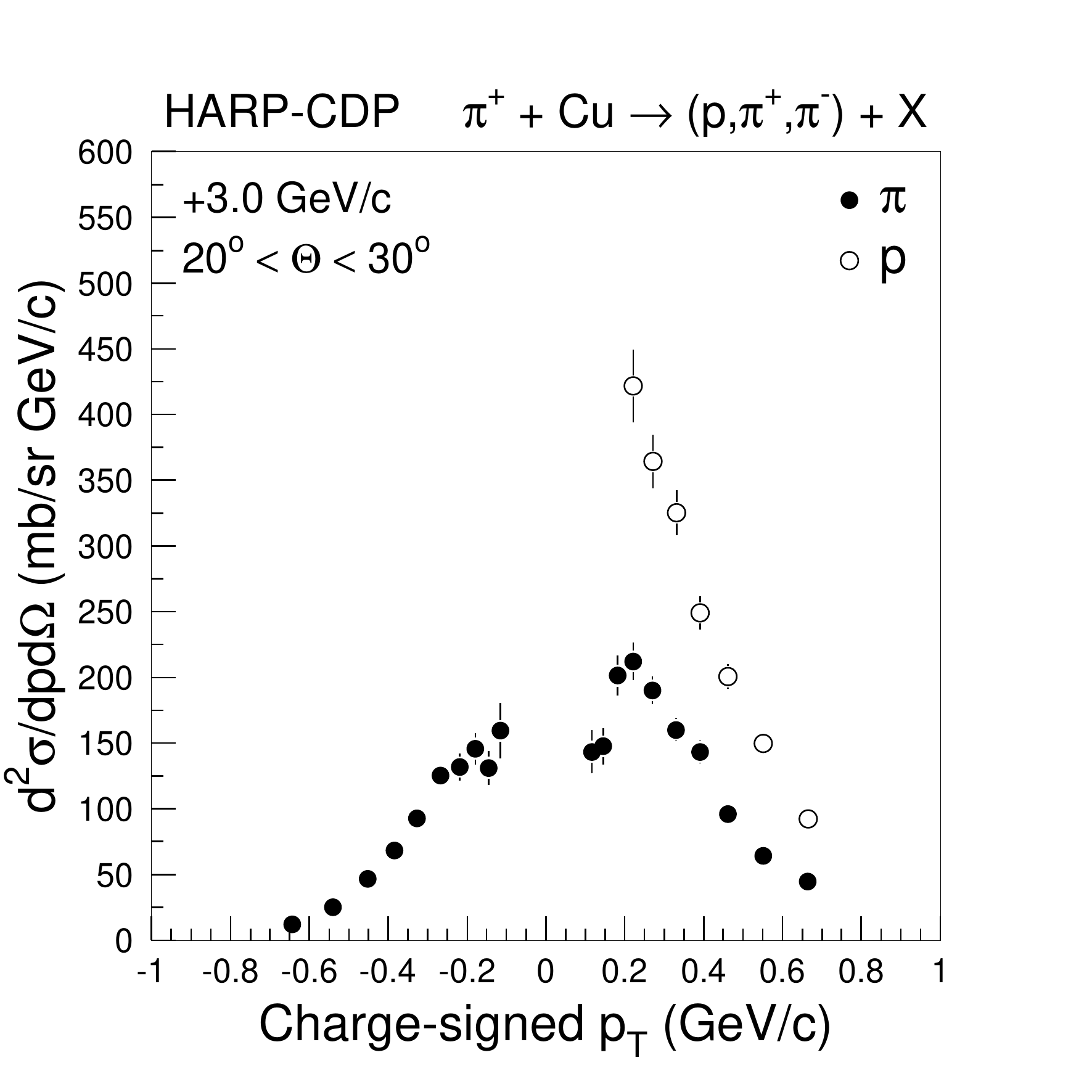} &
\includegraphics[height=0.30\textheight]{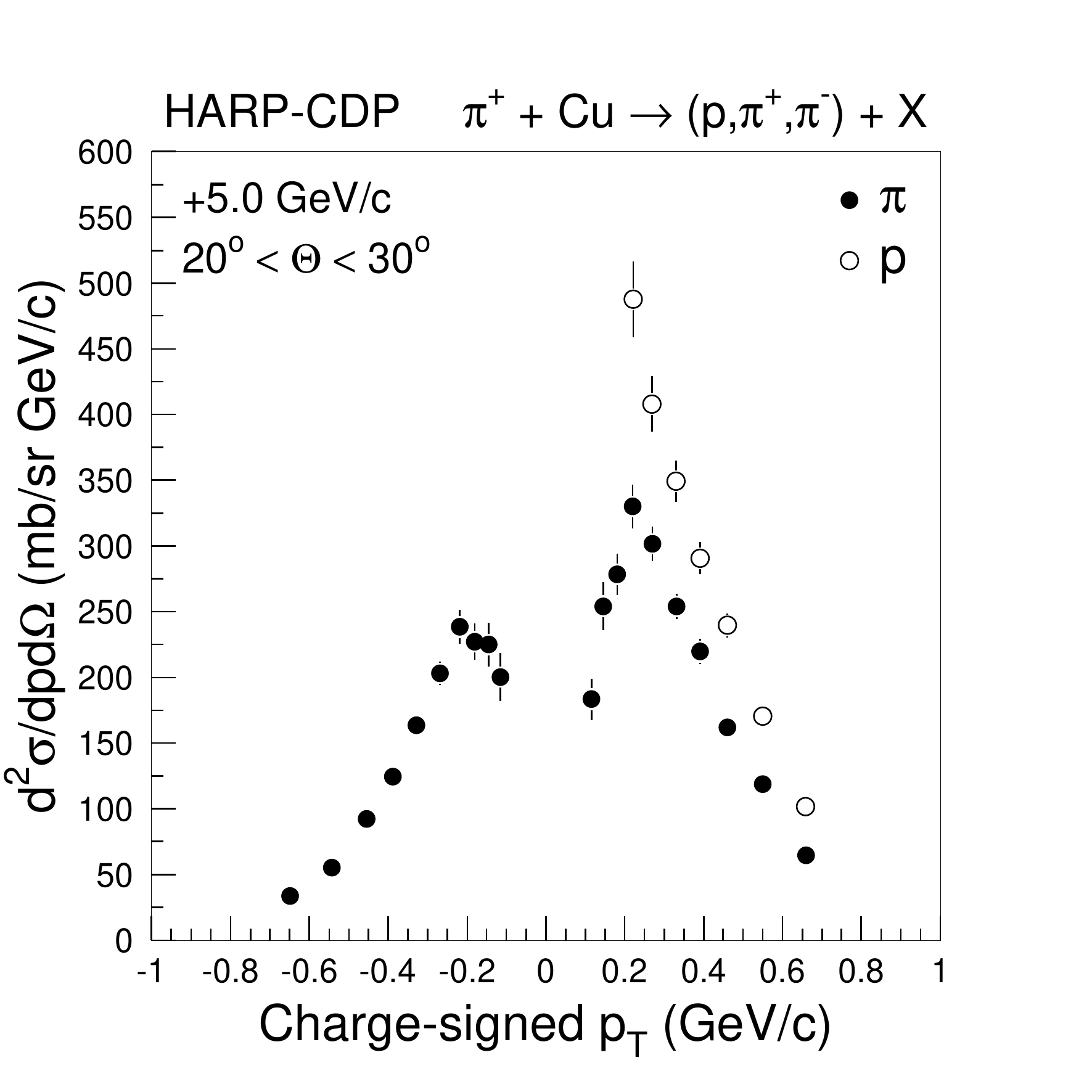} \\
\includegraphics[height=0.30\textheight]{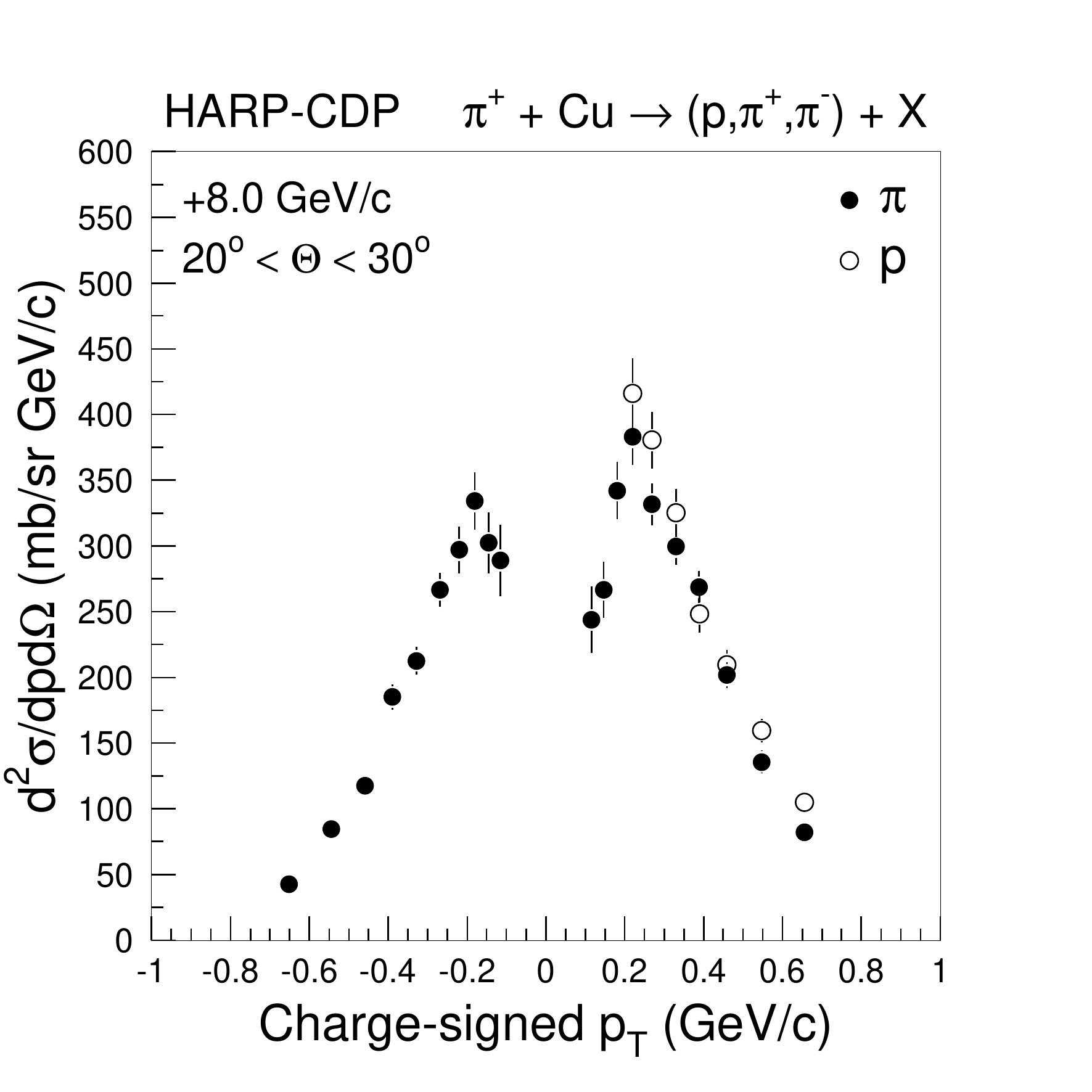} &
\includegraphics[height=0.30\textheight]{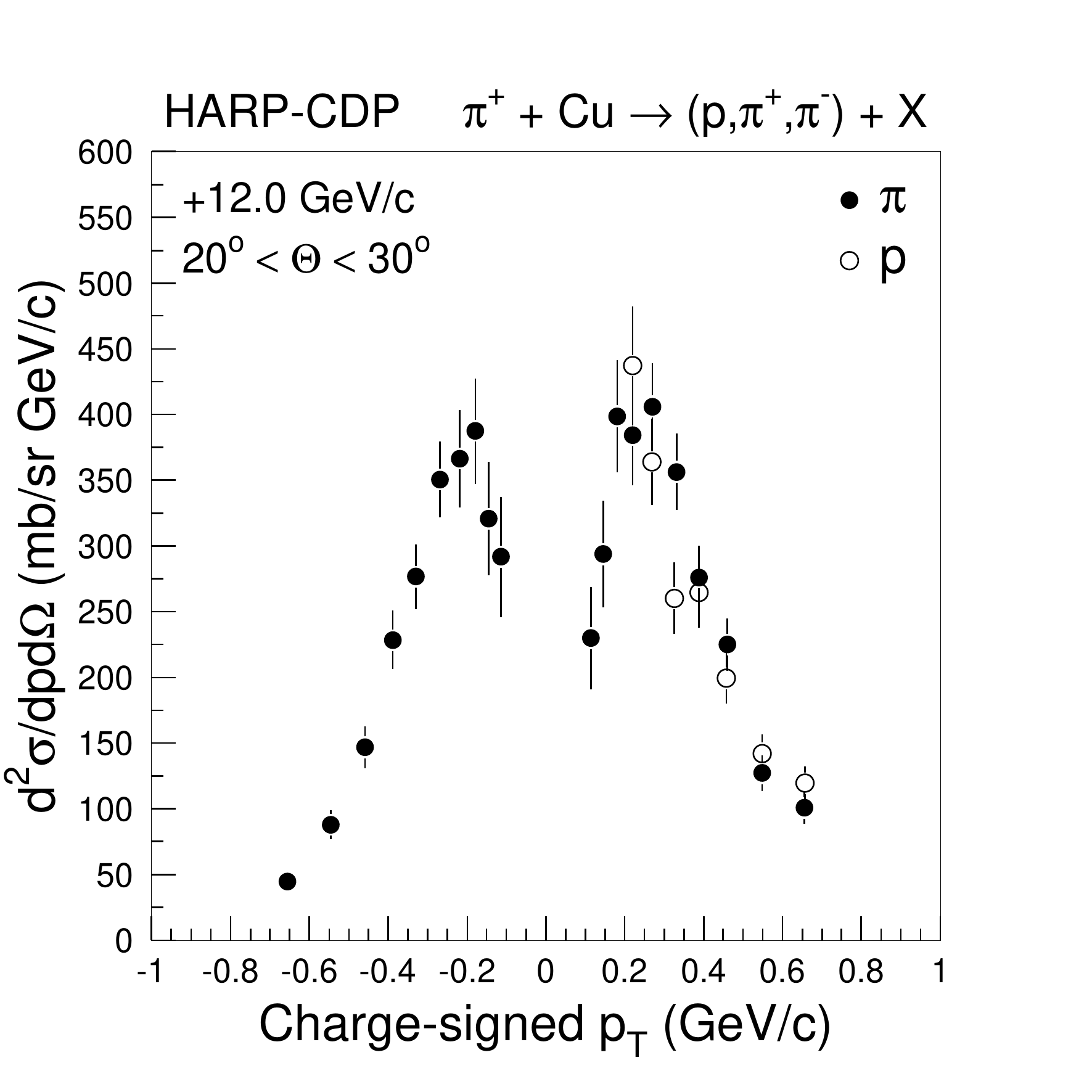} \\
\includegraphics[height=0.30\textheight]{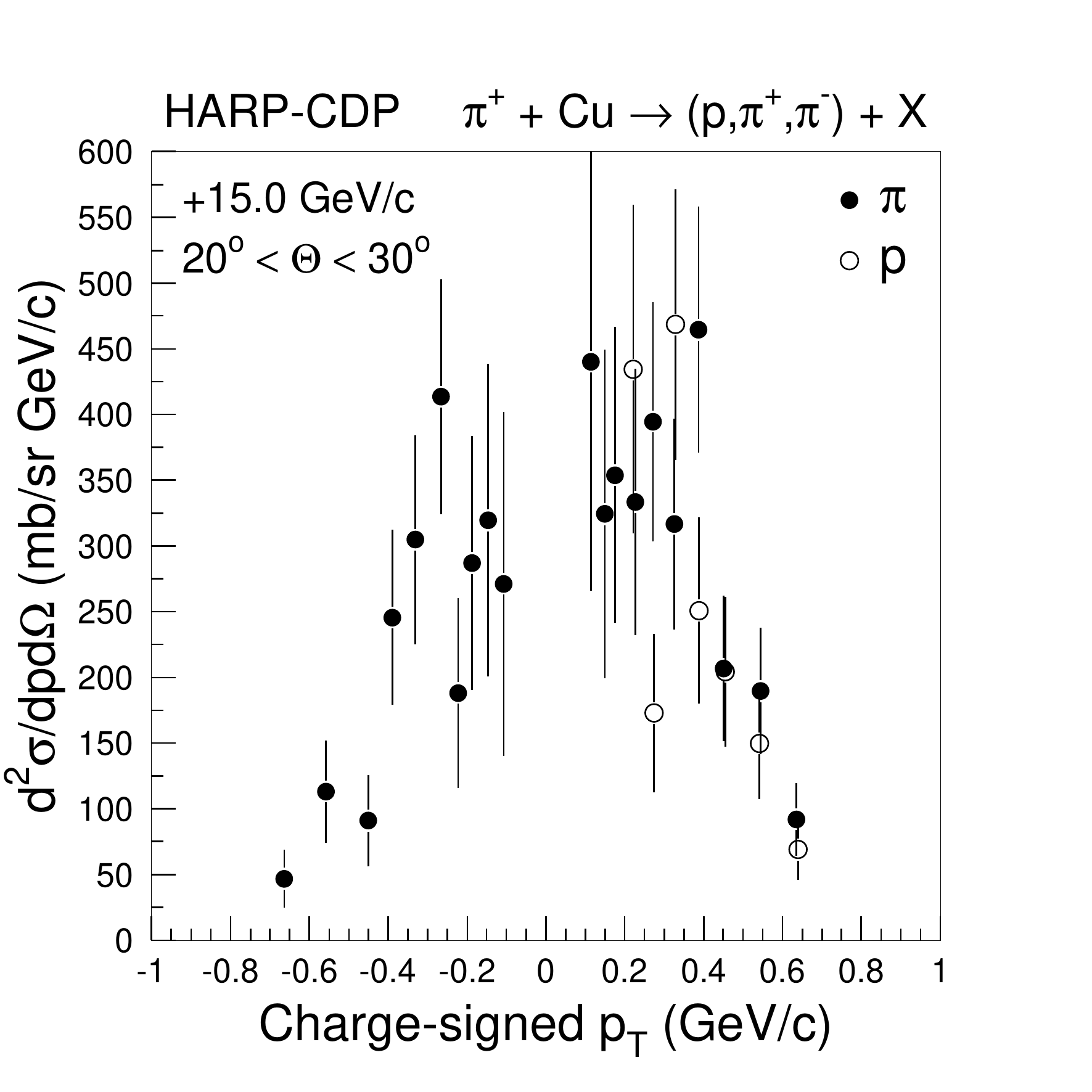} &  \\
\end{tabular}
\caption{Inclusive cross-sections of the production of secondary
protons, $\pi^+$'s, and $\pi^-$'s, by $\pi^+$'s on copper nuclei, 
in the polar-angle range $20^\circ < \theta < 30^\circ$, for
different $\pi^+$ beam momenta, as a function of the charge-signed 
$p_{\rm T}$ of the secondaries; the shown errors are total errors.}  
\label{xsvsmompip}
\end{center}
\end{figure*}

\begin{figure*}[h]
\begin{center}
\begin{tabular}{cc}
\includegraphics[height=0.30\textheight]{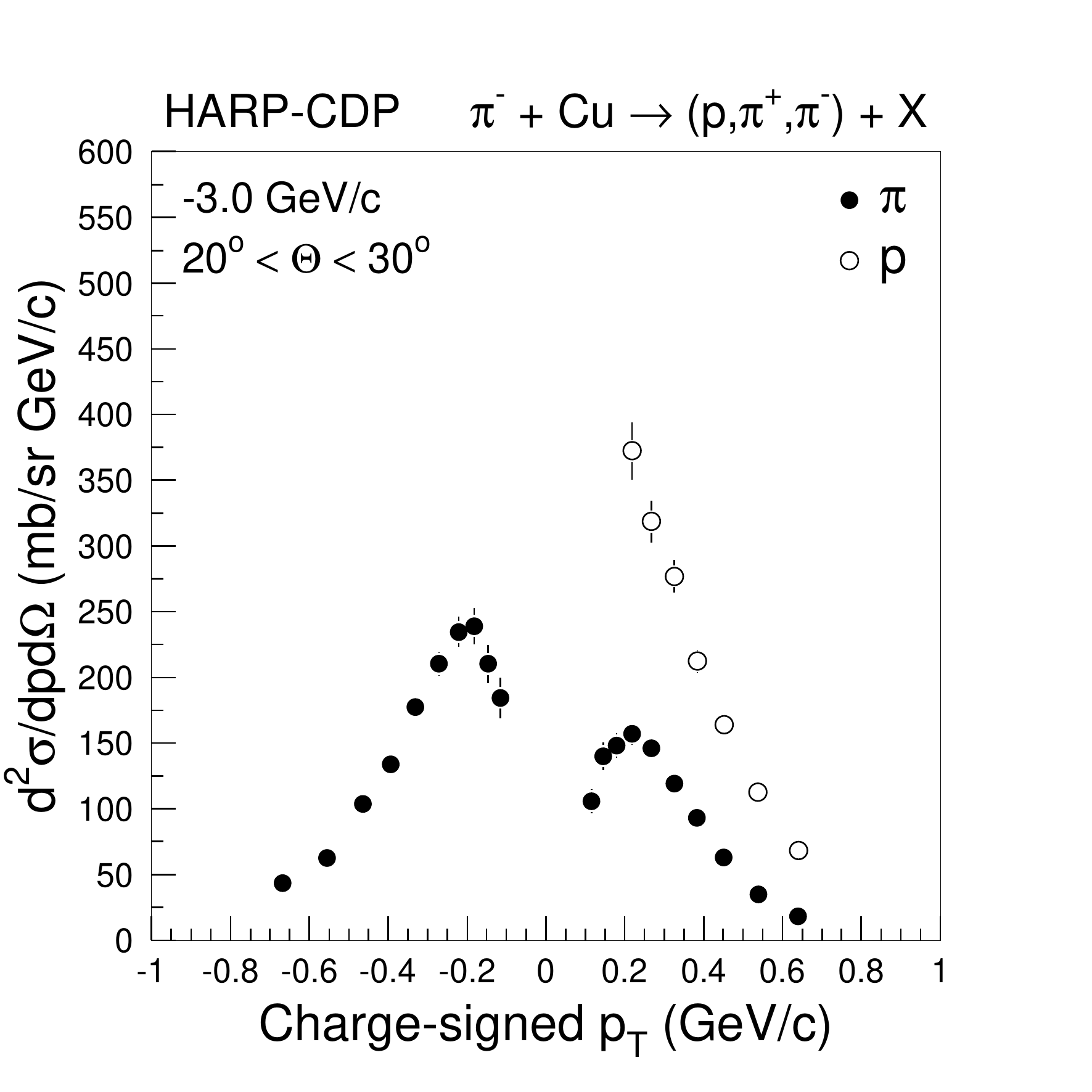} &
\includegraphics[height=0.30\textheight]{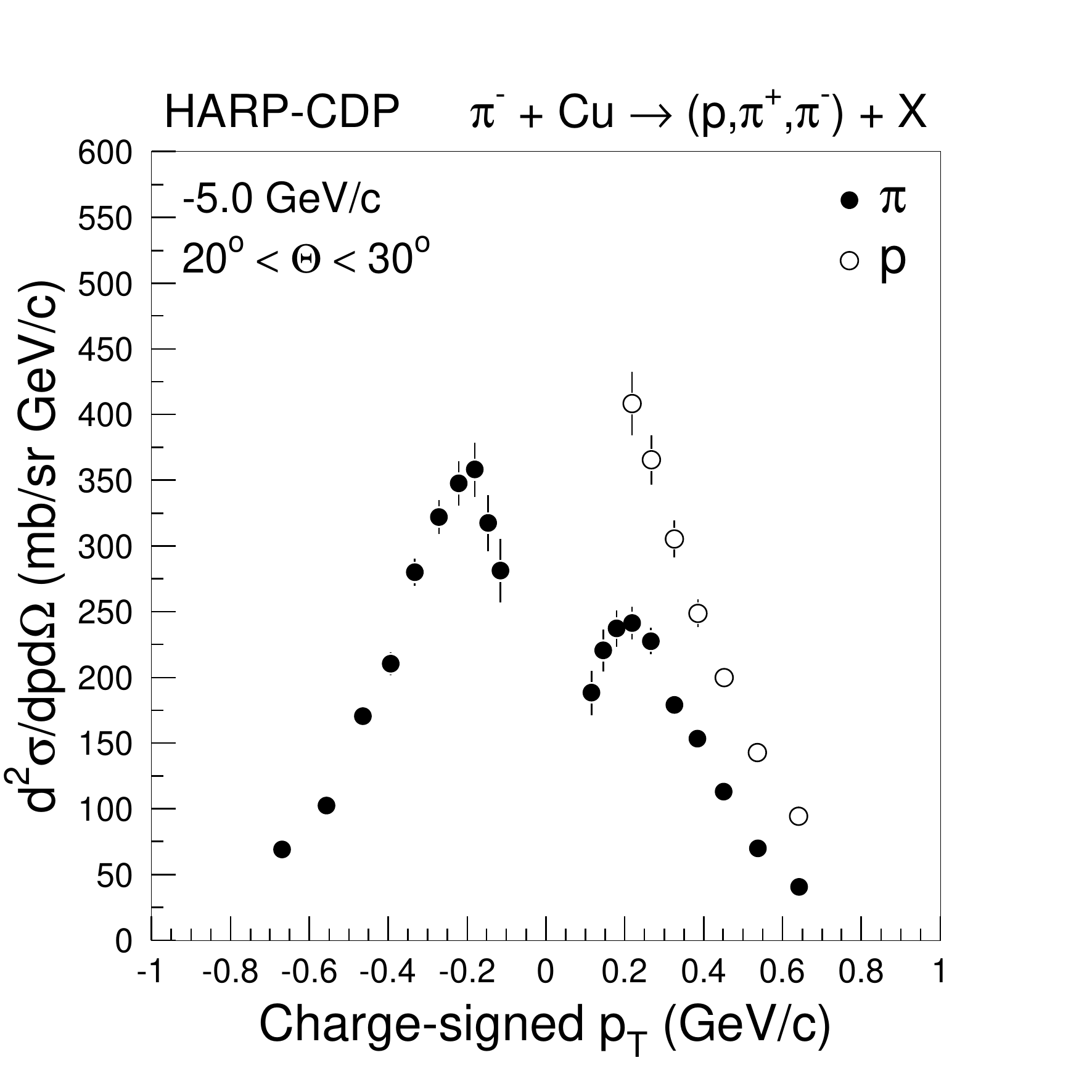} \\
\includegraphics[height=0.30\textheight]{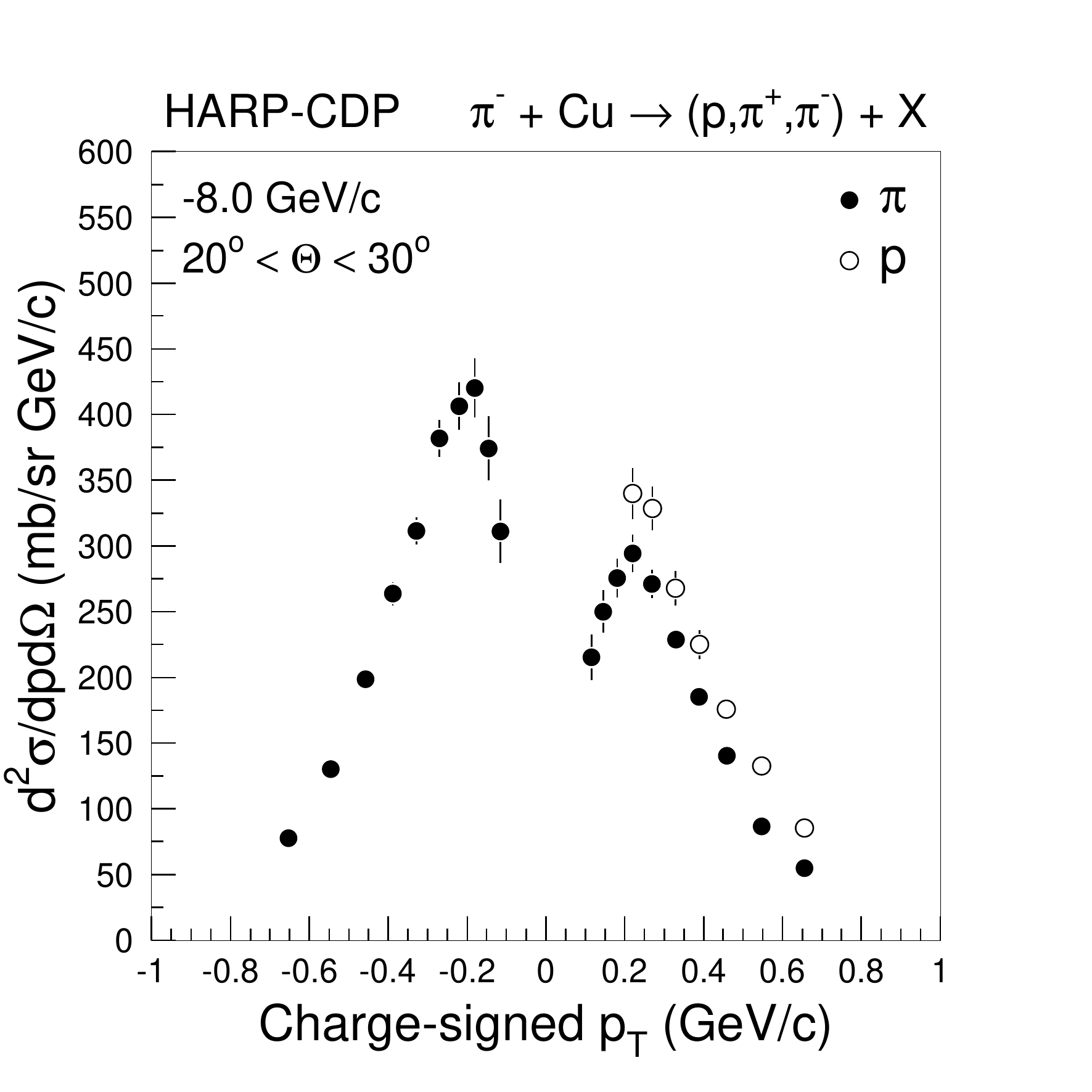} &
\includegraphics[height=0.30\textheight]{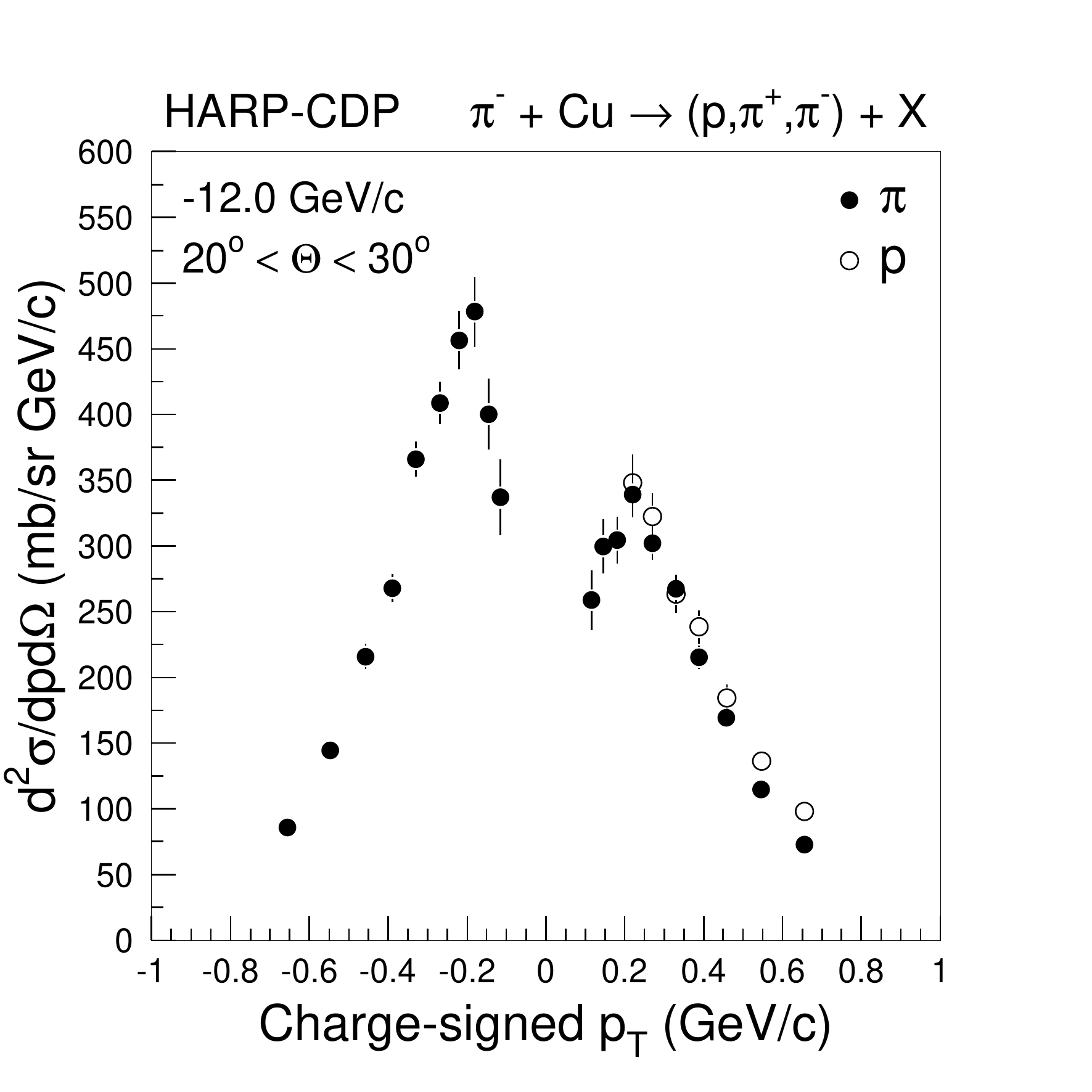} \\
\includegraphics[height=0.30\textheight]{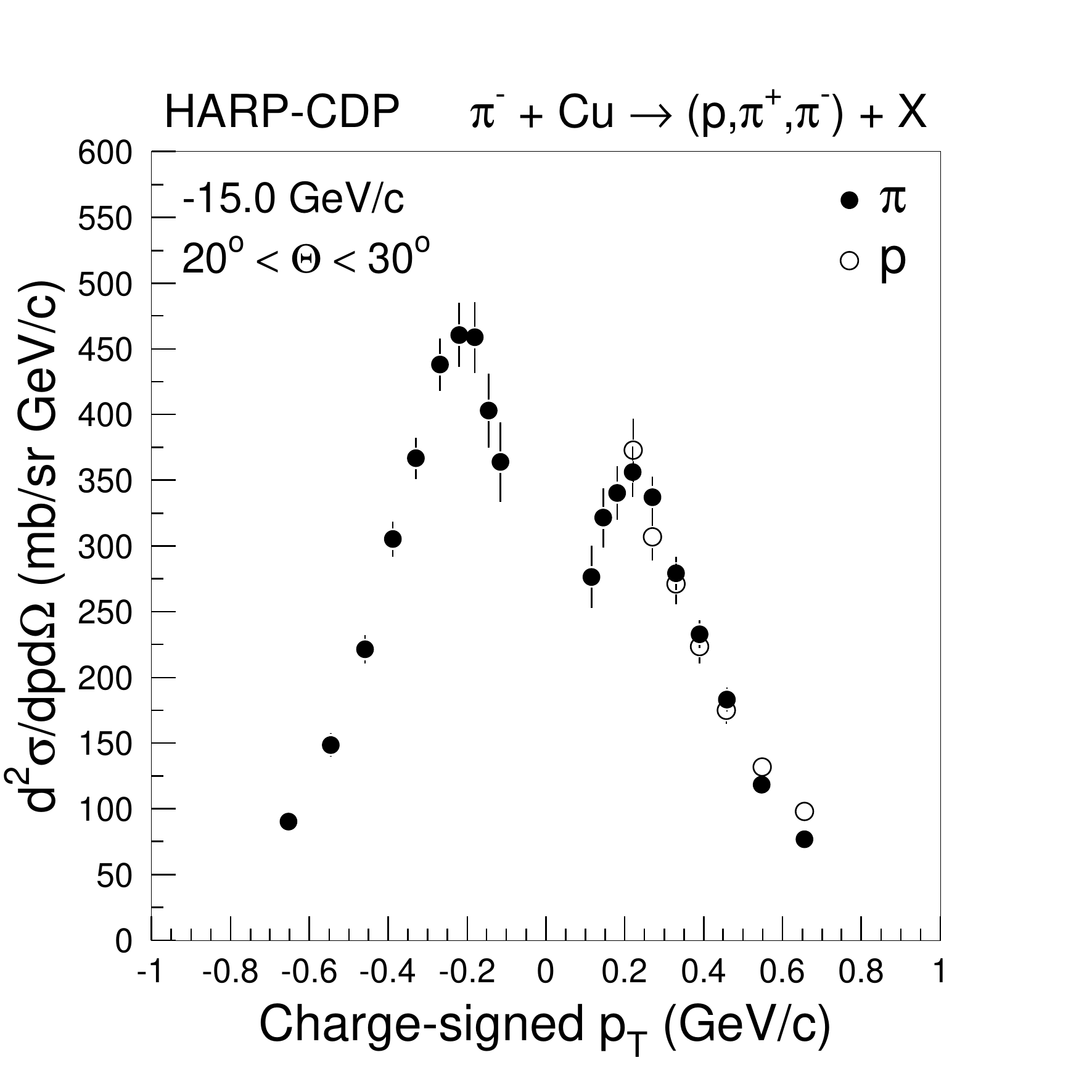} &  \\
\end{tabular}
\caption{Inclusive cross-sections of the production of secondary
protons, $\pi^+$'s, and $\pi^-$'s, by $\pi^-$'s on copper nuclei, 
in the polar-angle range $20^\circ < \theta < 30^\circ$, for
different $\pi^-$ beam momenta, as a function of the charge-signed 
$p_{\rm T}$ of the secondaries; the shown errors are total errors.} 
\label{xsvsmompim}
\end{center}
\end{figure*}

\begin{figure*}[h]
\begin{center}
\begin{tabular}{cc}
\includegraphics[height=0.30\textheight]{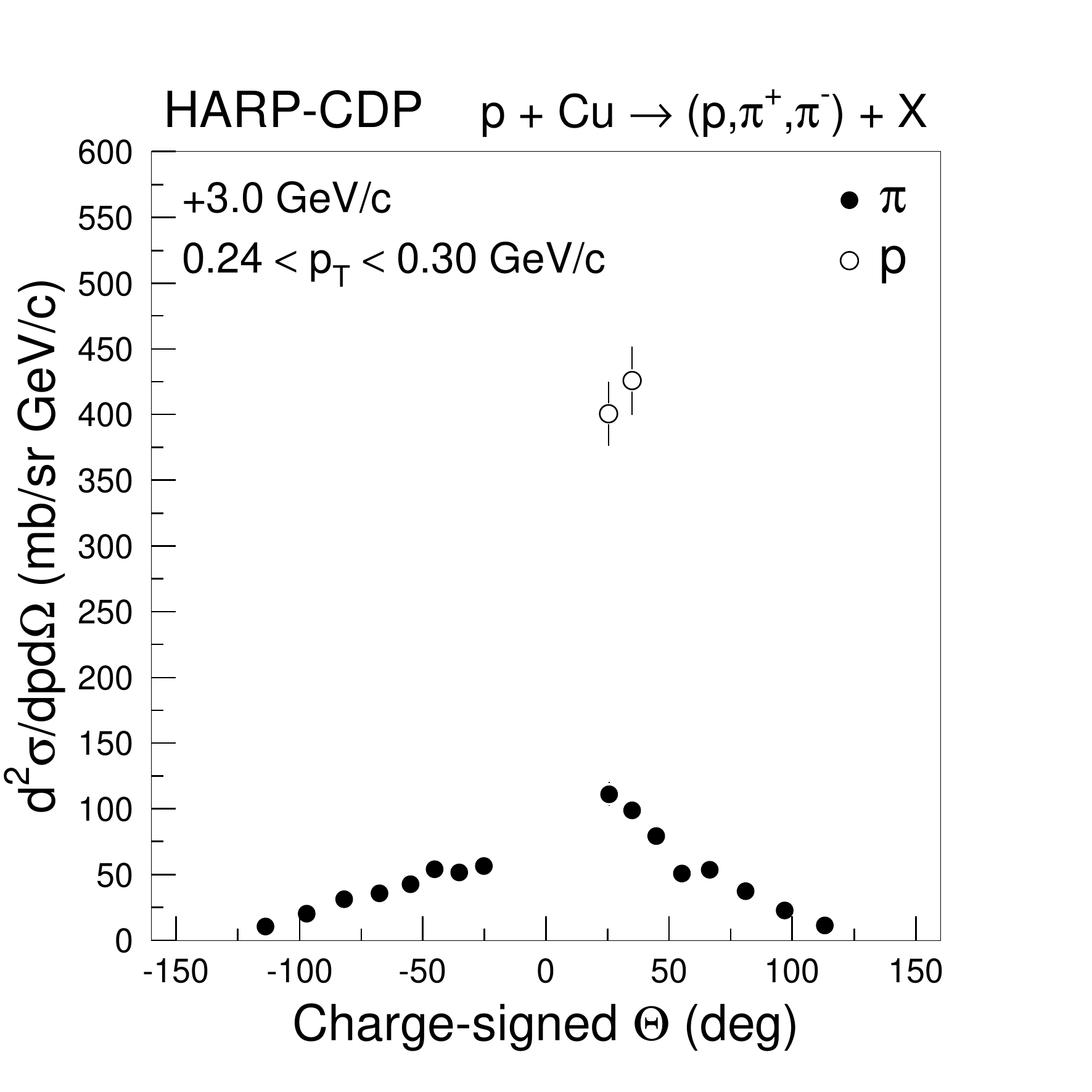} &
\includegraphics[height=0.30\textheight]{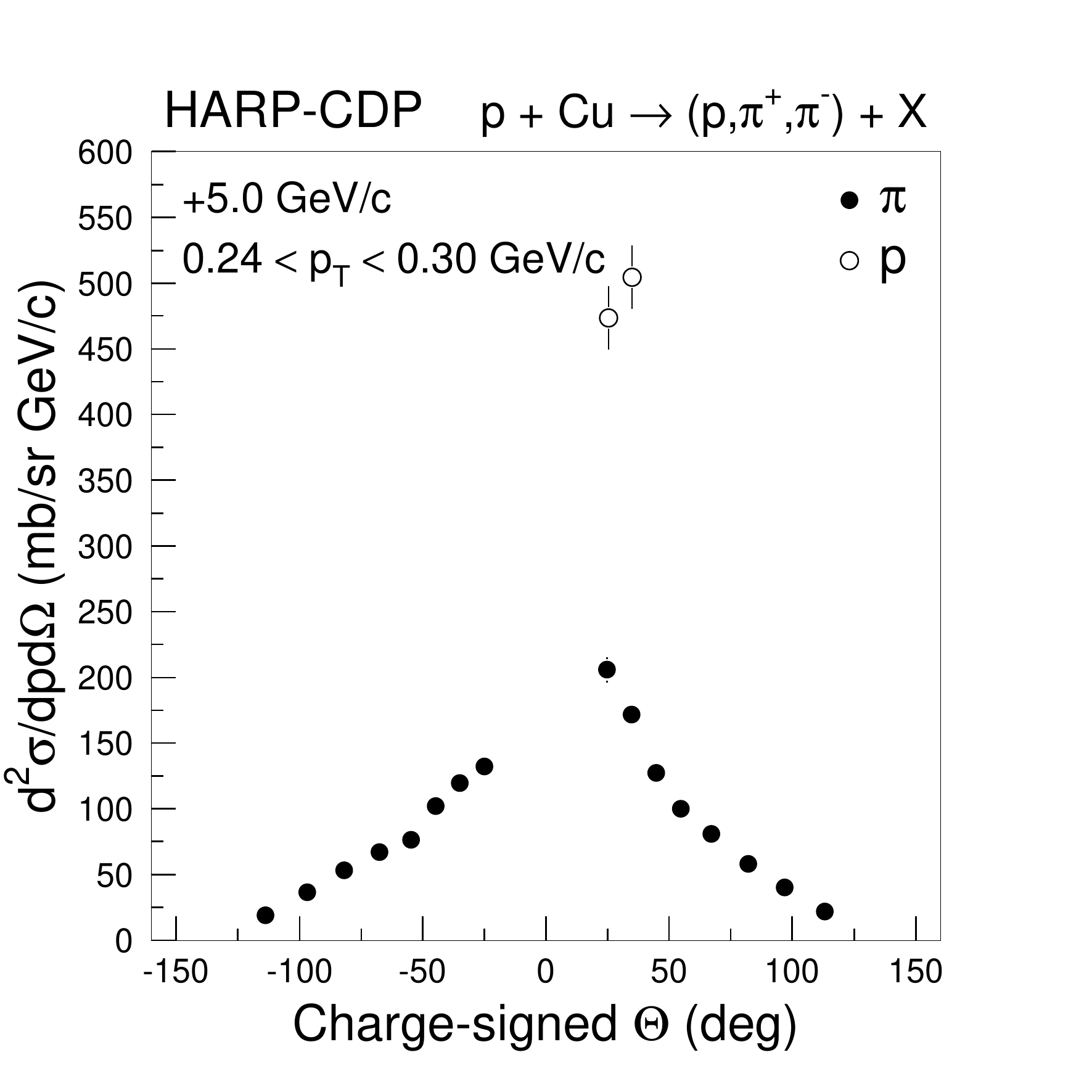} \\
\includegraphics[height=0.30\textheight]{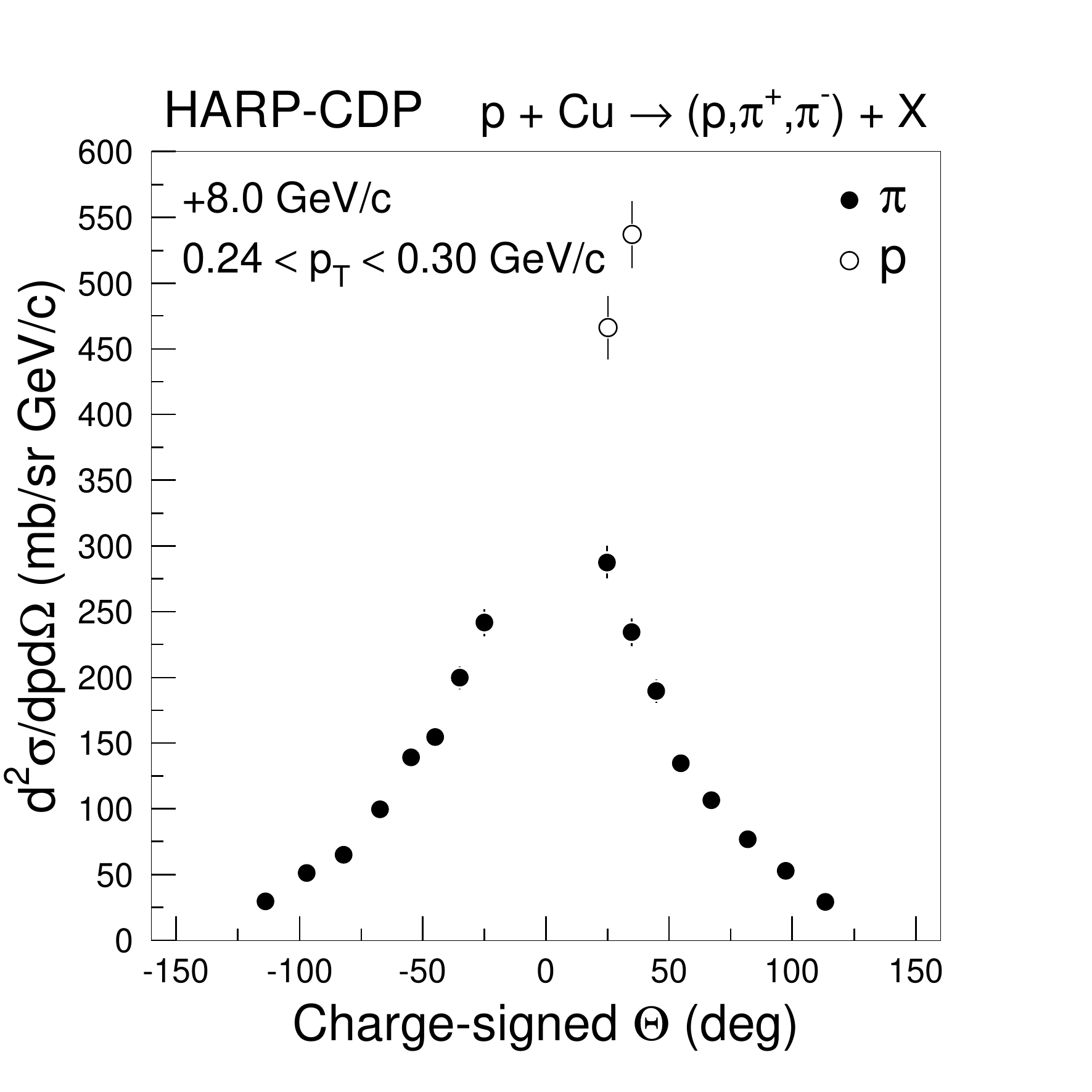} &
\includegraphics[height=0.30\textheight]{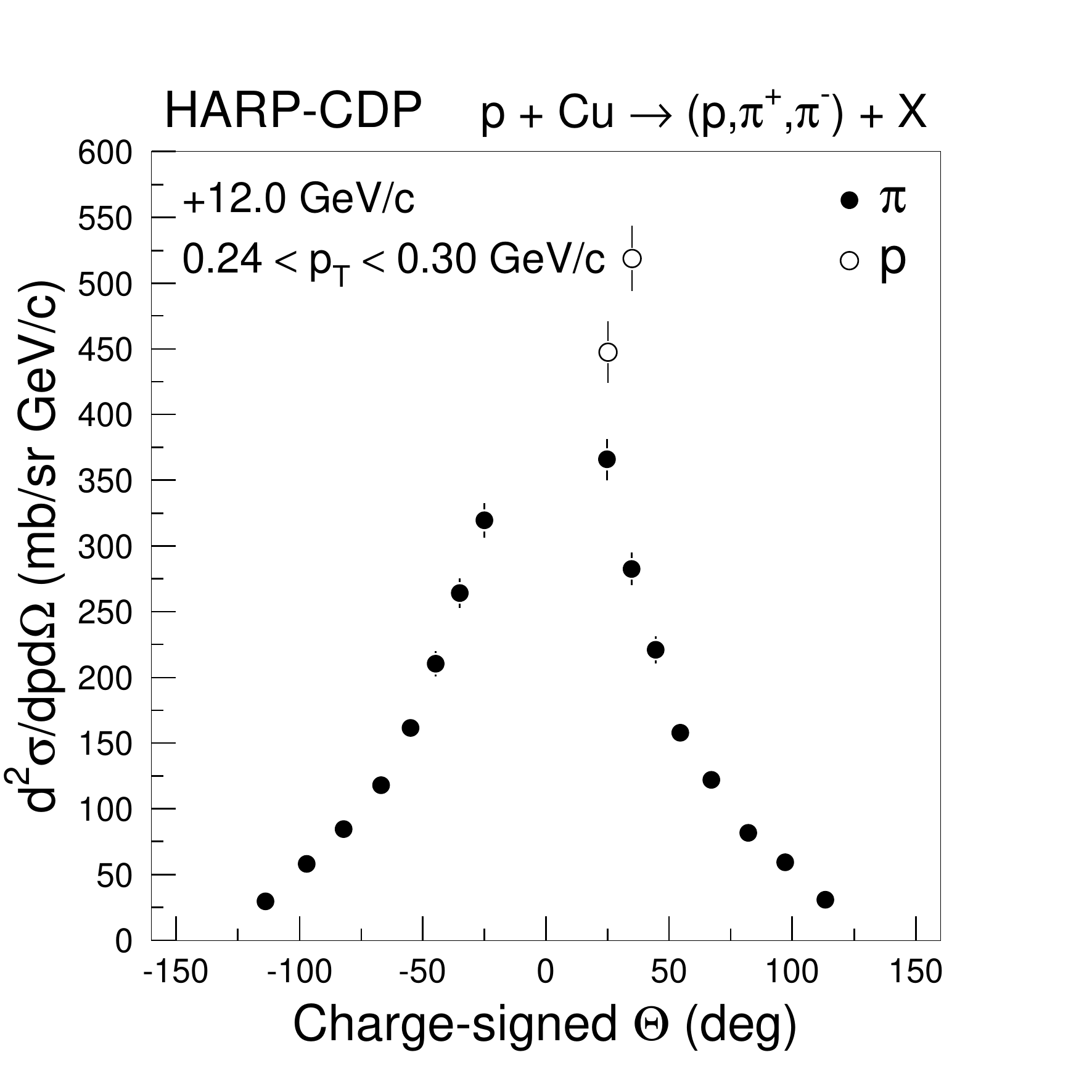} \\
\includegraphics[height=0.30\textheight]{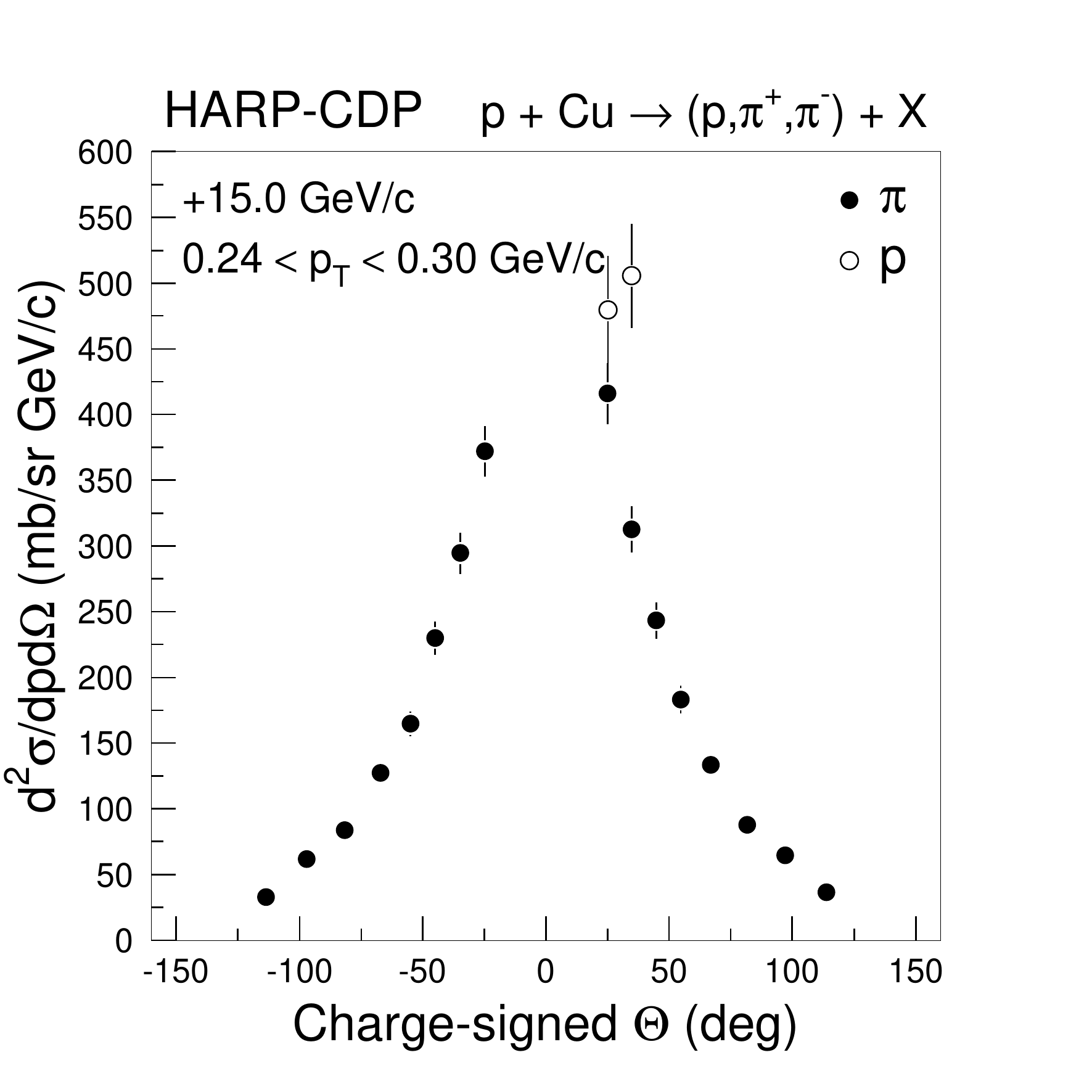} &  \\
\end{tabular}
\caption{Inclusive cross-sections of the production of secondary
protons, $\pi^+$'s, and $\pi^-$'s, with $p_{\rm T}$ in the range 
0.24--0.30~GeV/{\it c}, by protons on copper nuclei, for
different proton beam momenta, as a function of the charge-signed 
polar angle $\theta$ of the secondaries; the shown errors are 
total errors.} 
\label{xsvsthetapro}
\end{center}
\end{figure*}

\begin{figure*}[h]
\begin{center}
\begin{tabular}{cc}
\includegraphics[height=0.30\textheight]{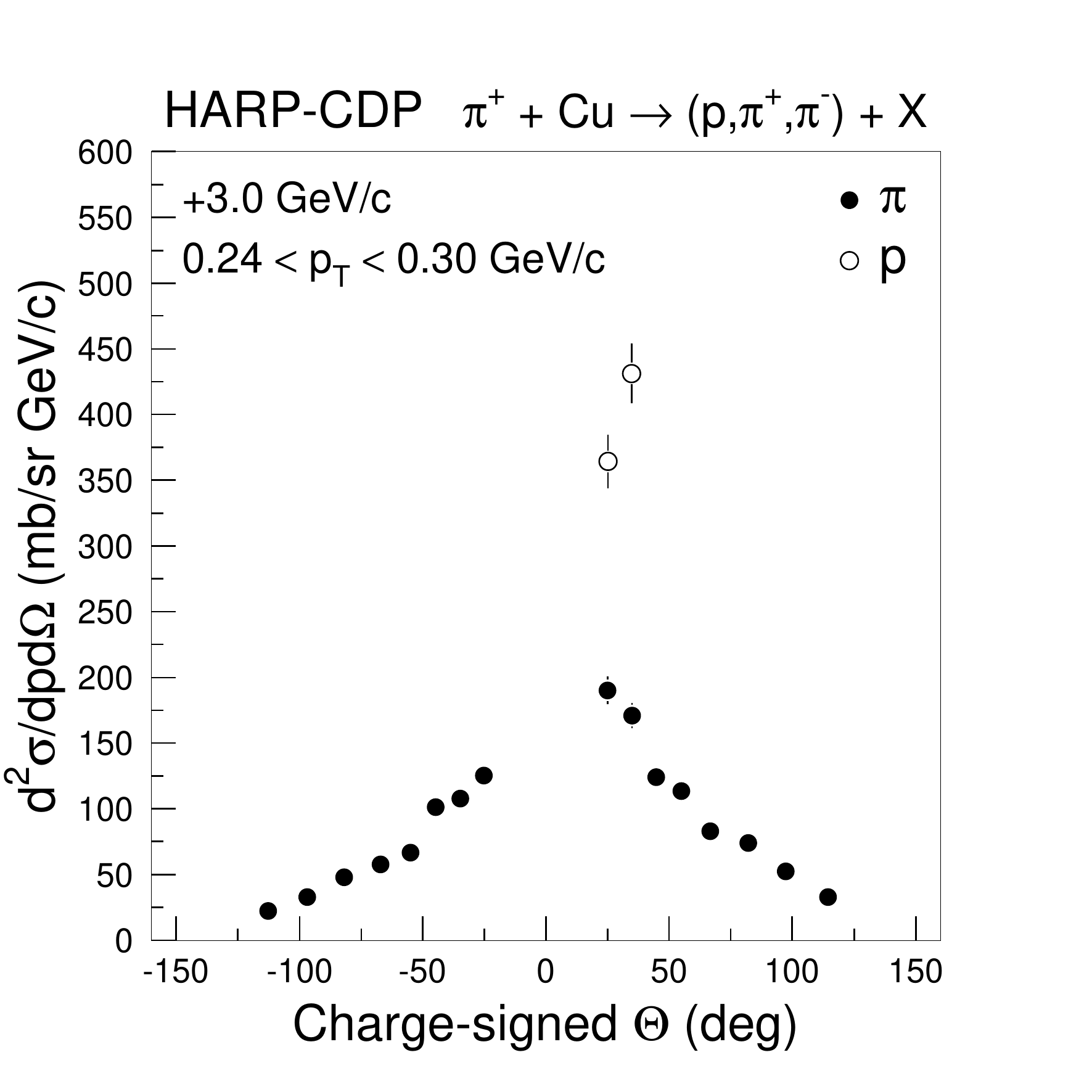} &
\includegraphics[height=0.30\textheight]{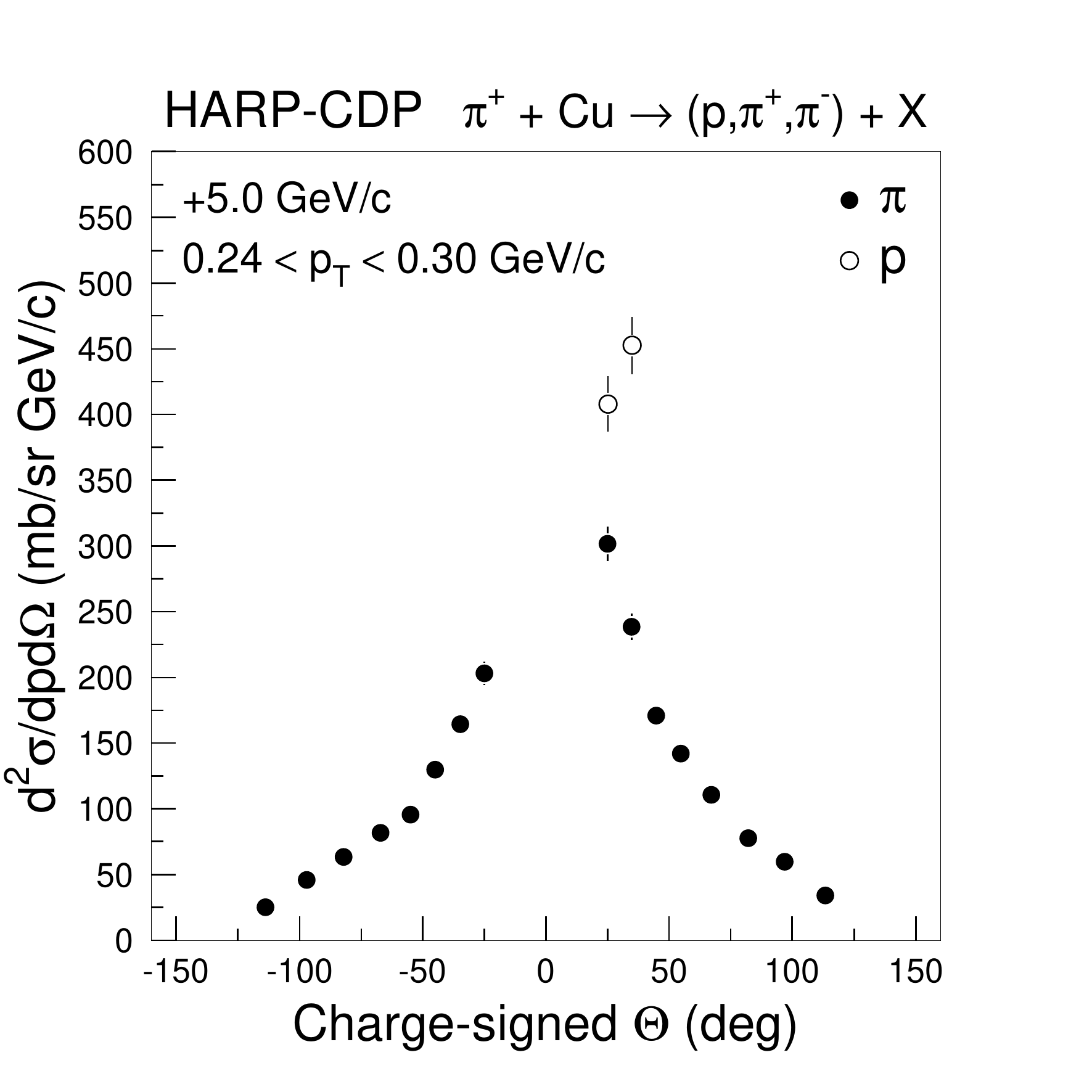} \\
\includegraphics[height=0.30\textheight]{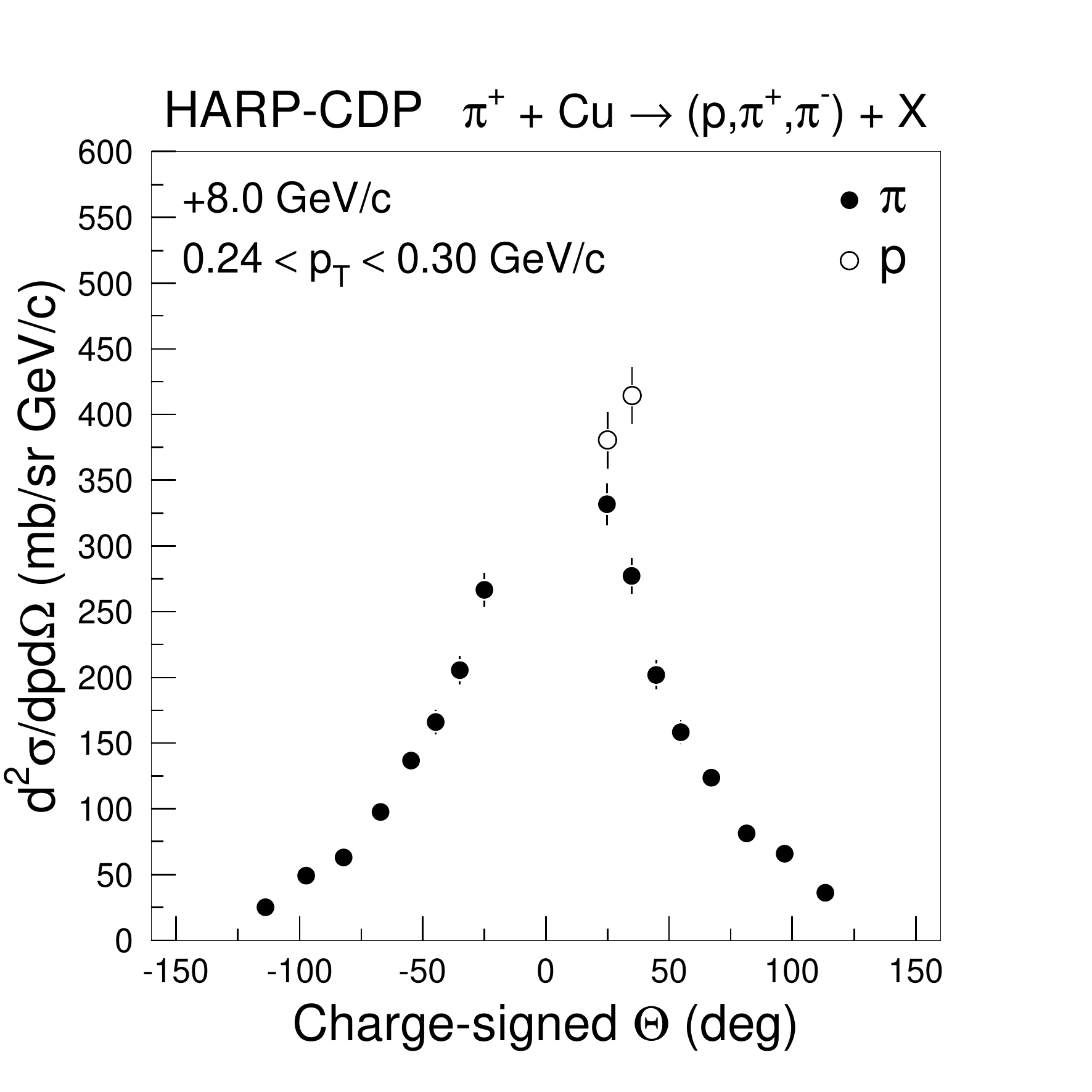} &
\includegraphics[height=0.30\textheight]{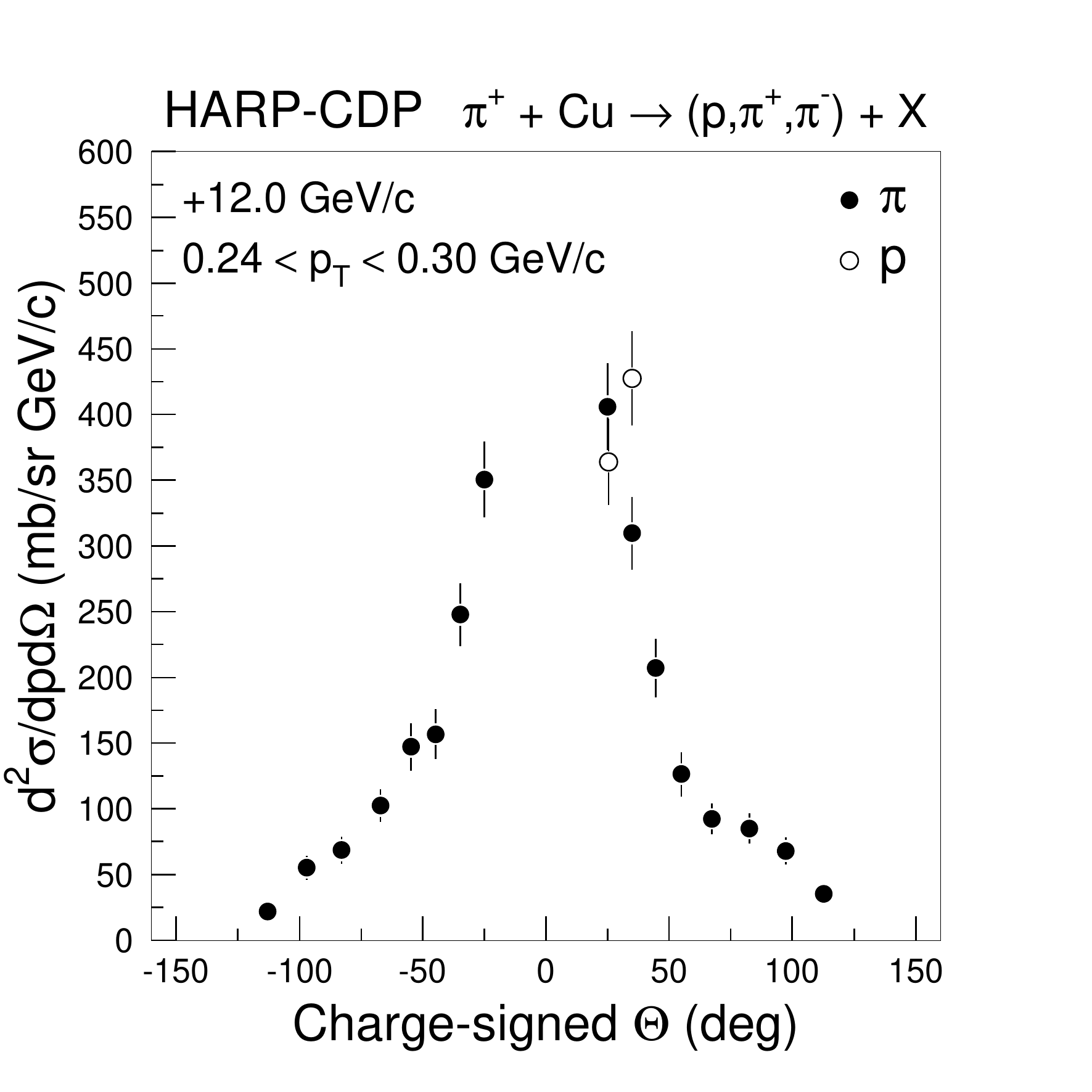} \\
\includegraphics[height=0.30\textheight]{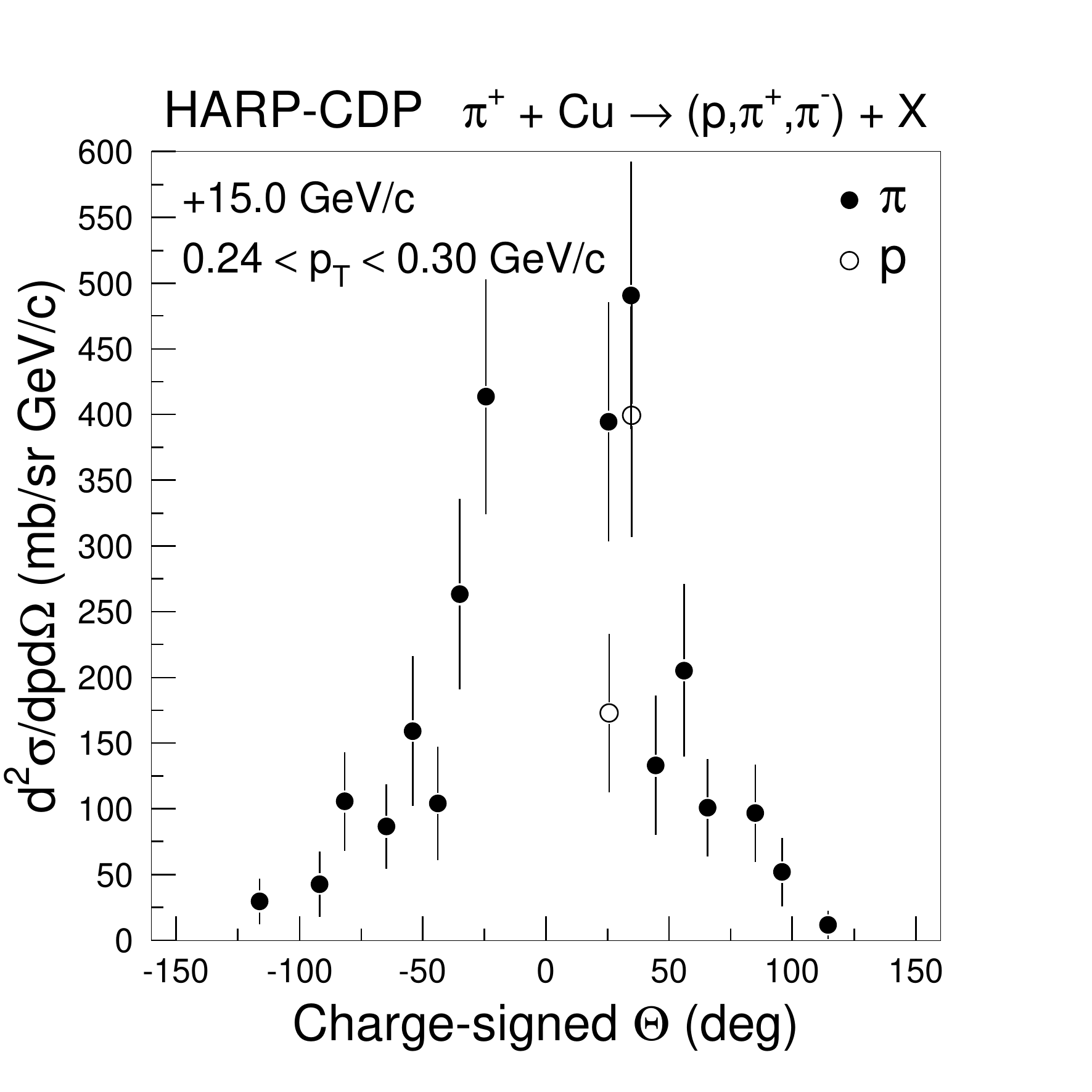} &  \\
\end{tabular}
\caption{Inclusive cross-sections of the production of secondary
protons, $\pi^+$'s, and $\pi^-$'s, with $p_{\rm T}$ in the range 
0.24--0.30~GeV/{\it c}, by $\pi^+$'s on copper nuclei, for
different $\pi^+$ beam momenta, as a function of the charge-signed 
polar angle $\theta$ of the secondaries; the shown errors are 
total errors.}  
\label{xsvsthetapip}
\end{center}
\end{figure*}

\begin{figure*}[h]
\begin{center}
\begin{tabular}{cc}
\includegraphics[height=0.30\textheight]{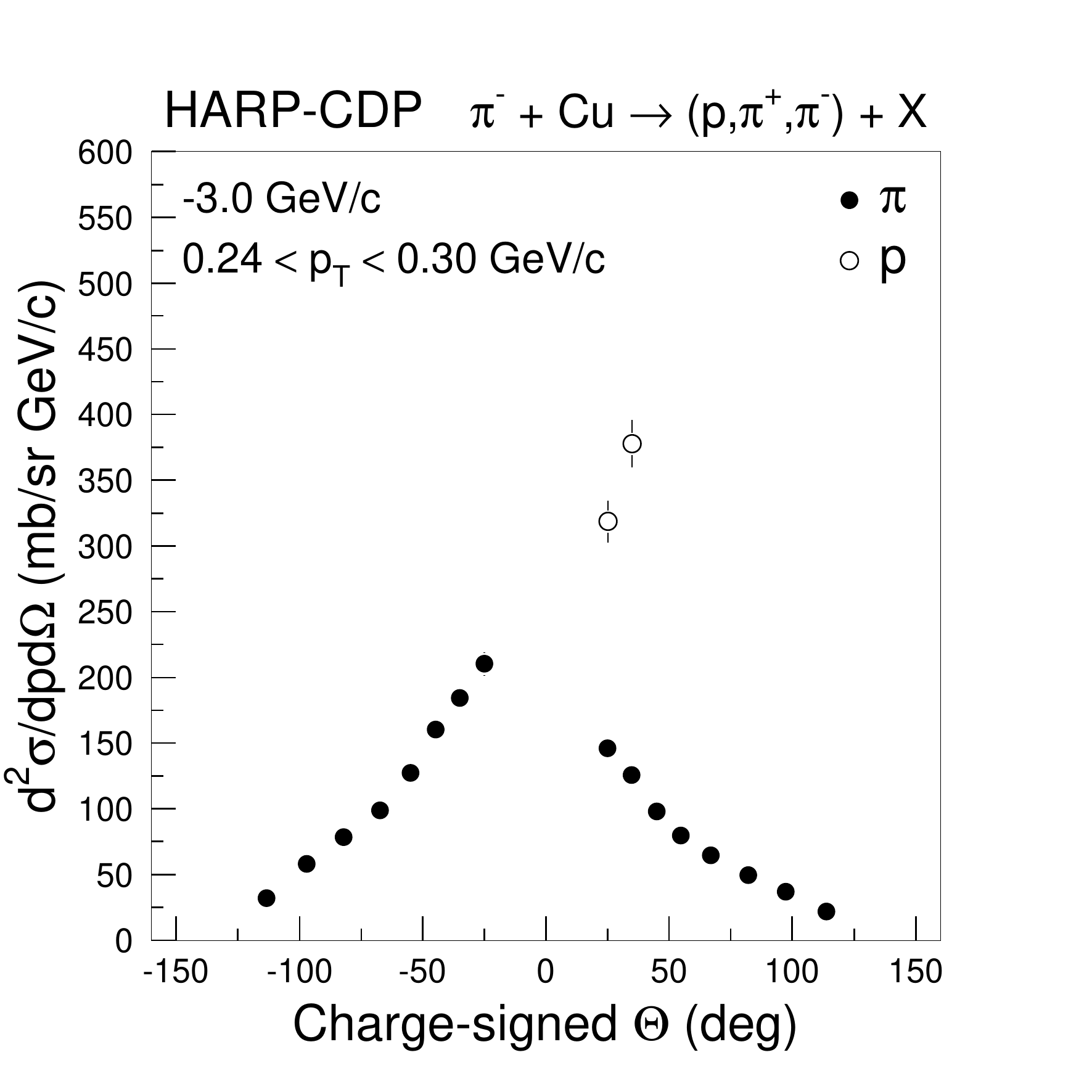} &
\includegraphics[height=0.30\textheight]{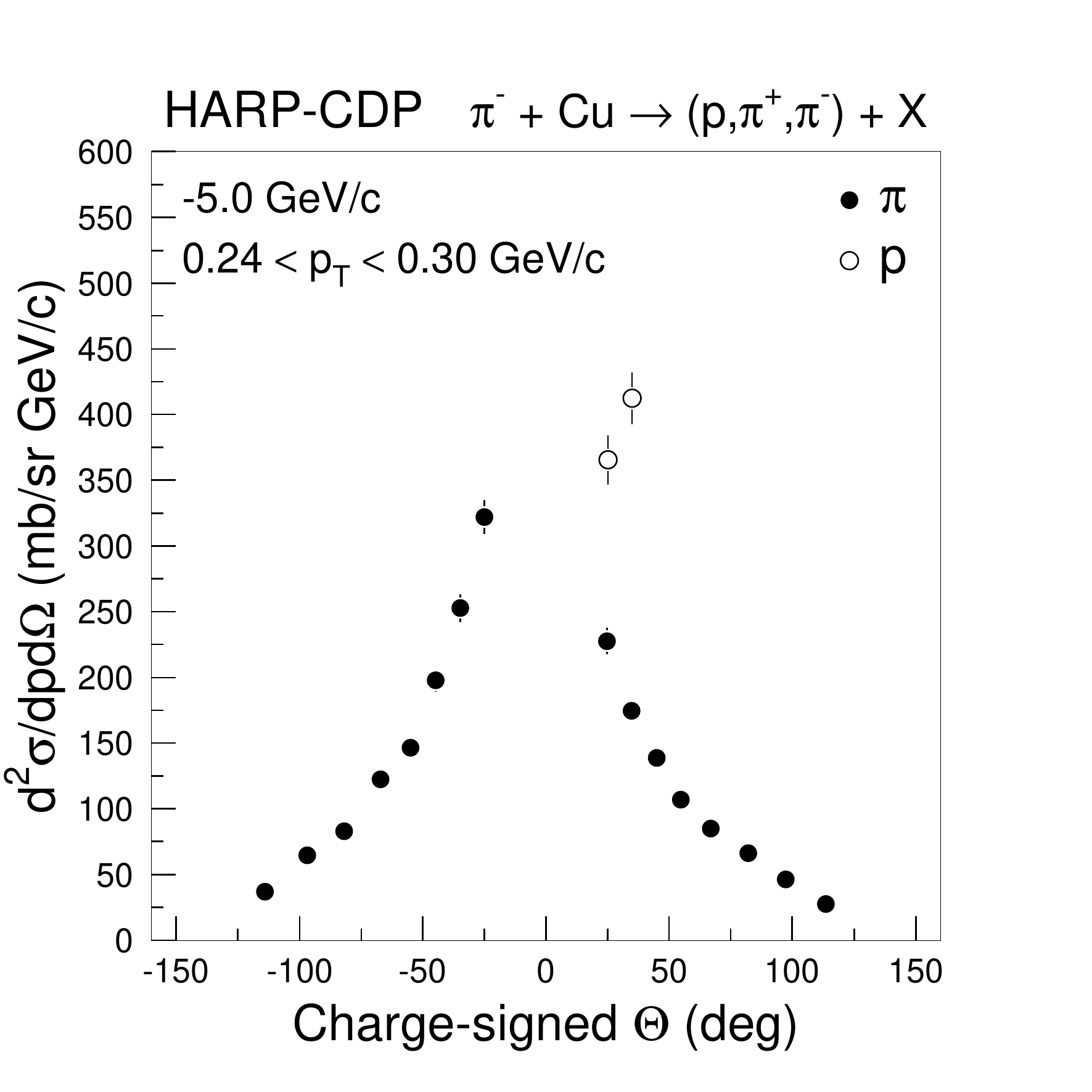} \\
\includegraphics[height=0.30\textheight]{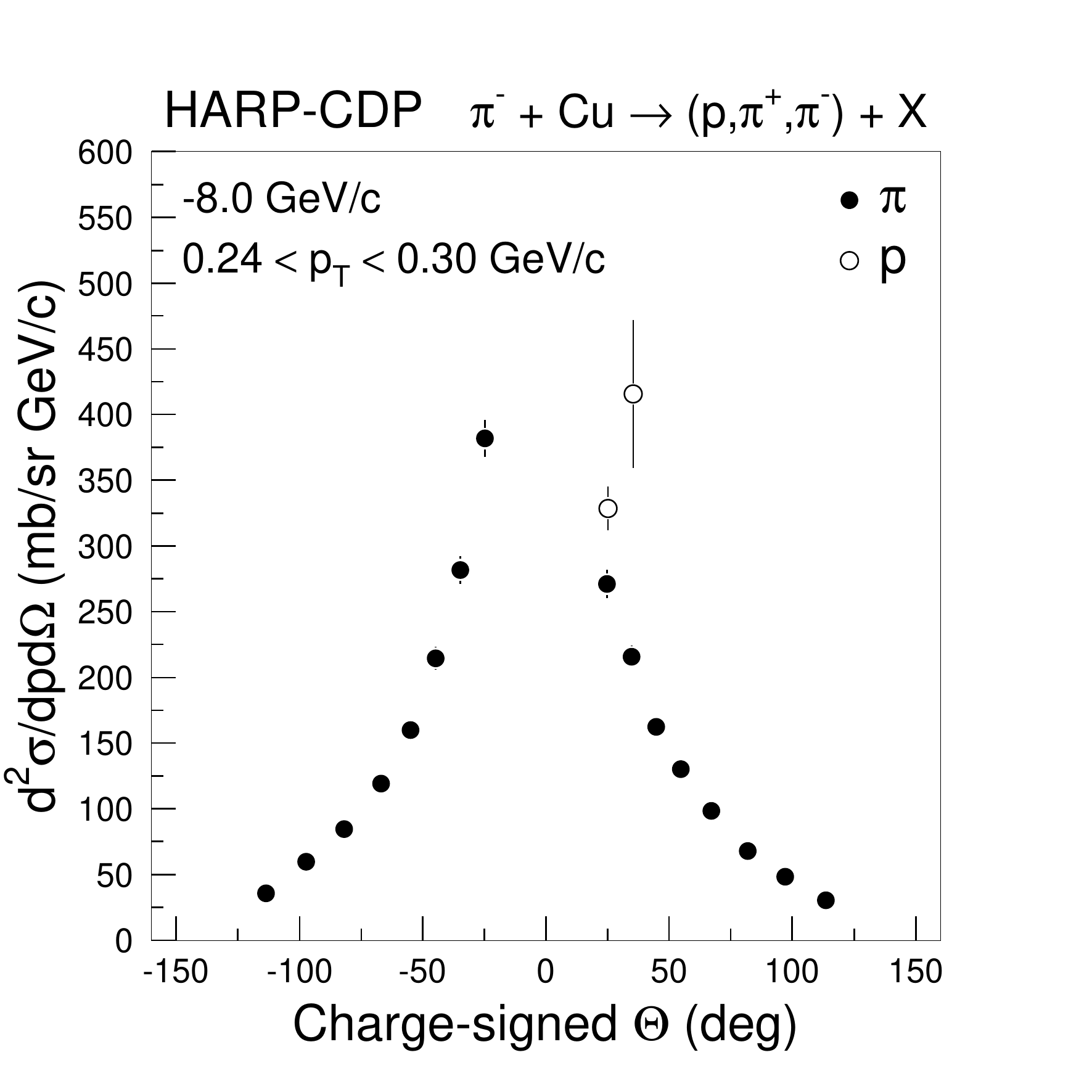} &
\includegraphics[height=0.30\textheight]{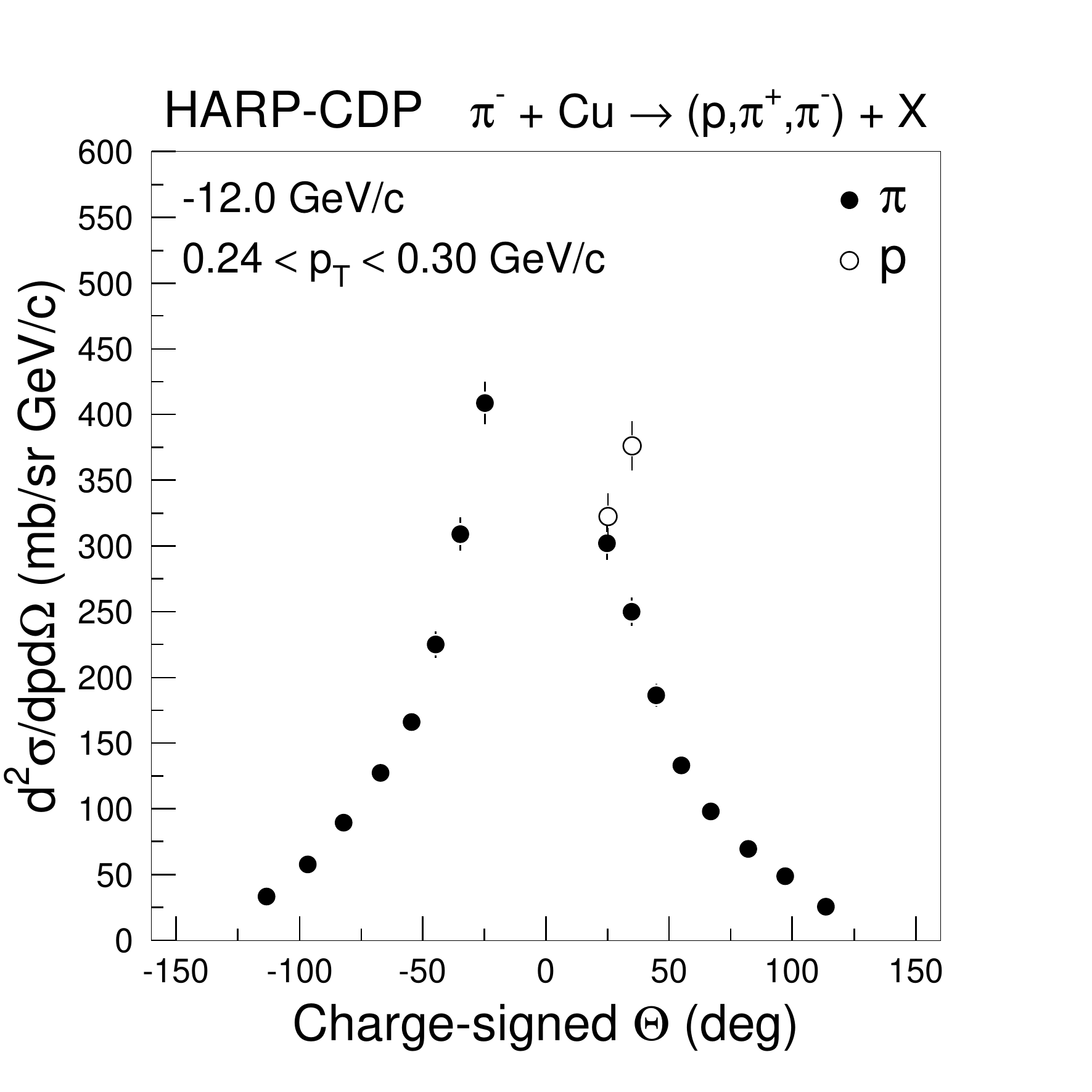} \\
\includegraphics[height=0.30\textheight]{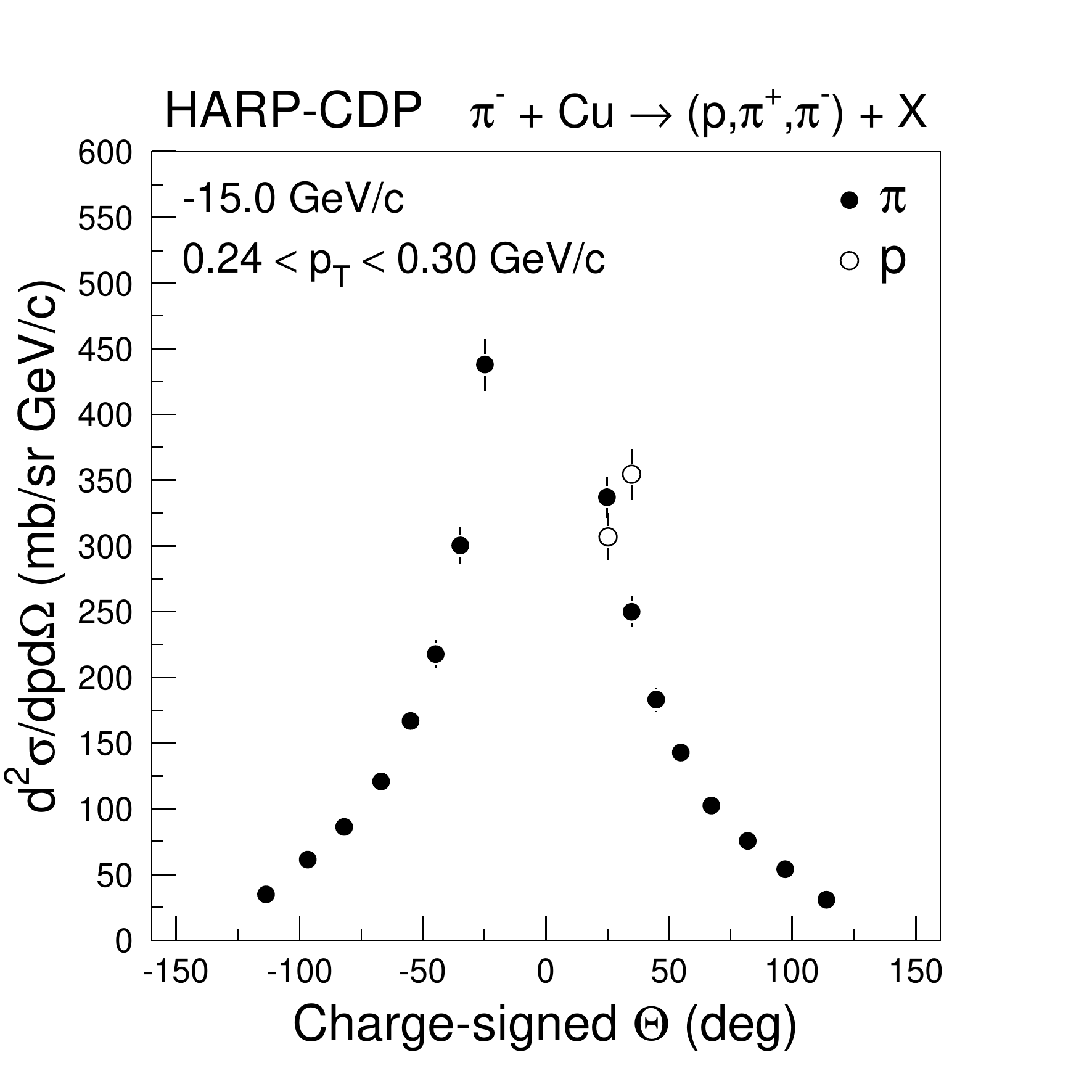} &  \\
\end{tabular}
\caption{Inclusive cross-sections of the production of secondary
protons, $\pi^+$'s, and $\pi^-$'s, with $p_{\rm T}$ in the range 
0.24--0.30~GeV/{\it c}, by $\pi^-$'s on copper nuclei, for
different $\pi^-$ beam momenta, as a function of the charge-signed 
polar angle $\theta$ of the secondaries; the shown errors are 
total errors.} 
\label{xsvsthetapim}
\end{center}
\end{figure*}

\begin{figure*}[h]
\begin{center}
\begin{tabular}{cc}
\includegraphics[height=0.30\textheight]{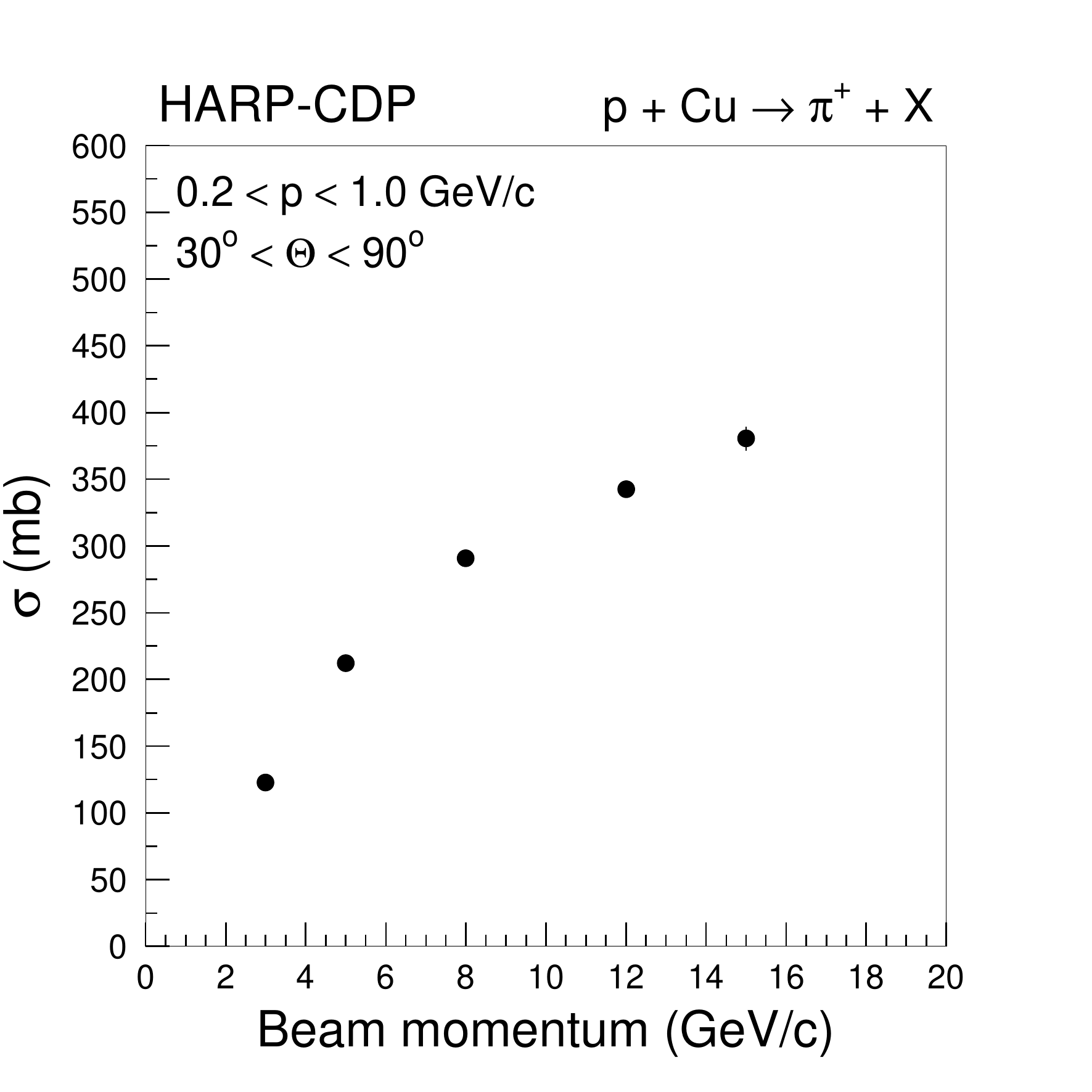} &
\includegraphics[height=0.30\textheight]{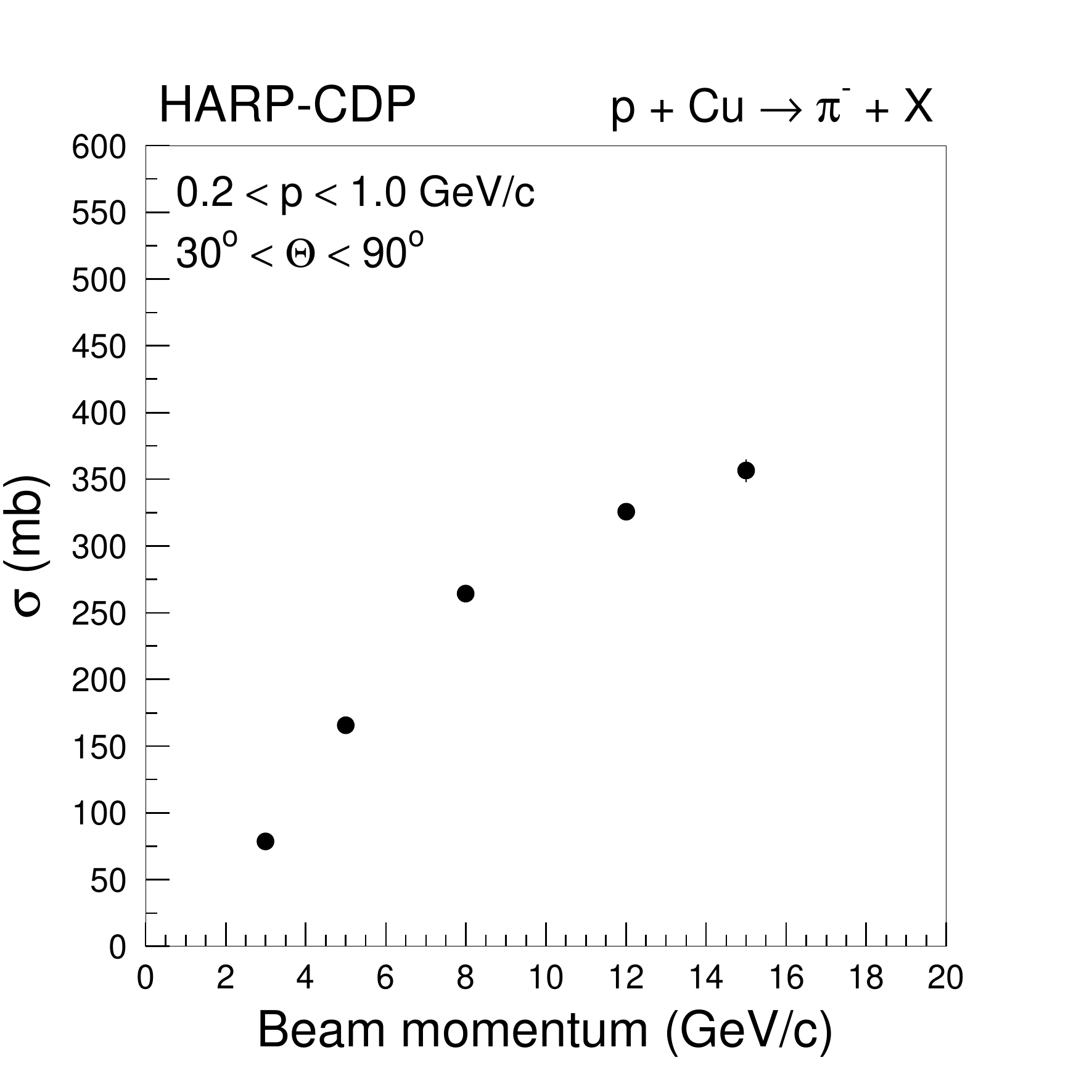} \\
\includegraphics[height=0.30\textheight]{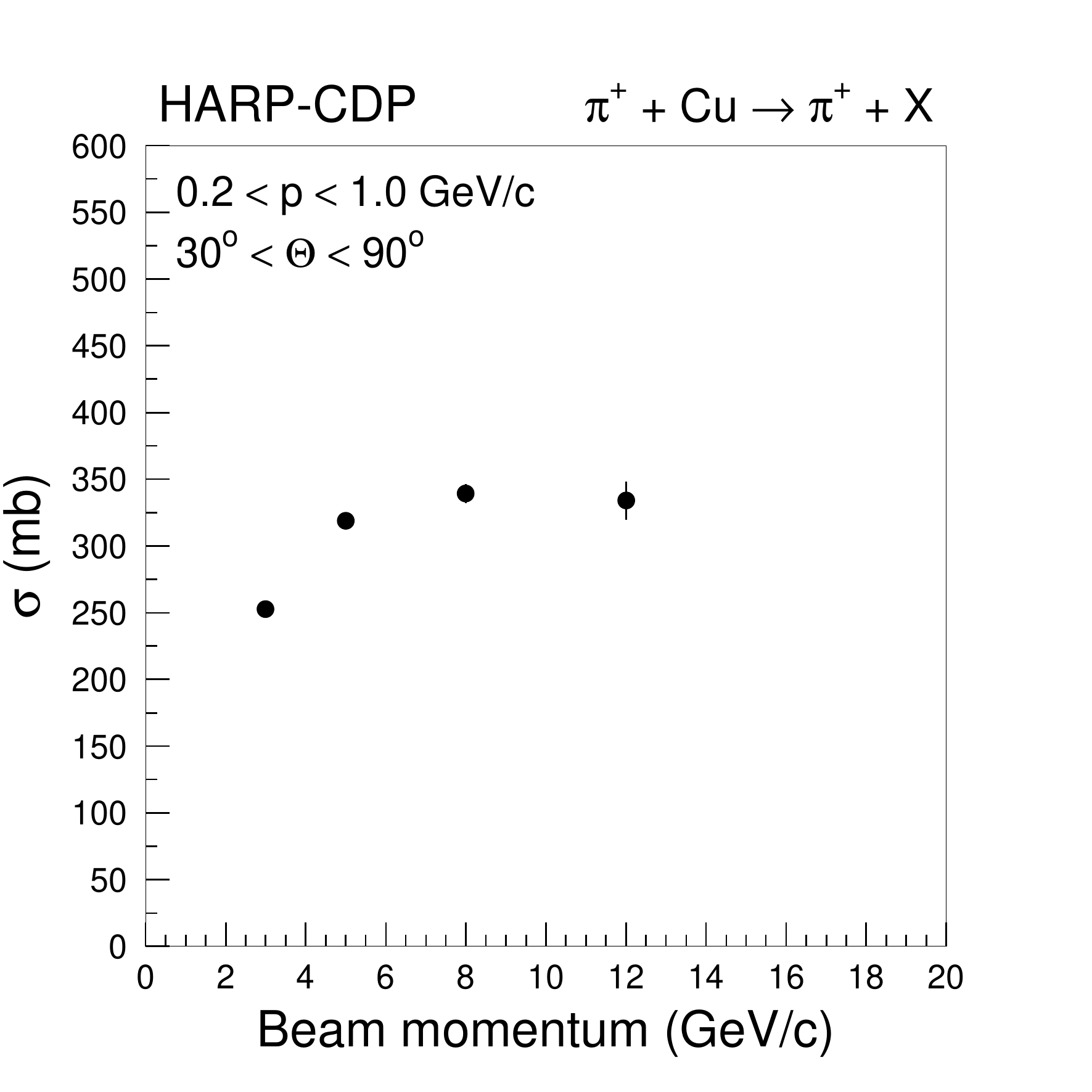} &
\includegraphics[height=0.30\textheight]{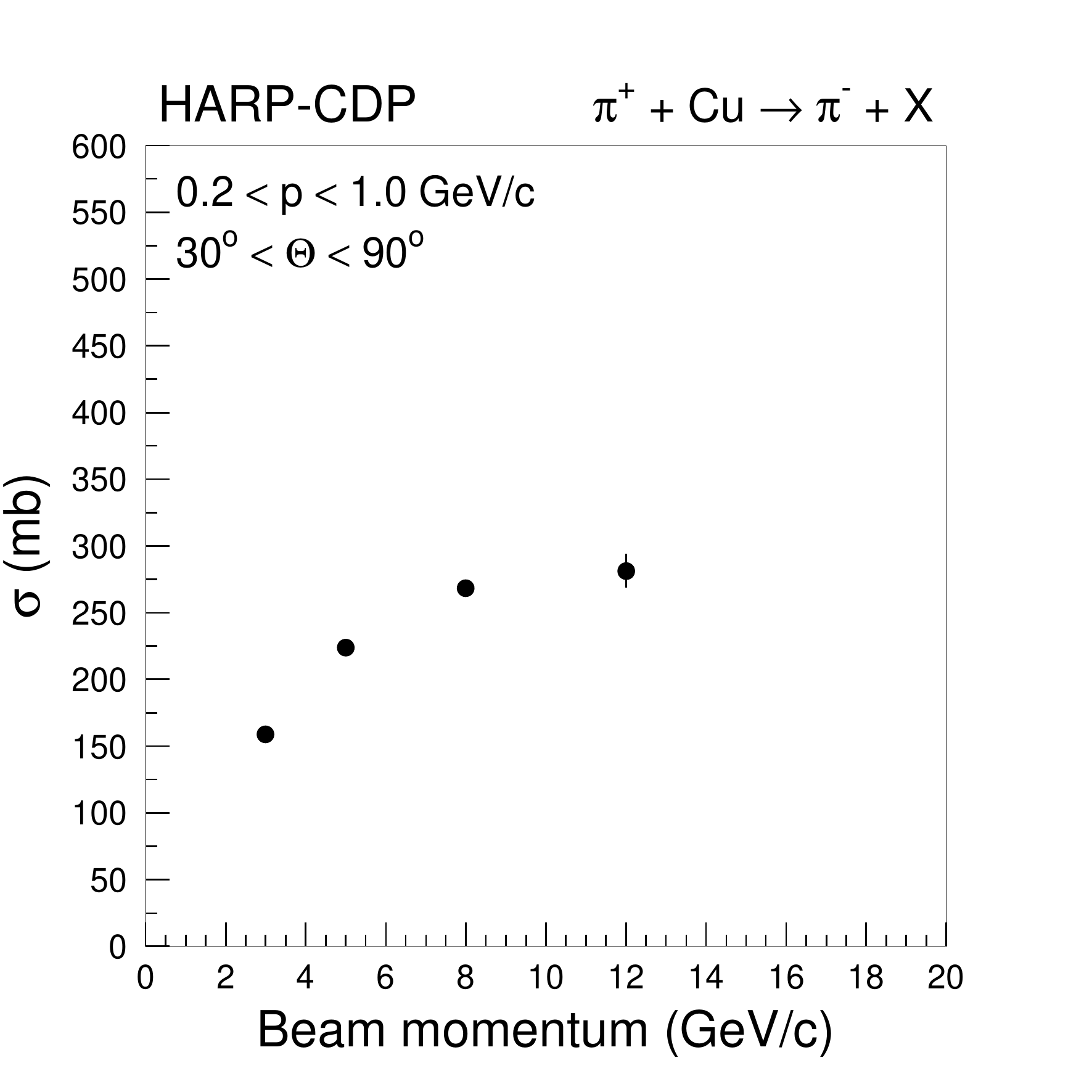} \\
\includegraphics[height=0.30\textheight]{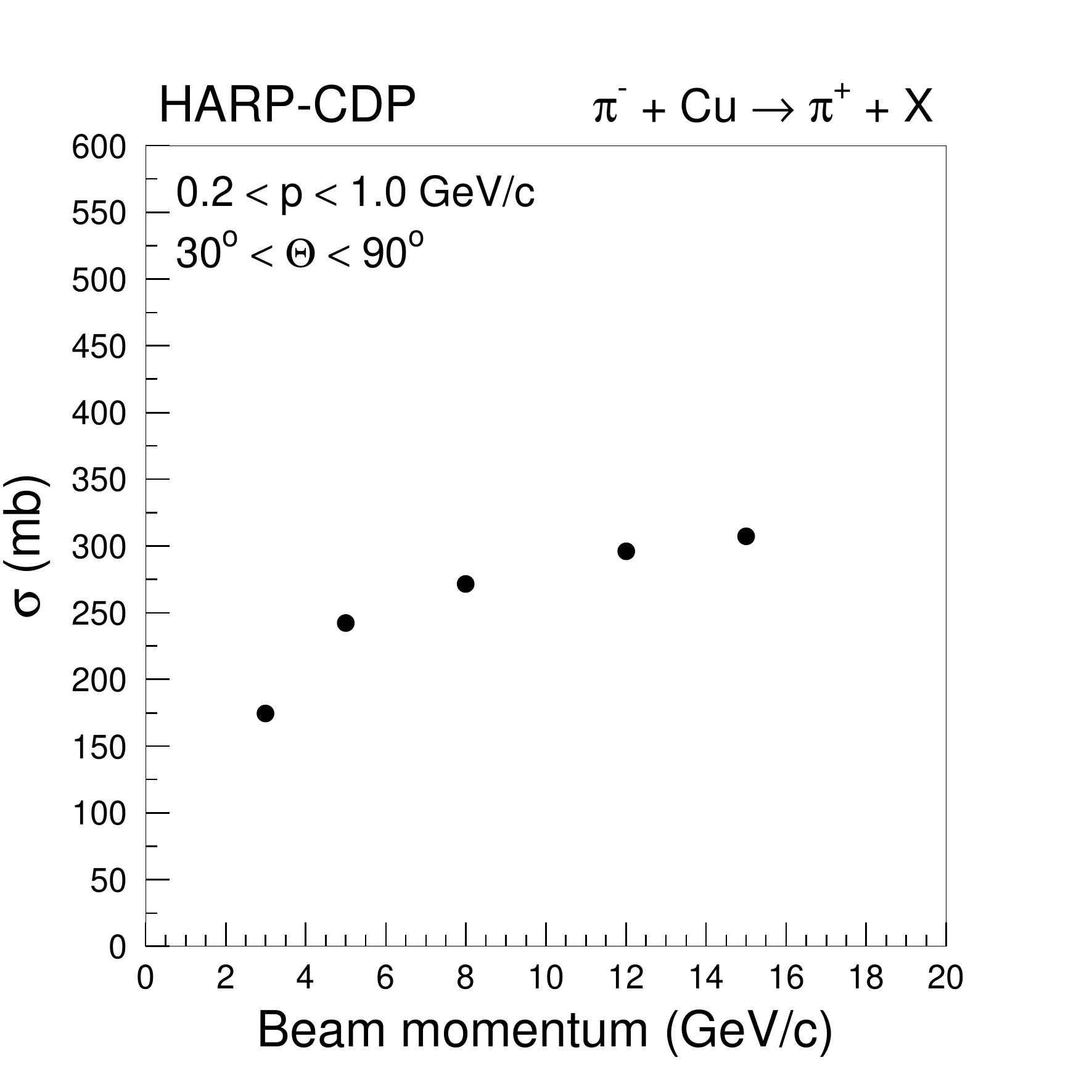} &  
\includegraphics[height=0.30\textheight]{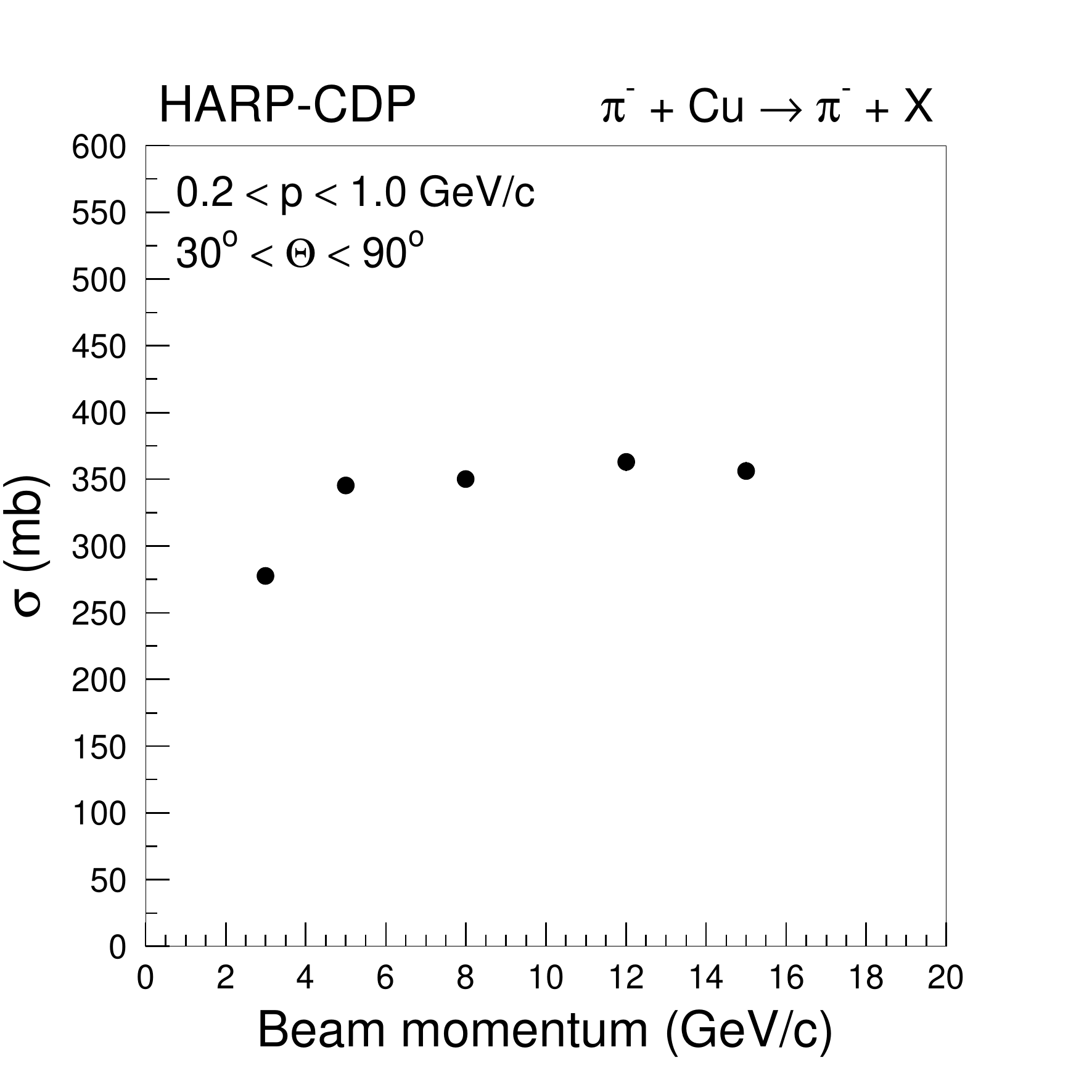} \\
\end{tabular}
\caption{Inclusive cross-sections of the production of 
secondary $\pi^+$'s and $\pi^-$'s, integrated over the momentum range 
$0.2 < p < 1.0$~GeV/{\it c} and the polar-angle range 
$30^\circ < \theta < 90^\circ$, from the interactions on copper nuclei
of protons (top row), $\pi^+$'s (middle row), and $\pi^-$'s (bottom row), 
as a function of the beam momentum; the shown errors are 
total errors and mostly smaller than the symbol size.} 
\label{fxscu}
\end{center}
\end{figure*}

We present in Figs.~\ref{xsvsmompro} to \ref{fxscu} what
we consider salient features of our cross-sections.

Figure~\ref{xsvsmompro} shows the inclusive cross-sections
of the production of protons, $\pi^+$'s, and $\pi^-$'s,
from incoming protons between 3~GeV/{\it c} and 15~GeV/{\it c}
momentum, as a function of their charge-signed $p_{\rm T}$.
The data refer to the polar-angle range 
$20^\circ < \theta < 30^\circ$.
Figures~\ref{xsvsmompip} and \ref{xsvsmompim} show the same
for incoming $\pi^+$'s and $\pi^-$'s.

Figure~\ref{xsvsthetapro} shows the inclusive cross-sections
of the production of protons, $\pi^+$'s, and $\pi^-$'s,
from incoming protons between 3~GeV/{\it c} and 15~GeV/{\it c}
momentum, this time as a function of their charge-signed 
polar angle $\theta$.
The data refer to the $p_{\rm T}$ range 
$0.24 < p_{\rm T} < 0.30$~GeV/{\it c}.
In this $p_{\rm T}$ range pions populate nearly all 
polar angles, whereas protons are absorbed at large polar angle 
and thus escape measurement. 
Figures~\ref{xsvsthetapip} and \ref{xsvsthetapim} show the same
for incoming $\pi^+$'s and $\pi^-$'s.  

In Fig.~\ref{fxscu}, we present the inclusive 
cross-sections of the production of 
secondary $\pi^+$'s and $\pi^-$'s, integrated over the momentum range
$0.2 < p < 1.0$~GeV/{\it c} and the polar-angle range  
$30^\circ < \theta < 90^\circ$ in the forward hemisphere, 
as a function of the beam momentum. 

\clearpage

\clearpage

\section{Deuteron production}

Besides pions and protons, also deuterons are produced in sizeable quantities on copper nuclei. Up to momenta of about 1~GeV/{\it c}, 
deuterons are easily separated from protons by \dedx . 

Table~\ref{deuteronsbypropippim} gives the ratio of deuteron to proton production as a function of the momentum at the vertex, for 8~GeV/{\it c} beam protons,
$\pi^+$'s, and $\pi^-$'s\footnote{We observe no appreciable dependence
of the deuteron to proton ratio on beam momentum.}. Cross-section ratios are not given if the data are scarce and the statistical error becomes comparable with the ratio itself---which is the case for deuterons at the high-momentum end of the spectrum.

\begin{figure*}[h]
\begin{center}
\includegraphics[height=0.5\textheight]{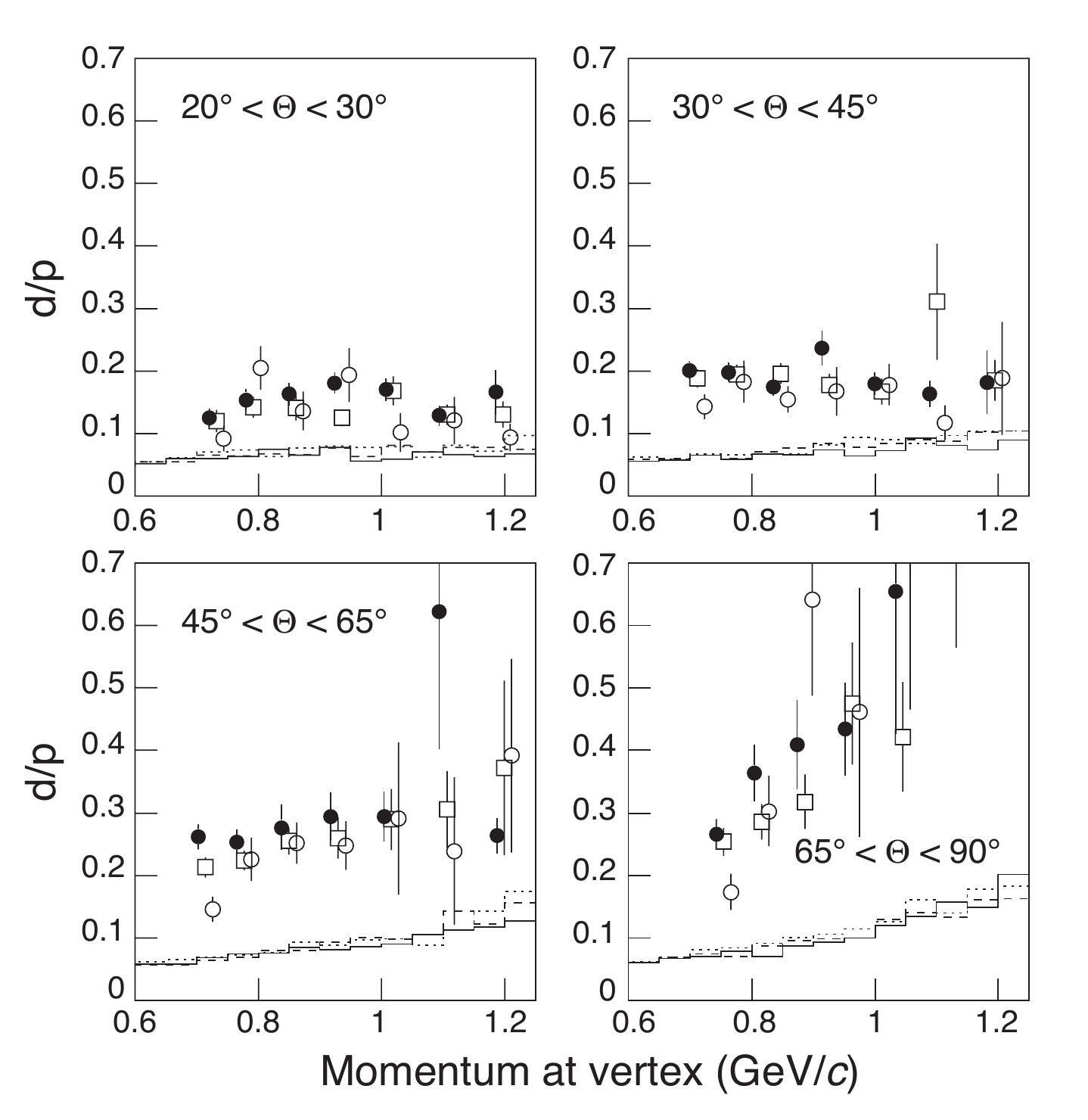} 
\caption{Deuteron to proton ratio for 8~GeV/{\it c} beam particles 
on copper nuclei, as a function of the momentum at the vertex,
for four polar-angle regions;
open squares denote beam protons, open circles beam $\pi^+$'s, and
full circles beam $\pi^-$'s; the full and broken lines denote 
predictions of Geant4's
FRITIOF model for the three different beam particles.} 
\label{dtopratio}
\end{center}
\end{figure*}

The measured deuteron to proton production ratios are illustrated in Fig.~\ref{dtopratio}, and compared with the predictions of Geant4's  
FRITIOF model. FRITIOF's predictions are shown for the same beam particles 
for which measured values are plotted. There is virtually no 
difference between its predictions for
incoming protons, $\pi^+$'s and $\pi^-$'s. FRITIOF underestimates 
deuteron production.

 \begin{table*}[h]
 \caption{Deuteron to proton ratio for beam
  protons, $\pi^+$'s, and $\pi^-$'s of 8~GeV/{\it c}
  momentum, as a function of the particle momentum $p$
  [GeV/{\it c}] at the vertex, for five polar-angle regions.}
 \label{deuteronsbypropippim}
 \begin{center}
 \begin{tabular}{|c|c||rcr|rcr|rcr|}
 \hline
 Polar angle & & \multicolumn{3}{c|}{Beam p} &
 \multicolumn{3}{c|}{Beam $\pi^+$} &\multicolumn{3}{c|}{Beam $\pi^-$} \\
 \cline{2-11}
 $\theta$ & $p$ &\multicolumn{3}{c|}{d/p} &\multicolumn{3}{c|}{d/p} &
 \multicolumn{3}{c|}{d/p} \\
 \hline
$20^\circ - 30^\circ$   & 0.732 & 0.120& $\!\!\!\pm\!\!\!$ &0.018 & 0.092& $\!\!\!\pm\!\!\!$ &0.020 & 0.125& $\!\!\!\pm\!\!\!$ &0.015 \\
                        & 0.792 & 0.142& $\!\!\!\pm\!\!\!$ &0.017 & 0.205& $\!\!\!\pm\!\!\!$ &0.035 & 0.154& $\!\!\!\pm\!\!\!$ &0.017 \\
                        & 0.861 & 0.141& $\!\!\!\pm\!\!\!$ &0.020 & 0.136& $\!\!\!\pm\!\!\!$ &0.031 & 0.163& $\!\!\!\pm\!\!\!$ &0.018 \\
                        & 0.936 & 0.125& $\!\!\!\pm\!\!\!$ &0.014 & 0.193& $\!\!\!\pm\!\!\!$ &0.043 & 0.181& $\!\!\!\pm\!\!\!$ &0.017 \\
                        & 1.019 & 0.168& $\!\!\!\pm\!\!\!$ &0.023 & 0.102& $\!\!\!\pm\!\!\!$ &0.031 & 0.170& $\!\!\!\pm\!\!\!$ &0.018 \\
                        & 1.106 & 0.130& $\!\!\!\pm\!\!\!$ &0.016 & 0.121& $\!\!\!\pm\!\!\!$ &0.038 & 0.129& $\!\!\!\pm\!\!\!$ &0.016 \\
                        & 1.197 & 0.130& $\!\!\!\pm\!\!\!$ &0.021 & 0.094& $\!\!\!\pm\!\!\!$ &0.022 & 0.166& $\!\!\!\pm\!\!\!$ &0.036 \\
 \hline
 \hline
$30^\circ - 45^\circ$   & 0.711 & 0.188& $\!\!\!\pm\!\!\!$ &0.015 & 0.143& $\!\!\!\pm\!\!\!$ &0.020 & 0.201& $\!\!\!\pm\!\!\!$ &0.014 \\
                        & 0.775 & 0.194& $\!\!\!\pm\!\!\!$ &0.016 & 0.183& $\!\!\!\pm\!\!\!$ &0.034 & 0.198& $\!\!\!\pm\!\!\!$ &0.015 \\
                        & 0.847 & 0.196& $\!\!\!\pm\!\!\!$ &0.016 & 0.155& $\!\!\!\pm\!\!\!$ &0.021 & 0.175& $\!\!\!\pm\!\!\!$ &0.014 \\
                        & 0.926 & 0.178& $\!\!\!\pm\!\!\!$ &0.018 & 0.167& $\!\!\!\pm\!\!\!$ &0.039 & 0.237& $\!\!\!\pm\!\!\!$ &0.028 \\
                        & 1.011 & 0.167& $\!\!\!\pm\!\!\!$ &0.021 & 0.178& $\!\!\!\pm\!\!\!$ &0.033 & 0.180& $\!\!\!\pm\!\!\!$ &0.018 \\
                        & 1.101 & 0.311& $\!\!\!\pm\!\!\!$ &0.093 & 0.117& $\!\!\!\pm\!\!\!$ &0.028 & 0.163& $\!\!\!\pm\!\!\!$ &0.021 \\
                        & 1.194 & 0.185& $\!\!\!\pm\!\!\!$ &0.033 & 0.188& $\!\!\!\pm\!\!\!$ &0.090 & 0.182& $\!\!\!\pm\!\!\!$ &0.051 \\
 \hline
 \hline
$45^\circ - 65^\circ$   & 0.714 & 0.213& $\!\!\!\pm\!\!\!$ &0.016 & 0.146& $\!\!\!\pm\!\!\!$ &0.020 & 0.262& $\!\!\!\pm\!\!\!$ &0.020 \\
                        & 0.777 & 0.224& $\!\!\!\pm\!\!\!$ &0.016 & 0.226& $\!\!\!\pm\!\!\!$ &0.035 & 0.253& $\!\!\!\pm\!\!\!$ &0.020 \\
                        & 0.850 & 0.255& $\!\!\!\pm\!\!\!$ &0.022 & 0.252& $\!\!\!\pm\!\!\!$ &0.033 & 0.277& $\!\!\!\pm\!\!\!$ &0.036 \\
                        & 0.930 & 0.260& $\!\!\!\pm\!\!\!$ &0.033 & 0.248& $\!\!\!\pm\!\!\!$ &0.039 & 0.294& $\!\!\!\pm\!\!\!$ &0.039 \\
                        & 1.016 & 0.290& $\!\!\!\pm\!\!\!$ &0.049 & 0.291& $\!\!\!\pm\!\!\!$ &0.122 & 0.294& $\!\!\!\pm\!\!\!$ &0.040 \\
                        & 1.106 & 0.306& $\!\!\!\pm\!\!\!$ &0.061 & 0.239& $\!\!\!\pm\!\!\!$ &0.118 & 0.622& $\!\!\!\pm\!\!\!$ &0.219 \\
                        & 1.199 & 0.372& $\!\!\!\pm\!\!\!$ &0.140 & 0.392& $\!\!\!\pm\!\!\!$ &0.155 & 0.264& $\!\!\!\pm\!\!\!$ &0.028 \\
 \hline
 \hline
$65^\circ - 90^\circ$   & 0.754 & 0.254& $\!\!\!\pm\!\!\!$ &0.023 & 0.174& $\!\!\!\pm\!\!\!$ &0.029 & 0.266& $\!\!\!\pm\!\!\!$ &0.024 \\
                        & 0.816 & 0.286& $\!\!\!\pm\!\!\!$ &0.028 & 0.303& $\!\!\!\pm\!\!\!$ &0.056 & 0.364& $\!\!\!\pm\!\!\!$ &0.045 \\
                        & 0.886 & 0.318& $\!\!\!\pm\!\!\!$ &0.044 & 0.641& $\!\!\!\pm\!\!\!$ &0.153 & 0.409& $\!\!\!\pm\!\!\!$ &0.071 \\
                        & 0.963 & 0.475& $\!\!\!\pm\!\!\!$ &0.098 & 0.461& $\!\!\!\pm\!\!\!$ &0.199 & 0.434& $\!\!\!\pm\!\!\!$ &0.075 \\
                        & 1.045 & 0.422& $\!\!\!\pm\!\!\!$ &0.088 & 0.789& $\!\!\!\pm\!\!\!$ &0.323 & 0.655& $\!\!\!\pm\!\!\!$ &0.229 \\
                         & 1.132 & 0.826& $\!\!\!\pm\!\!\!$ &0.262 & \multicolumn{3}{c|}{ } & \multicolumn{3}{c|}{ }\\
 \hline
 \hline
$90^\circ - 125^\circ$  & 0.743 & 0.316& $\!\!\!\pm\!\!\!$ &0.038 & 0.277& $\!\!\!\pm\!\!\!$ &0.042 & 0.255& $\!\!\!\pm\!\!\!$ &0.028 \\
                        & 0.809 & 0.418& $\!\!\!\pm\!\!\!$ &0.055 & 0.367& $\!\!\!\pm\!\!\!$ &0.077 & 0.399& $\!\!\!\pm\!\!\!$ &0.063 \\
                        & 0.883 & 0.788& $\!\!\!\pm\!\!\!$ &0.194 & 0.575& $\!\!\!\pm\!\!\!$ &0.230 & 0.903& $\!\!\!\pm\!\!\!$ &0.297 \\
                         & 0.963 & \multicolumn{3}{c|}{ } & 0.466& $\!\!\!\pm\!\!\!$ &0.180 & 0.810& $\!\!\!\pm\!\!\!$ &0.250 \\
 \hline
 \end{tabular}
 \end{center}
 \end{table*}

\clearpage

\section{Comparison of charged-pion production on beryllium, copper,
and tantalum }

\begin{figure*}[h]
\begin{center}
\begin{tabular}{cc}
\includegraphics[height=0.30\textheight]{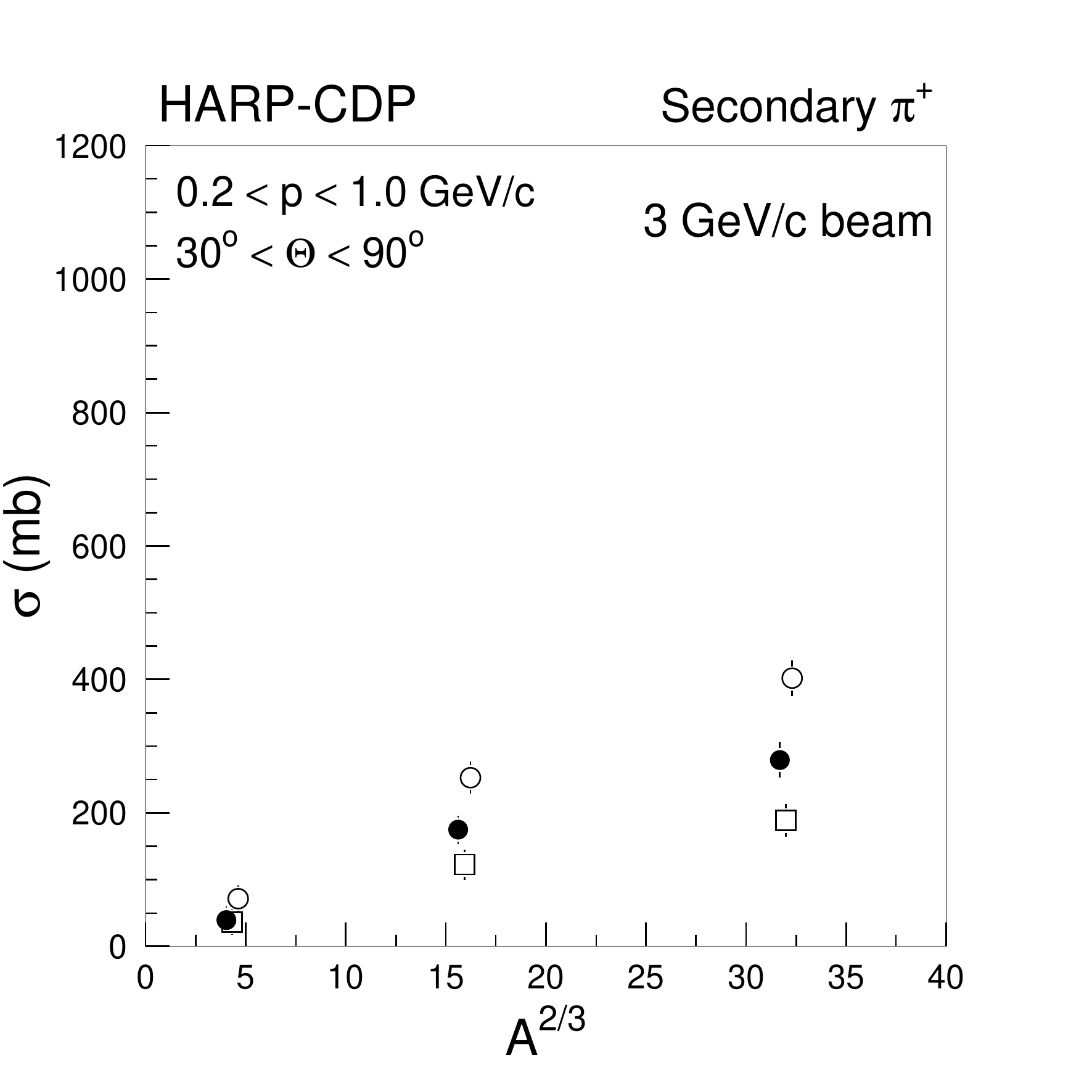} &
\includegraphics[height=0.30\textheight]{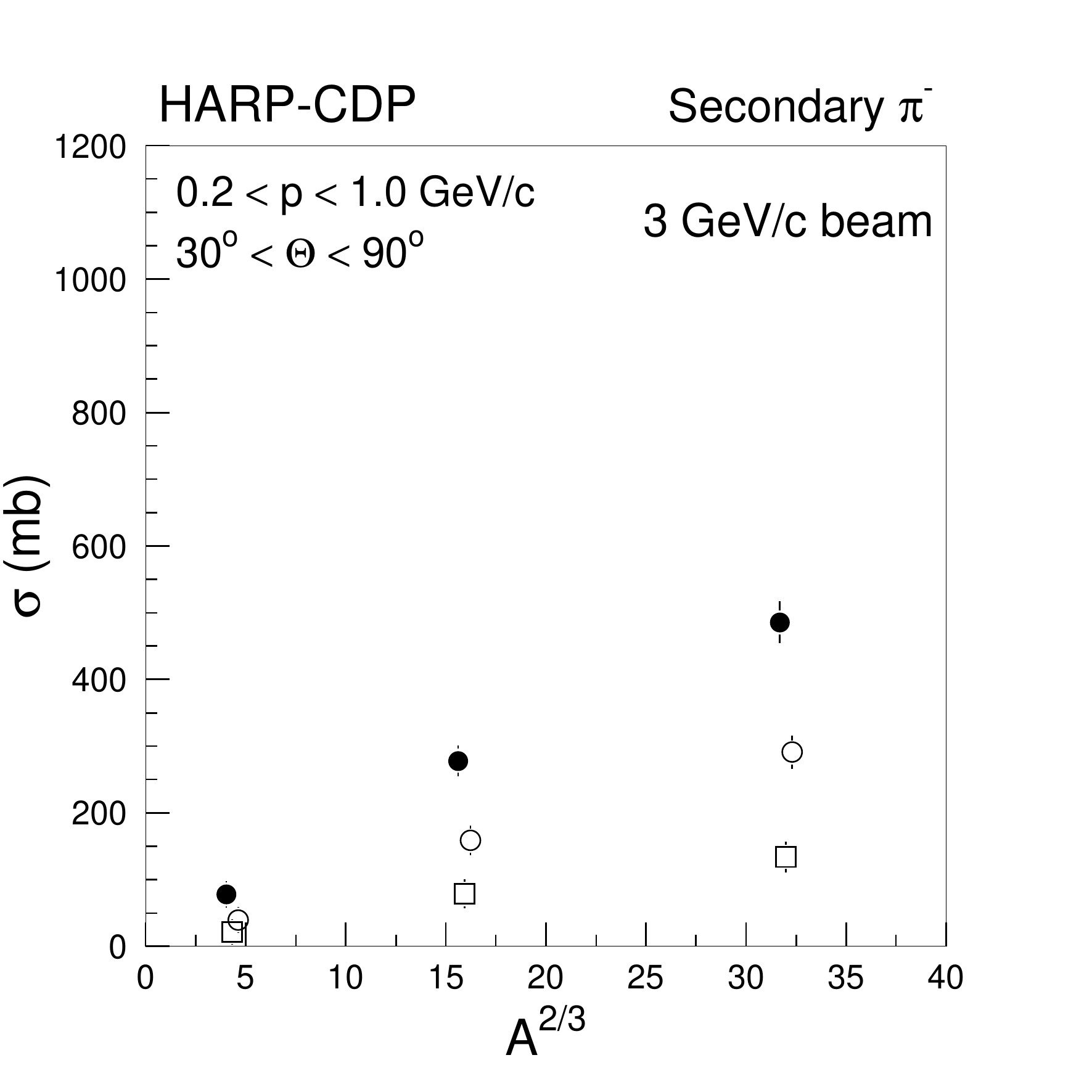} \\
\includegraphics[height=0.30\textheight]{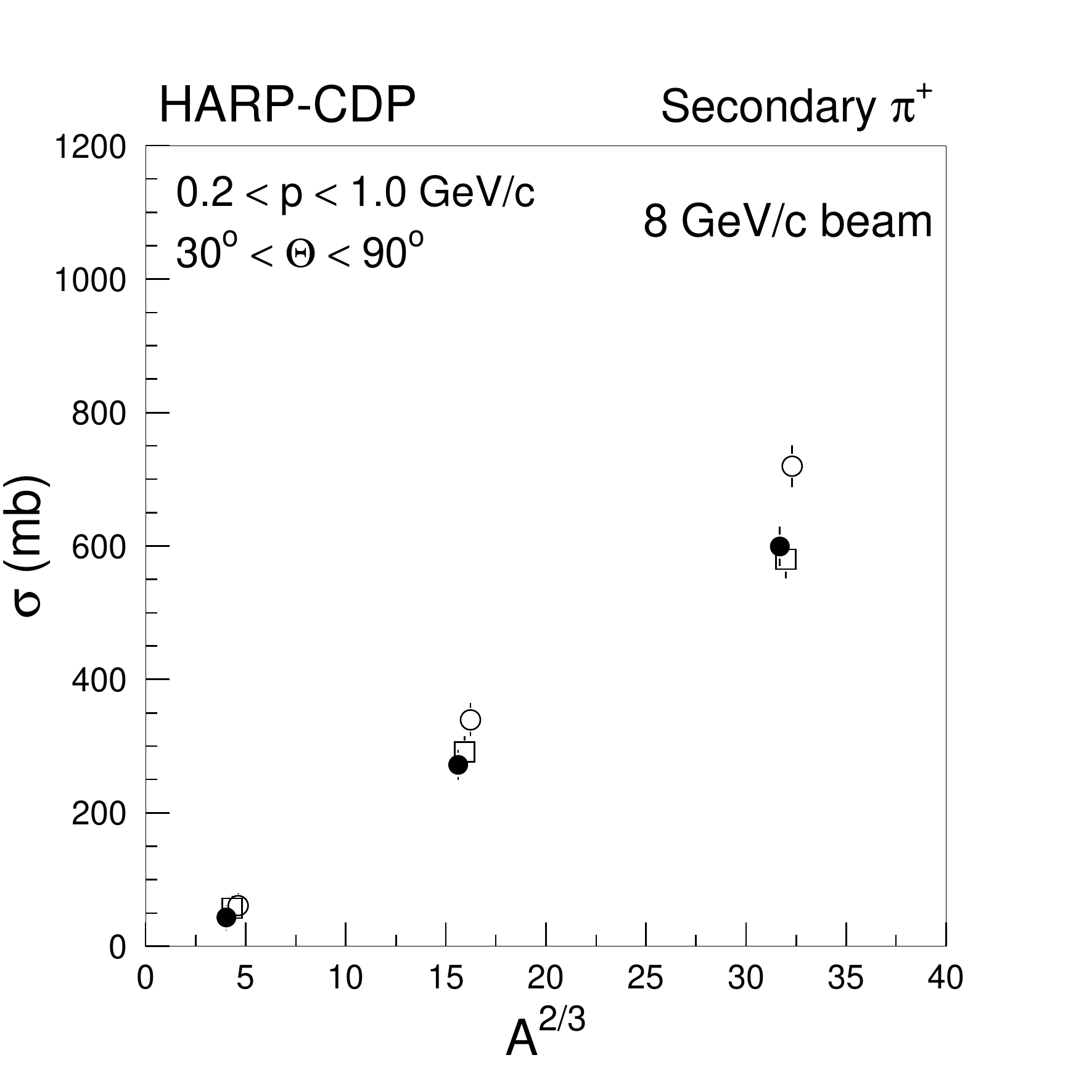} &
\includegraphics[height=0.30\textheight]{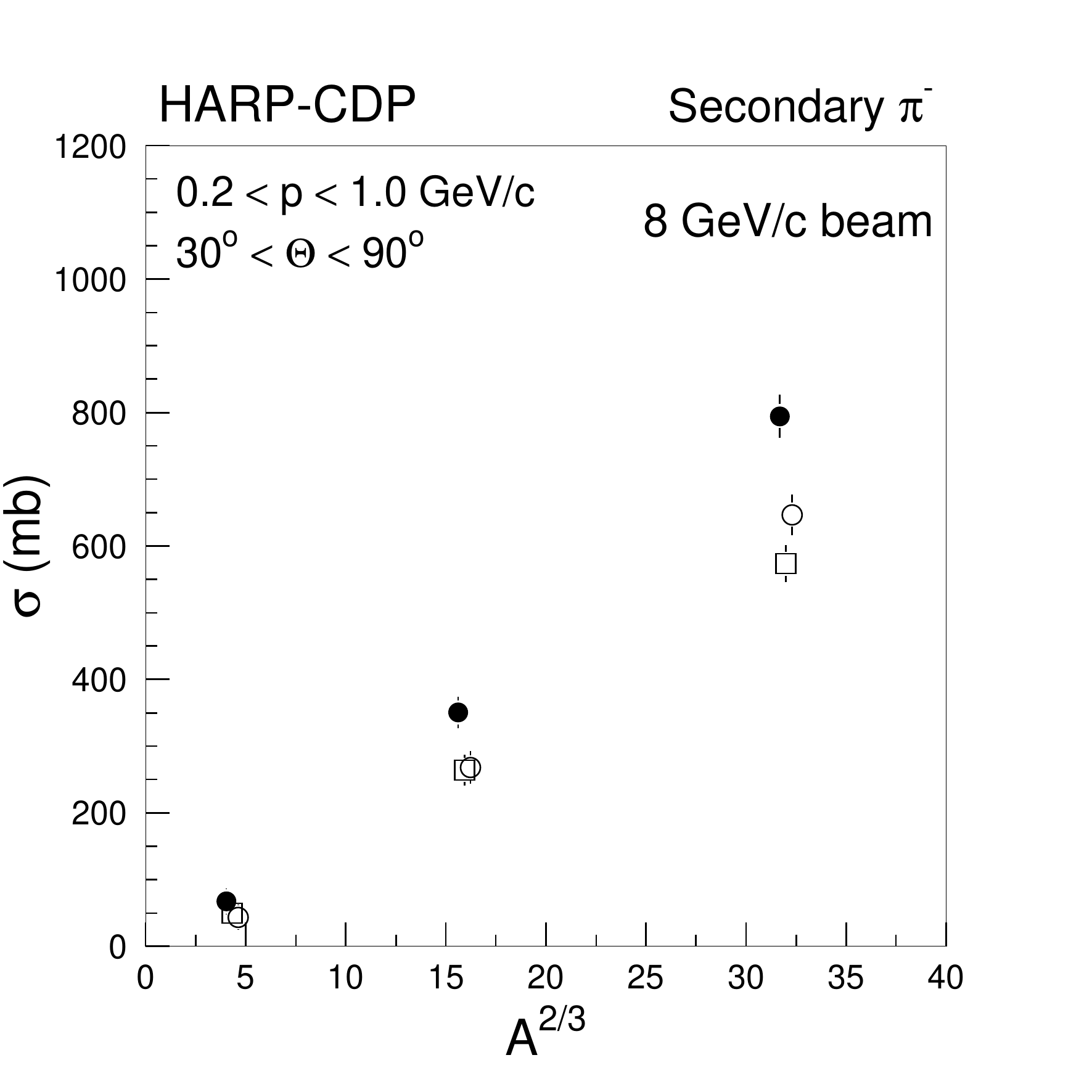} \\
\includegraphics[height=0.30\textheight]{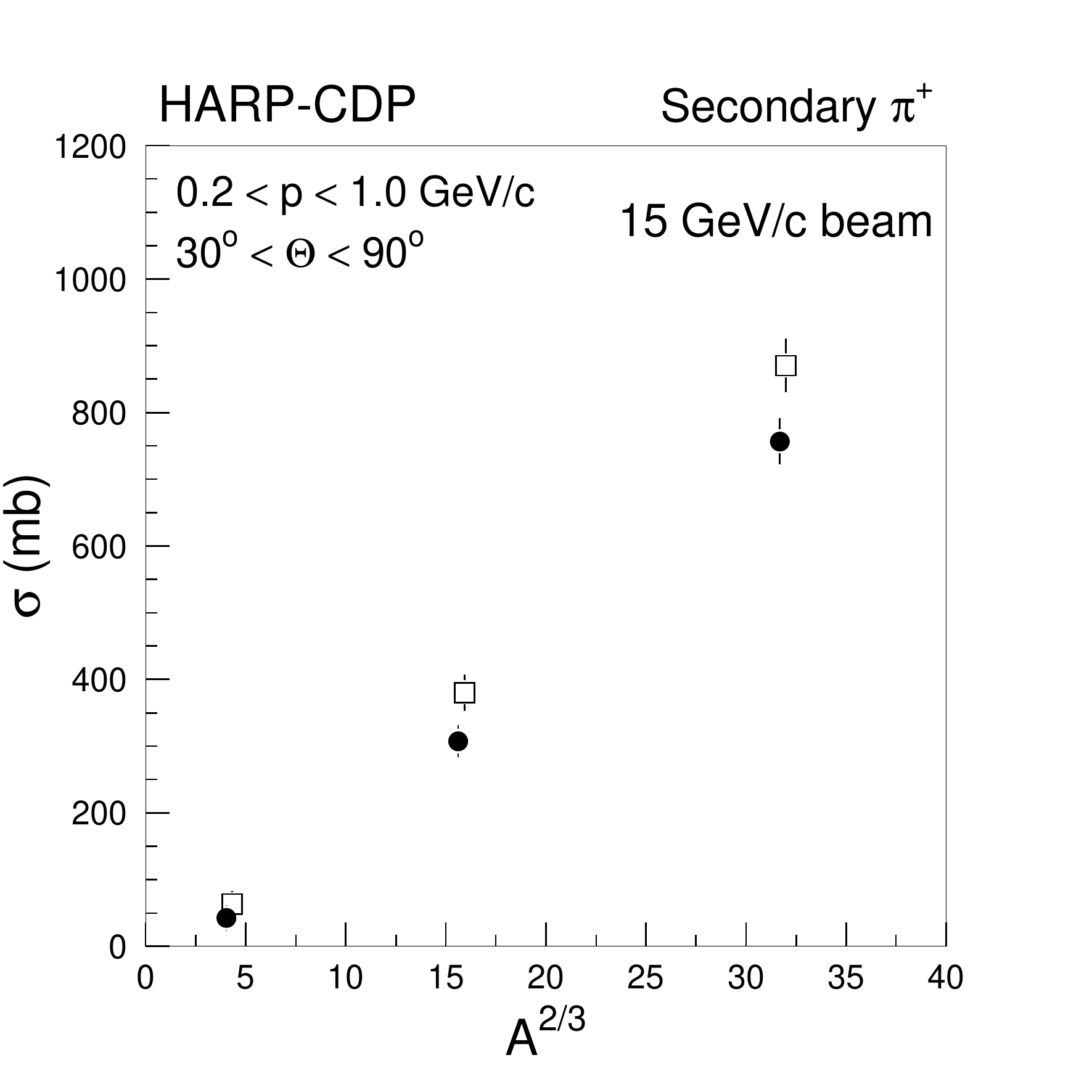} &  
\includegraphics[height=0.30\textheight]{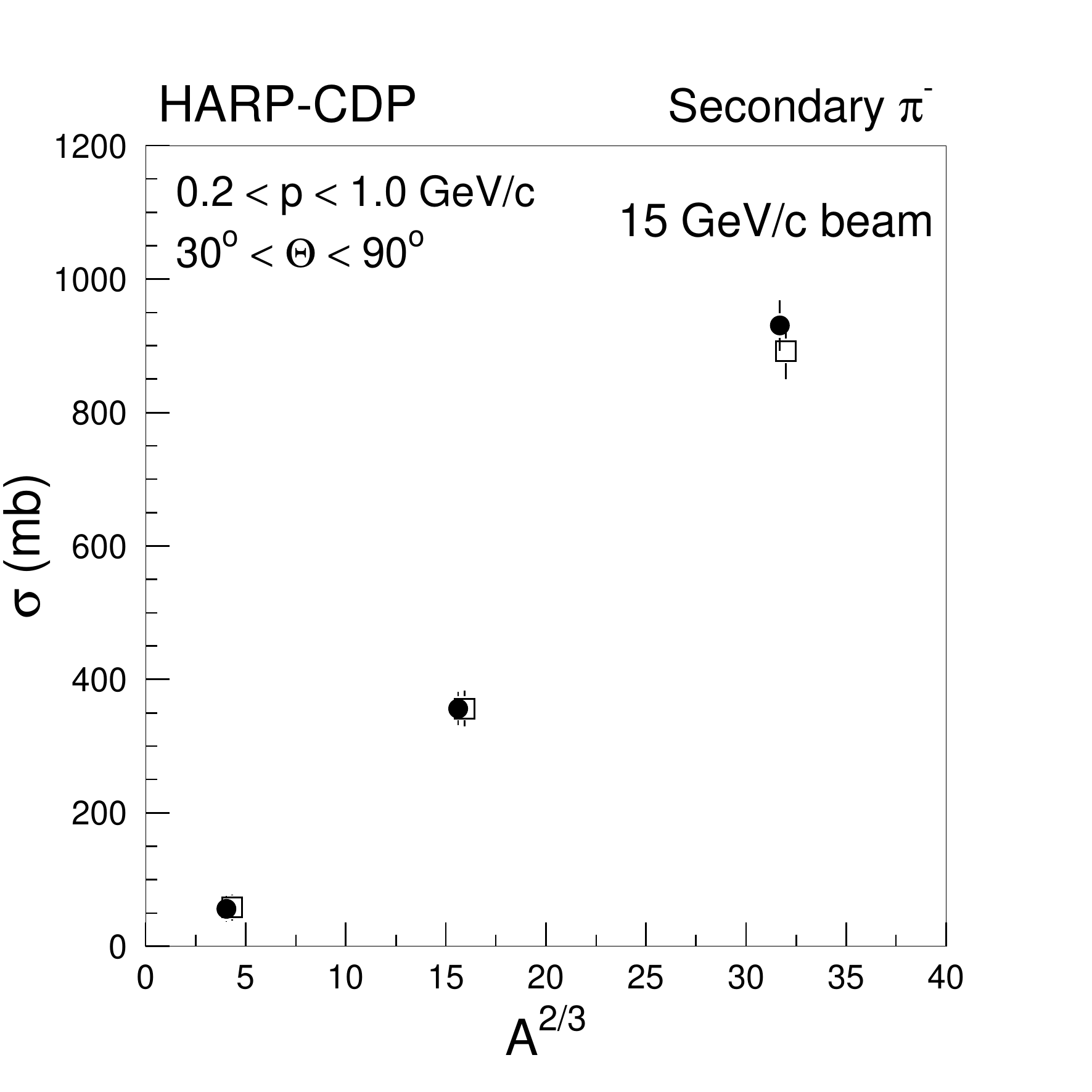} \\
\end{tabular}
\caption{Inclusive cross-sections of $\pi^+$ and $\pi^-$ production by 
protons (open squares), $\pi^+$'s (open circles),
and $\pi^-$'s (black circles), as a function of 
$A^{2/3}$ for, from left to right, beryllium, copper, and 
tantalum nuclei;
the cross-sections are integrated over the momentum range 
$0.2 < p < 1.0$~GeV/{\it c} and the polar-angle range 
$30^\circ < \theta < 90^\circ$; the shown errors are 
total errors and often smaller than the symbol size.} 
\label{ComparisonBeCuTa}
\end{center}
\end{figure*}

Figure~\ref{ComparisonBeCuTa} presents a comparison between 
the inclusive cross-sections of $\pi^+$ and $\pi^-$ 
production, integrated over the secondaries' momentum range 
$0.2 < p < 1.0$~GeV/{\it c} and polar-angle range 
$30^\circ < \theta < 90^\circ$,
in the interactions of protons, $\pi^+$ and $\pi^-$, with 
beryllium (A~=~9.01), copper (A~=~63.55), and
tantalum (A~=~181.0) nuclei\footnote{The beryllium data with 
$+8.9$~GeV/{\it c} beam momentum~\cite{Beryllium1,Beryllium2} 
have been scaled, by interpolation, to a beam momentum of 
$8.0$~GeV/{\it c}.}. 
The comparison employs the
scaling variable $A^{2/3}$ where $A$ is the atomic number of
the respective nucleus. 
We note the approximately linear dependence on this scaling
variable. At low beam momentum, the slope exhibits a strong
dependence on beam particle type, which tends to disappear 
with higher beam momentum. 

\clearpage

\section{Comparison of our results with results from 
other experiments}

\subsection{Comparison with E802 results}
Experiment E802~\cite{E802} at Brookhaven National 
Laboratory (BNL) measured
secondary $\pi^+$'s in the polar-angle
range $5^\circ < \theta < 58^\circ$ from the interactions of
$+14.6$~GeV/{\it c} protons with copper nuclei.

\begin{figure*}[ht]
\begin{center}
\vspace*{8mm}
\includegraphics[width=0.6\textwidth]{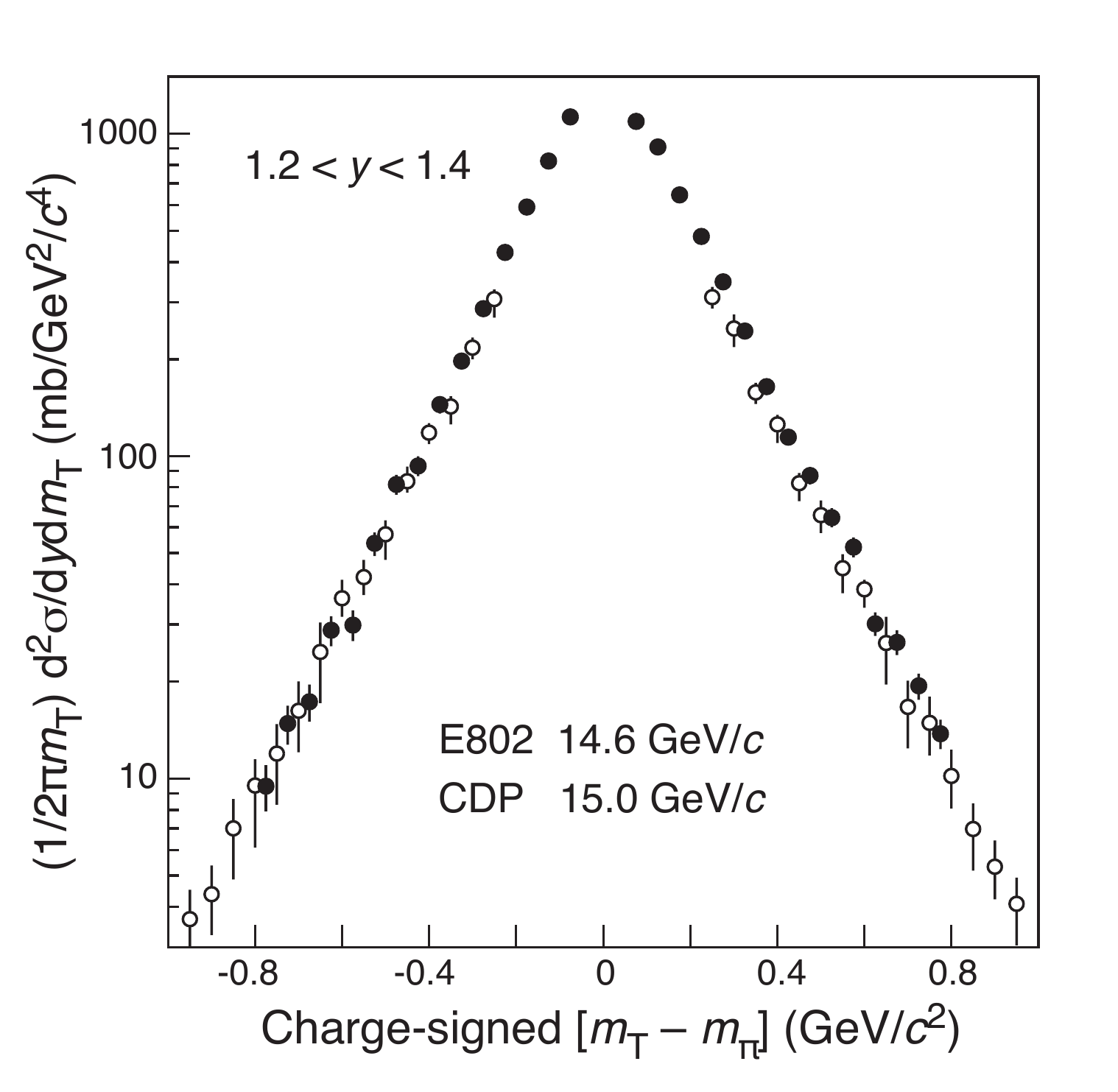} 
\caption{Comparison of our cross-sections (black circles) 
of $\pi^\pm$ production by $+15.0$~GeV/{\it c} 
protons off copper nuclei, 
with the cross-sections on copper nuclei 
published by the E802 Collaboration for the proton beam 
momentum of $+14.6$~GeV/{\it c} (open circles); all errors are 
statistical only.}
\label{comparisonwithE802}
\end{center}
\end{figure*}

Figure~\ref{comparisonwithE802} shows their published Lorentz-invariant 
cross-section of $\pi^+$ and $\pi^-$ production by
$+14.6$~GeV/{\it c} protons, in the rapidity range $1.2 < y < 1.4$,
as a function of $m_{\rm T} - m_{\pi}$, where $m_{\rm T}$ denotes
the pion transverse mass. Their data are compared 
with our cross-sections from the interactions of $+15.0$~GeV/{\it c} 
protons with copper nuclei, expressed in 
the same unit as used by E802. 
Since E802 quoted only statistical
errors, our data in Fig.~\ref{comparisonwithE802} are
also shown with their statistical errors.

The E802 $\pi^\pm$ cross-sections are in good agreement 
with our cross-sections measured nearly at the same proton 
beam momentum, taking into account 
the normalization uncertainty of (10--15)\% quoted by E802.   
We draw attention to the good agreement of the slopes 
of the cross-sections over two orders of magnitude.  

\subsection{Comparison with E910 results}

\begin{figure*}[ht]
\begin{center}
\vspace*{8mm}
\includegraphics[width=0.6\textwidth]{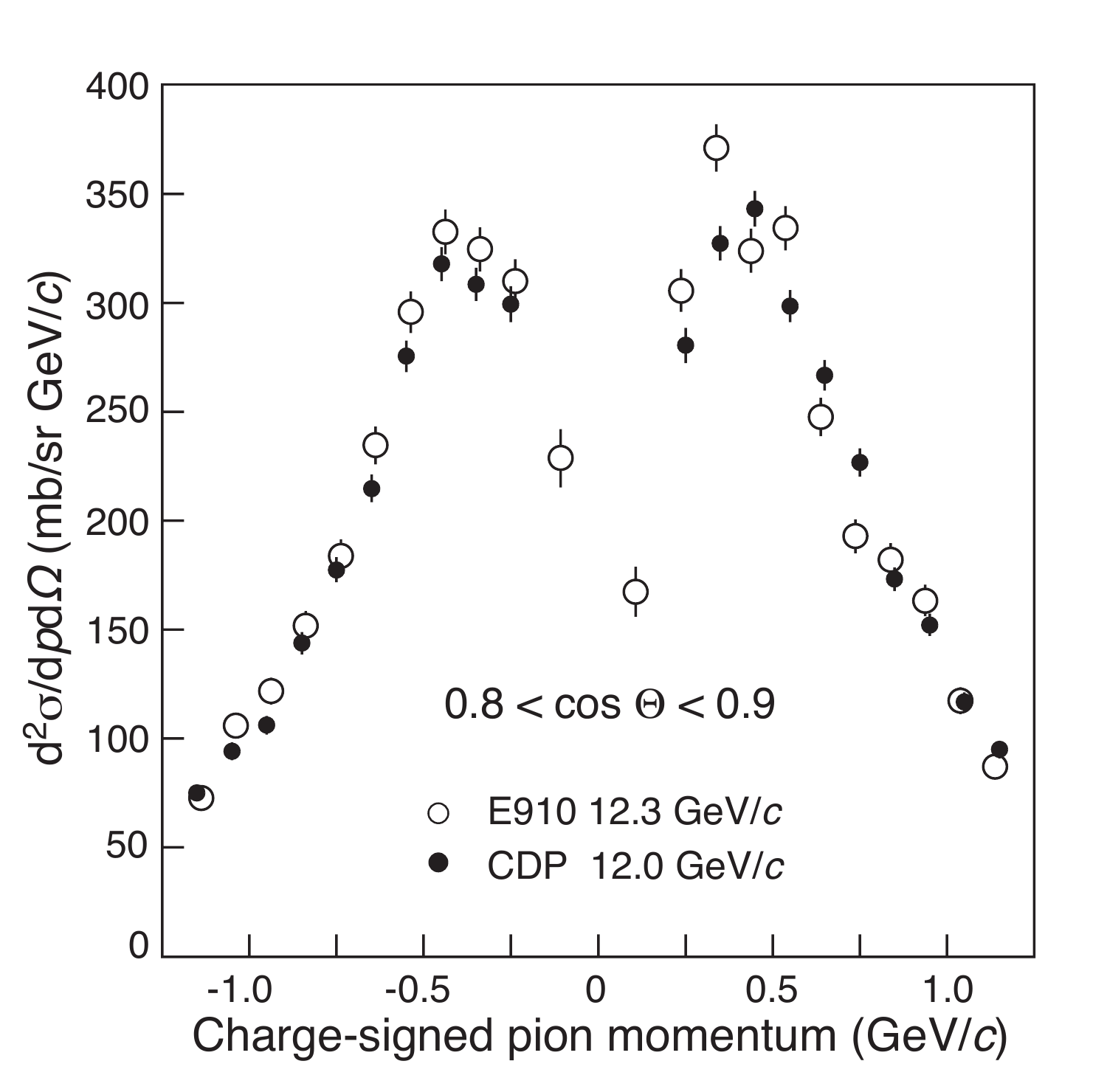} 
\caption{Comparison of our cross-sections (black circles)  
of $\pi^\pm$ production by $+12.0$~GeV/{\it c} 
protons off copper nuclei, 
with the cross-sections 
published by the E910 Collaboration for the proton beam 
momentum of $+12.3$~GeV/{\it c} (open circles); 
all errors are statistical only.}
\label{comparisonwithE910}
\end{center}
\end{figure*}

BNL experiment E910~\cite{E910}  
measured secondary charged pions in the momentum
range 0.1--6~GeV/{\it c} from the interactions of $+12.3$~GeV/{\it c} 
protons with copper nuclei.
This experiment used a TPC for the measurement of secondaries,
with a comfortably large track length of $\sim$1.5~m. This feature,
together with a magnetic field strength of 0.5~T, 
is of particular significance, since it permits  
considerably better charge identification and proton--pion 
separation by \dedx\ than is possible in the HARP detector.
Figure~\ref{comparisonwithE910} shows their published  
cross-section ${\rm d}^2 \sigma / {\rm d}p {\rm d}\Omega$ 
of $\pi^\pm$ production by $+12.3$~GeV/{\it c} protons,
in the polar-angle range $0.8 < \cos\theta < 0.9$. 
Since E910 quoted only statistical
errors, our data in Fig.~\ref{comparisonwithE910} from the
interactions of $+12.0$~GeV/{\it c} protons with copper 
are also shown with their statistical errors. The normalization 
uncertainty quoted by E910 is $\leq$5\%. 

Also here, the E910 data are shown as published, and our data
are expressed in the same unit as used by E910.  
We draw attention to the good agreement 
in the $\pi^+ / \pi^-$ ratio 
between the cross-sections from E910 and our cross-sections. 

\subsection{Comparison with results from the HARP Collaboration}

\begin{figure*}[ht]
\begin{center}
\begin{tabular}{cc}
\includegraphics[height=0.45\textwidth]{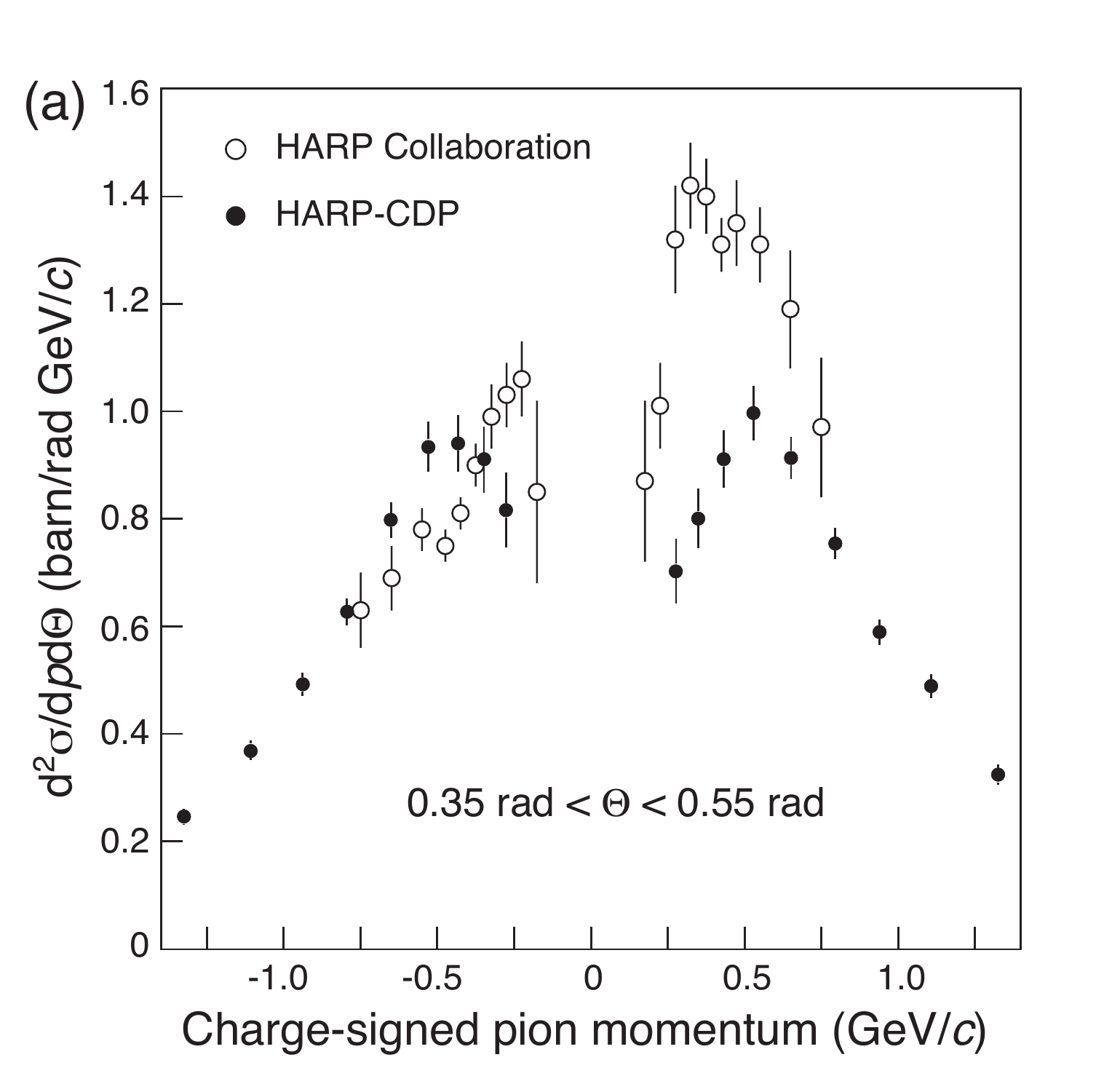} & 
\includegraphics[height=0.45\textwidth]{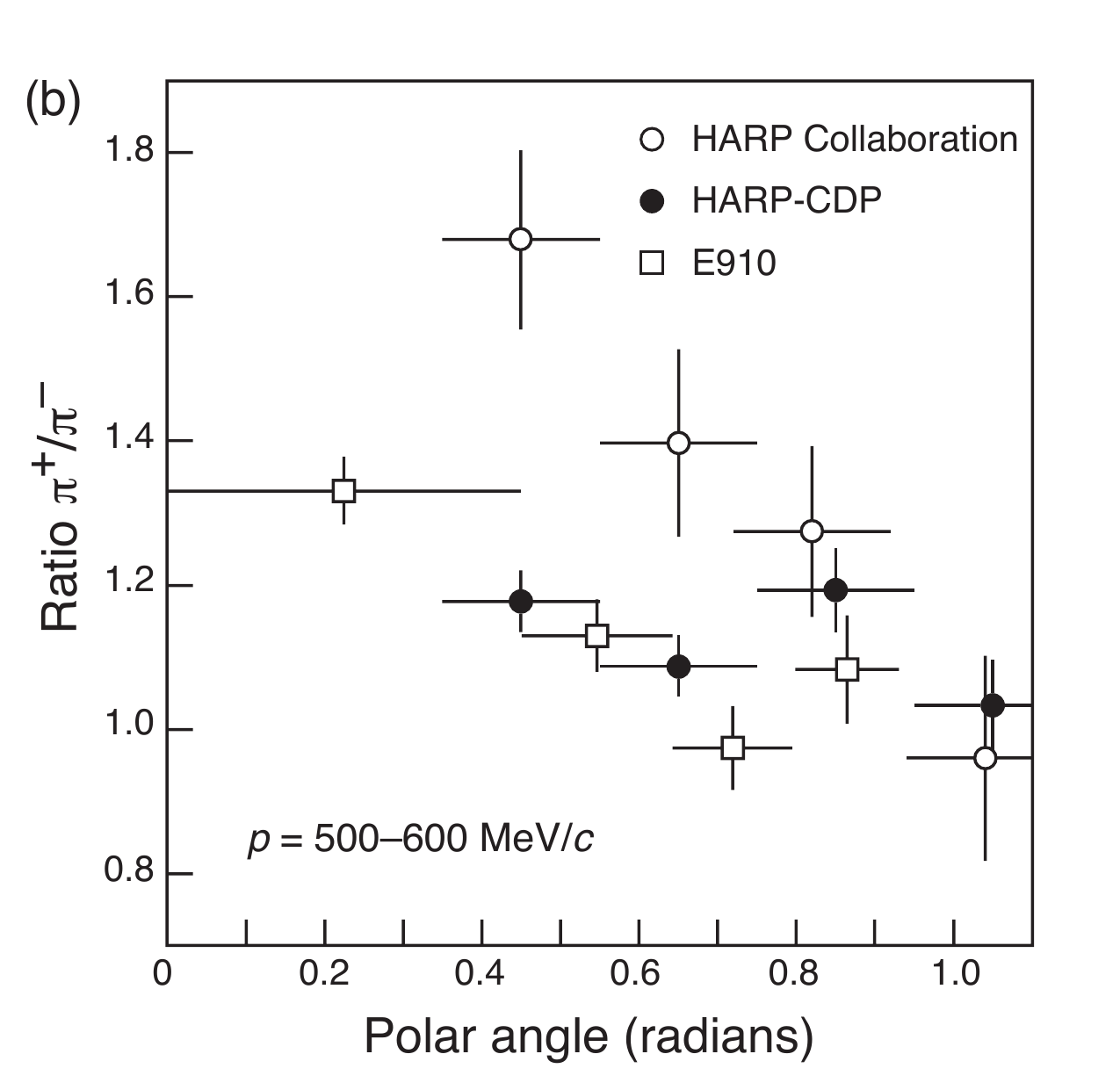}  \\
\end{tabular}
\caption{(a) Comparison of our cross-sections (black circles) of $\pi^\pm$ production by $+12.0$~GeV/{\it c} protons off copper nuclei with the cross-sections published by the HARP Collaboration (open circles); (b) Comparison of our ratio $\pi^+/\pi^-$ at $+12.0$~GeV/{\it c} 
beam momentum as a function of the polar angle $\theta$ with the ratios published by the 
HARP Collaboration; also shown are the ratios $\pi^+/\pi^-$ published by the E910 Collaboration for a $+12.3$~GeV/{\it c} beam momentum; 
in (a) total errors are shown; in (b) for the HARP Collaboration total errors are shown (only those are published), for E910 and our group, the errors are statistical only.}
\label{comparisonwithOH}
\end{center}
\end{figure*}

Figure~\ref{comparisonwithOH} (a) shows the 
comparison of our
cross-sections of pion production by $+12.0$~GeV/{\it c} 
protons off copper nuclei with the ones 
published by the HARP Collaboration~\cite{OffLApaper},
in the polar-angle range $0.35 < \theta < 0.55$~rad.
The latter cross-sections are plotted as published, 
while we expressed our cross-sections in 
the unit used by the HARP Collaboration. 
Figure~\ref{comparisonwithOH} (b)
shows our ratio $\pi^+/\pi^-$ as a function of the
polar angle $\theta$ in comparison with the ratios published by the 
E910 Collaboration (at the slightly different proton beam 
momentum of $+12.3$~GeV/{\it c})
and by the HARP Collaboration.

The discrepancy between our results and those published by the HARP Collaboration is evident. We note the difference especially of the $\pi^+$ cross-section, and the difference in the reported momentum range. The discrepancy is even more serious as the same data set has been analysed by both groups.

We hold that the discrepancy is caused by problems in the HARP 
Collaboration's data analysis. They result primarily, but not 
exclusively, from a lack of understanding TPC track distortions 
and RPC timing signals. These problems, together with others that 
affect the HARP Collaboration's data analysis, are discussed in 
detail in Refs~\cite{JINSTpub,EPJCpub,WhiteBookseries} and summarized in the 
Appendix of Ref.~\cite{Beryllium1}. 

\section{Summary}

From the analysis of data from the HARP large-angle spectrometer
(polar angle $\theta$ in the range $20^\circ < \theta < 125^\circ$), double-differential 
cross-sections ${\rm d}^2 \sigma / {\rm d}p {\rm d}\Omega$ 
of the production of secondary protons, $\pi^+$'s, and $\pi^-$'s,
and of deuterons, have been obtained. The incoming beam particles were protons 
and pions with momenta from $\pm 3$ to $\pm 15$~GeV/{\it c}, 
impinging on a 5\% $\lambda_{\rm abs}$ thick stationary 
copper target. 
Our cross-sections for $\pi^+$ and $\pi^-$ production 
agree with results from BNL experiments E802 and E910 but disagree with 
the results of the HARP Collaboration that were obtained 
from the same raw data.

\section*{Acknowledgements}

We are greatly indebted to many technical collaborators whose 
diligent and hard work made the HARP detector a well-functioning 
instrument. We thank all HARP colleagues who devoted time and 
effort to the design and construction of the detector, to data taking, 
and to setting up the computing and software infrastructure. 
We express our sincere gratitude to HARP's funding agencies 
for their support.  


\clearpage

\appendix

\section{Cross-section Tables}


  
 \begin{table*}[h]
 \begin{scriptsize}
 \caption{Double-differential inclusive
  cross-section ${\rm d}^2 \sigma /{\rm d}p{\rm d}\Omega$
  [mb/(GeV/{\it c} sr)] of the production of protons
  in p + Cu $\rightarrow$ p + X interactions
  with $+3.0$~GeV/{\it c} beam momentum;
  the first error is statistical, the second systematic; 
 $p_{\rm T}$ in GeV/{\it c}, polar angle $\theta$ in degrees.}
 \label{pro.procu3}
 \begin{center}
 \begin{tabular}{|c||c|c|rcrcr||c|c|rcrcr|}
 \hline
   & \multicolumn{7}{c||}{$20<\theta<30$}
  & \multicolumn{7}{c|}{$30<\theta<40$} \\
 \hline
 $p_{\rm T}$ & $\langle p_{\rm T} \rangle$ & $\langle \theta \rangle$
  & \multicolumn{5}{c||}{${\rm d}^2 \sigma /{\rm d}p{\rm d}\Omega$}
  &$\langle p_{\rm T} \rangle$ & $\langle \theta \rangle$
  & \multicolumn{5}{c|}{${\rm d}^2 \sigma /{\rm d}p{\rm d}\Omega$} \\
 \hline
0.20--0.24 & 0.222 &  25.1 &  454.76 & $\!\!\pm\!\!$ &  20.41 & $\!\!\pm\!\!$ &  25.27 &  &  &  \multicolumn{5}{c|}{ } \\
0.24--0.30 & 0.271 &  25.4 &  400.56 & $\!\!\pm\!\!$ &  14.75 & $\!\!\pm\!\!$ &  19.36 & 0.271 &  35.0 &  425.92 & $\!\!\pm\!\!$ &  17.82 & $\!\!\pm\!\!$ &  19.16 \\
0.30--0.36 & 0.331 &  25.1 &  353.84 & $\!\!\pm\!\!$ &  14.06 & $\!\!\pm\!\!$ &  14.96 & 0.331 &  35.0 &  396.34 & $\!\!\pm\!\!$ &  14.24 & $\!\!\pm\!\!$ &  15.27 \\
0.36--0.42 & 0.393 &  25.1 &  300.58 & $\!\!\pm\!\!$ &  12.77 & $\!\!\pm\!\!$ &  11.05 & 0.392 &  34.7 &  349.89 & $\!\!\pm\!\!$ &  13.67 & $\!\!\pm\!\!$ &  11.68 \\
0.42--0.50 & 0.461 &  25.1 &  236.99 & $\!\!\pm\!\!$ &   9.69 & $\!\!\pm\!\!$ &   7.73 & 0.462 &  35.0 &  276.40 & $\!\!\pm\!\!$ &  10.44 & $\!\!\pm\!\!$ &   8.13 \\
0.50--0.60 & 0.553 &  25.3 &  199.75 & $\!\!\pm\!\!$ &   8.09 & $\!\!\pm\!\!$ &   6.61 & 0.551 &  35.2 &  214.18 & $\!\!\pm\!\!$ &   8.27 & $\!\!\pm\!\!$ &   6.46 \\
\hline
 \end{tabular}
 \end{center}
 \end{scriptsize}
 \end{table*}
  
 \begin{table*}[h]
 \begin{scriptsize}
 \caption{Double-differential inclusive
  cross-section ${\rm d}^2 \sigma /{\rm d}p{\rm d}\Omega$
  [mb/(GeV/{\it c} sr)] of the production of $\pi^+$'s
  in p + Cu $\rightarrow$ $\pi^+$ + X interactions
  with $+3.0$~GeV/{\it c} beam momentum;
  the first error is statistical, the second systematic; 
 $p_{\rm T}$ in GeV/{\it c}, polar angle $\theta$ in degrees.}
 \label{pip.procu3}
 \begin{center}
 \begin{tabular}{|c||c|c|rcrcr||c|c|rcrcr|}
 \hline
   & \multicolumn{7}{c||}{$20<\theta<30$}
  & \multicolumn{7}{c|}{$30<\theta<40$} \\
 \hline
 $p_{\rm T}$ & $\langle p_{\rm T} \rangle$ & $\langle \theta \rangle$
  & \multicolumn{5}{c||}{${\rm d}^2 \sigma /{\rm d}p{\rm d}\Omega$}
  &$\langle p_{\rm T} \rangle$ & $\langle \theta \rangle$
  & \multicolumn{5}{c|}{${\rm d}^2 \sigma /{\rm d}p{\rm d}\Omega$} \\
 \hline
0.10--0.13 & 0.116 &  25.2 &  115.62 & $\!\!\pm\!\!$ &  13.74 & $\!\!\pm\!\!$ &  10.13 & 0.115 &  35.1 &  112.03 & $\!\!\pm\!\!$ &  13.00 & $\!\!\pm\!\!$ &   9.14 \\
0.13--0.16 & 0.145 &  25.3 &  113.77 & $\!\!\pm\!\!$ &  12.16 & $\!\!\pm\!\!$ &   7.91 & 0.144 &  35.1 &  121.45 & $\!\!\pm\!\!$ &  12.24 & $\!\!\pm\!\!$ &   8.73 \\
0.16--0.20 & 0.180 &  25.1 &  151.41 & $\!\!\pm\!\!$ &  12.15 & $\!\!\pm\!\!$ &   9.05 & 0.180 &  35.0 &  102.01 & $\!\!\pm\!\!$ &   9.31 & $\!\!\pm\!\!$ &   5.97 \\
0.20--0.24 & 0.221 &  24.9 &  131.06 & $\!\!\pm\!\!$ &  10.88 & $\!\!\pm\!\!$ &   6.84 & 0.223 &  34.5 &  120.40 & $\!\!\pm\!\!$ &  10.02 & $\!\!\pm\!\!$ &   6.17 \\
0.24--0.30 & 0.271 &  25.6 &  111.15 & $\!\!\pm\!\!$ &   7.97 & $\!\!\pm\!\!$ &   4.64 & 0.270 &  35.0 &   98.69 & $\!\!\pm\!\!$ &   7.33 & $\!\!\pm\!\!$ &   4.19 \\
0.30--0.36 & 0.332 &  24.7 &   71.19 & $\!\!\pm\!\!$ &   6.14 & $\!\!\pm\!\!$ &   2.89 & 0.329 &  34.5 &   78.67 & $\!\!\pm\!\!$ &   6.53 & $\!\!\pm\!\!$ &   3.19 \\
\hline
 \end{tabular}
 \end{center}
 \end{scriptsize}
 \end{table*}
  
 \begin{table*}[h]
 \begin{scriptsize}
 \caption{Double-differential inclusive
  cross-section ${\rm d}^2 \sigma /{\rm d}p{\rm d}\Omega$
  [mb/(GeV/{\it c} sr)] of the production of $\pi^-$'s
  in p + Cu $\rightarrow$ $\pi^-$ + X interactions
  with $+3.0$~GeV/{\it c} beam momentum;
  the first error is statistical, the second systematic; 
 $p_{\rm T}$ in GeV/{\it c}, polar angle $\theta$ in degrees.}
 \label{pim.procu3}
 \begin{center}
 \begin{tabular}{|c||c|c|rcrcr||c|c|rcrcr|}
 \hline
   & \multicolumn{7}{c||}{$20<\theta<30$}
  & \multicolumn{7}{c|}{$30<\theta<40$} \\
 \hline
 $p_{\rm T}$ & $\langle p_{\rm T} \rangle$ & $\langle \theta \rangle$
  & \multicolumn{5}{c||}{${\rm d}^2 \sigma /{\rm d}p{\rm d}\Omega$}
  &$\langle p_{\rm T} \rangle$ & $\langle \theta \rangle$
  & \multicolumn{5}{c|}{${\rm d}^2 \sigma /{\rm d}p{\rm d}\Omega$} \\
 \hline
0.10--0.13 & 0.117 &  24.5 &  102.15 & $\!\!\pm\!\!$ &  12.97 & $\!\!\pm\!\!$ &   9.74 & 0.115 &  34.2 &   94.31 & $\!\!\pm\!\!$ &  11.53 & $\!\!\pm\!\!$ &   8.13 \\
0.13--0.16 & 0.145 &  25.6 &   81.94 & $\!\!\pm\!\!$ &  10.38 & $\!\!\pm\!\!$ &   6.12 & 0.144 &  34.8 &   71.57 & $\!\!\pm\!\!$ &   9.16 & $\!\!\pm\!\!$ &   5.57 \\
0.16--0.20 & 0.179 &  25.0 &   82.63 & $\!\!\pm\!\!$ &   8.74 & $\!\!\pm\!\!$ &   5.25 & 0.176 &  34.8 &   71.03 & $\!\!\pm\!\!$ &   7.48 & $\!\!\pm\!\!$ &   4.72 \\
0.20--0.24 & 0.219 &  25.1 &   82.87 & $\!\!\pm\!\!$ &   8.47 & $\!\!\pm\!\!$ &   4.75 & 0.220 &  34.4 &   77.33 & $\!\!\pm\!\!$ &   8.22 & $\!\!\pm\!\!$ &   4.69 \\
0.24--0.30 & 0.268 &  25.3 &   56.50 & $\!\!\pm\!\!$ &   5.68 & $\!\!\pm\!\!$ &   2.80 & 0.264 &  35.2 &   51.48 & $\!\!\pm\!\!$ &   5.35 & $\!\!\pm\!\!$ &   2.51 \\
0.30--0.36 & 0.327 &  24.9 &   39.58 & $\!\!\pm\!\!$ &   4.68 & $\!\!\pm\!\!$ &   2.17 & 0.328 &  34.7 &   37.06 & $\!\!\pm\!\!$ &   4.47 & $\!\!\pm\!\!$ &   1.90 \\
\hline
 \end{tabular}
 \end{center}
 \end{scriptsize}
 \end{table*}
  
 \begin{table*}[h]
 \begin{scriptsize}
 \caption{Double-differential inclusive
  cross-section ${\rm d}^2 \sigma /{\rm d}p{\rm d}\Omega$
  [mb/(GeV/{\it c} sr)] of the production of protons
  in $\pi^+$ + Cu $\rightarrow$ p + X interactions
  with $+3.0$~GeV/{\it c} beam momentum;
  the first error is statistical, the second systematic; 
 $p_{\rm T}$ in GeV/{\it c}, polar angle $\theta$ in degrees.}
 \label{pro.pipcu3}
 \begin{center}
 \begin{tabular}{|c||c|c|rcrcr||c|c|rcrcr|}
 \hline
   & \multicolumn{7}{c||}{$20<\theta<30$}
  & \multicolumn{7}{c|}{$30<\theta<40$} \\
 \hline
 $p_{\rm T}$ & $\langle p_{\rm T} \rangle$ & $\langle \theta \rangle$
  & \multicolumn{5}{c||}{${\rm d}^2 \sigma /{\rm d}p{\rm d}\Omega$}
  &$\langle p_{\rm T} \rangle$ & $\langle \theta \rangle$
  & \multicolumn{5}{c|}{${\rm d}^2 \sigma /{\rm d}p{\rm d}\Omega$} \\
 \hline
0.20--0.24 & 0.221 &  25.3 &  421.79 & $\!\!\pm\!\!$ &  14.72 & $\!\!\pm\!\!$ &  23.52 &  &  &  \multicolumn{5}{c|}{ } \\
0.24--0.30 & 0.271 &  25.2 &  364.28 & $\!\!\pm\!\!$ &  10.24 & $\!\!\pm\!\!$ &  17.75 & 0.272 &  34.8 &  431.30 & $\!\!\pm\!\!$ &  11.49 & $\!\!\pm\!\!$ &  19.50 \\
0.30--0.36 & 0.331 &  25.3 &  325.25 & $\!\!\pm\!\!$ &   9.91 & $\!\!\pm\!\!$ &  13.88 & 0.332 &  35.1 &  374.11 & $\!\!\pm\!\!$ &  10.14 & $\!\!\pm\!\!$ &  14.53 \\
0.36--0.42 & 0.391 &  25.3 &  249.09 & $\!\!\pm\!\!$ &   8.50 & $\!\!\pm\!\!$ &   9.41 & 0.391 &  35.1 &  303.36 & $\!\!\pm\!\!$ &   9.36 & $\!\!\pm\!\!$ &  10.27 \\
0.42--0.50 & 0.462 &  25.4 &  200.66 & $\!\!\pm\!\!$ &   6.50 & $\!\!\pm\!\!$ &   6.80 & 0.463 &  35.1 &  244.71 & $\!\!\pm\!\!$ &   7.19 & $\!\!\pm\!\!$ &   7.36 \\
0.50--0.60 & 0.552 &  25.4 &  149.71 & $\!\!\pm\!\!$ &   5.05 & $\!\!\pm\!\!$ &   5.13 & 0.552 &  35.3 &  190.54 & $\!\!\pm\!\!$ &   5.69 & $\!\!\pm\!\!$ &   5.89 \\
\hline
 \end{tabular}
 \end{center}
 \end{scriptsize}
 \end{table*}
  
 \begin{table*}[h]
 \begin{scriptsize}
 \caption{Double-differential inclusive
  cross-section ${\rm d}^2 \sigma /{\rm d}p{\rm d}\Omega$
  [mb/(GeV/{\it c} sr)] of the production of $\pi^+$'s
  in $\pi^+$ + Cu $\rightarrow$ $\pi^+$ + X interactions
  with $+3.0$~GeV/{\it c} beam momentum;
  the first error is statistical, the second systematic; 
 $p_{\rm T}$ in GeV/{\it c}, polar angle $\theta$ in degrees.}
 \label{pip.pipcu3}
 \begin{center}
 \begin{tabular}{|c||c|c|rcrcr||c|c|rcrcr|}
 \hline
   & \multicolumn{7}{c||}{$20<\theta<30$}
  & \multicolumn{7}{c|}{$30<\theta<40$} \\
 \hline
 $p_{\rm T}$ & $\langle p_{\rm T} \rangle$ & $\langle \theta \rangle$
  & \multicolumn{5}{c||}{${\rm d}^2 \sigma /{\rm d}p{\rm d}\Omega$}
  &$\langle p_{\rm T} \rangle$ & $\langle \theta \rangle$
  & \multicolumn{5}{c|}{${\rm d}^2 \sigma /{\rm d}p{\rm d}\Omega$} \\
 \hline
0.10--0.13 & 0.116 &  24.9 &  143.38 & $\!\!\pm\!\!$ &  10.93 & $\!\!\pm\!\!$ &  12.40 & 0.115 &  35.2 &  138.39 & $\!\!\pm\!\!$ &  10.44 & $\!\!\pm\!\!$ &  11.11 \\
0.13--0.16 & 0.146 &  25.1 &  147.65 & $\!\!\pm\!\!$ &   9.98 & $\!\!\pm\!\!$ &   9.64 & 0.146 &  35.2 &  150.54 & $\!\!\pm\!\!$ &   9.67 & $\!\!\pm\!\!$ &  10.04 \\
0.16--0.20 & 0.182 &  24.9 &  201.29 & $\!\!\pm\!\!$ &   9.96 & $\!\!\pm\!\!$ &  11.46 & 0.181 &  35.0 &  149.48 & $\!\!\pm\!\!$ &   8.16 & $\!\!\pm\!\!$ &   8.08 \\
0.20--0.24 & 0.222 &  24.7 &  212.13 & $\!\!\pm\!\!$ &  10.00 & $\!\!\pm\!\!$ &  10.53 & 0.220 &  34.7 &  167.65 & $\!\!\pm\!\!$ &   8.54 & $\!\!\pm\!\!$ &   7.81 \\
0.24--0.30 & 0.270 &  24.9 &  190.05 & $\!\!\pm\!\!$ &   7.52 & $\!\!\pm\!\!$ &   7.26 & 0.272 &  34.9 &  170.94 & $\!\!\pm\!\!$ &   6.98 & $\!\!\pm\!\!$ &   6.54 \\
0.30--0.36 & 0.330 &  25.0 &  159.98 & $\!\!\pm\!\!$ &   6.83 & $\!\!\pm\!\!$ &   5.54 & 0.332 &  34.9 &  157.11 & $\!\!\pm\!\!$ &   6.79 & $\!\!\pm\!\!$ &   5.41 \\
\hline
 \end{tabular}
 \end{center}
 \end{scriptsize}
 \end{table*}
  
 \begin{table*}[h]
 \begin{scriptsize}
 \caption{Double-differential inclusive
  cross-section ${\rm d}^2 \sigma /{\rm d}p{\rm d}\Omega$
  [mb/(GeV/{\it c} sr)] of the production of $\pi^-$'s
  in $\pi^+$ + Cu $\rightarrow$ $\pi^-$ + X interactions
  with $+3.0$~GeV/{\it c} beam momentum;
  the first error is statistical, the second systematic; 
 $p_{\rm T}$ in GeV/{\it c}, polar angle $\theta$ in degrees.}
 \label{pim.pipcu3}
 \begin{center}
 \begin{tabular}{|c||c|c|rcrcr||c|c|rcrcr|}
 \hline
   & \multicolumn{7}{c||}{$20<\theta<30$}
  & \multicolumn{7}{c|}{$30<\theta<40$} \\
 \hline
 $p_{\rm T}$ & $\langle p_{\rm T} \rangle$ & $\langle \theta \rangle$
  & \multicolumn{5}{c||}{${\rm d}^2 \sigma /{\rm d}p{\rm d}\Omega$}
  &$\langle p_{\rm T} \rangle$ & $\langle \theta \rangle$
  & \multicolumn{5}{c|}{${\rm d}^2 \sigma /{\rm d}p{\rm d}\Omega$} \\
 \hline
0.10--0.13 & 0.115 &  24.8 &  159.43 & $\!\!\pm\!\!$ &  12.24 & $\!\!\pm\!\!$ &  17.20 & 0.115 &  34.9 &  144.77 & $\!\!\pm\!\!$ &  10.72 & $\!\!\pm\!\!$ &  12.50 \\
0.13--0.16 & 0.145 &  24.7 &  131.14 & $\!\!\pm\!\!$ &   9.31 & $\!\!\pm\!\!$ &   9.03 & 0.145 &  35.4 &  117.09 & $\!\!\pm\!\!$ &   8.29 & $\!\!\pm\!\!$ &   8.22 \\
0.16--0.20 & 0.179 &  24.6 &  145.75 & $\!\!\pm\!\!$ &   8.28 & $\!\!\pm\!\!$ &   8.40 & 0.179 &  34.8 &  122.57 & $\!\!\pm\!\!$ &   7.07 & $\!\!\pm\!\!$ &   7.14 \\
0.20--0.24 & 0.219 &  25.2 &  131.80 & $\!\!\pm\!\!$ &   7.82 & $\!\!\pm\!\!$ &   6.49 & 0.219 &  34.9 &  124.90 & $\!\!\pm\!\!$ &   7.46 & $\!\!\pm\!\!$ &   6.48 \\
0.24--0.30 & 0.267 &  25.2 &  125.20 & $\!\!\pm\!\!$ &   6.12 & $\!\!\pm\!\!$ &   4.96 & 0.267 &  34.8 &  107.95 & $\!\!\pm\!\!$ &   5.62 & $\!\!\pm\!\!$ &   4.30 \\
0.30--0.36 & 0.327 &  25.2 &   92.77 & $\!\!\pm\!\!$ &   5.26 & $\!\!\pm\!\!$ &   3.51 & 0.326 &  35.0 &   92.05 & $\!\!\pm\!\!$ &   5.09 & $\!\!\pm\!\!$ &   3.45 \\
\hline
 \end{tabular}
 \end{center}
 \end{scriptsize}
 \end{table*}
  
 \begin{table*}[h]
 \begin{scriptsize}
 \caption{Double-differential inclusive
  cross-section ${\rm d}^2 \sigma /{\rm d}p{\rm d}\Omega$
  [mb/(GeV/{\it c} sr)] of the production of protons
  in $\pi^-$ + Cu $\rightarrow$ p + X interactions
  with $-3.0$~GeV/{\it c} beam momentum;
  the first error is statistical, the second systematic; 
 $p_{\rm T}$ in GeV/{\it c}, polar angle $\theta$ in degrees.}
 \label{pro.pimcu3}
 \begin{center}
 \begin{tabular}{|c||c|c|rcrcr||c|c|rcrcr|}
 \hline
   & \multicolumn{7}{c||}{$20<\theta<30$}
  & \multicolumn{7}{c|}{$30<\theta<40$} \\
 \hline
 $p_{\rm T}$ & $\langle p_{\rm T} \rangle$ & $\langle \theta \rangle$
  & \multicolumn{5}{c||}{${\rm d}^2 \sigma /{\rm d}p{\rm d}\Omega$}
  &$\langle p_{\rm T} \rangle$ & $\langle \theta \rangle$
  & \multicolumn{5}{c|}{${\rm d}^2 \sigma /{\rm d}p{\rm d}\Omega$} \\
 \hline
0.20--0.24 & 0.219 &  25.2 &  372.39 & $\!\!\pm\!\!$ &   7.97 & $\!\!\pm\!\!$ &  20.22 &  &  &  \multicolumn{5}{c|}{ } \\
0.24--0.30 & 0.267 &  25.3 &  318.53 & $\!\!\pm\!\!$ &   5.72 & $\!\!\pm\!\!$ &  15.16 & 0.269 &  35.0 &  377.71 & $\!\!\pm\!\!$ &   6.36 & $\!\!\pm\!\!$ &  16.73 \\
0.30--0.36 & 0.326 &  25.3 &  276.93 & $\!\!\pm\!\!$ &   5.34 & $\!\!\pm\!\!$ &  11.32 & 0.326 &  35.1 &  343.56 & $\!\!\pm\!\!$ &   5.83 & $\!\!\pm\!\!$ &  12.92 \\
0.36--0.42 & 0.384 &  25.2 &  212.26 & $\!\!\pm\!\!$ &   4.59 & $\!\!\pm\!\!$ &   7.61 & 0.385 &  35.1 &  260.18 & $\!\!\pm\!\!$ &   5.03 & $\!\!\pm\!\!$ &   8.41 \\
0.42--0.50 & 0.452 &  25.3 &  164.01 & $\!\!\pm\!\!$ &   3.50 & $\!\!\pm\!\!$ &   5.26 & 0.451 &  35.2 &  219.72 & $\!\!\pm\!\!$ &   4.05 & $\!\!\pm\!\!$ &   6.24 \\
0.50--0.60 & 0.538 &  25.4 &  112.87 & $\!\!\pm\!\!$ &   2.55 & $\!\!\pm\!\!$ &   3.67 & 0.537 &  35.2 &  151.30 & $\!\!\pm\!\!$ &   2.97 & $\!\!\pm\!\!$ &   4.42 \\
\hline
 \end{tabular}
 \end{center}
 \end{scriptsize}
 \end{table*}
  
 \begin{table*}[h]
 \begin{scriptsize}
 \caption{Double-differential inclusive
  cross-section ${\rm d}^2 \sigma /{\rm d}p{\rm d}\Omega$
  [mb/(GeV/{\it c} sr)] of the production of $\pi^+$'s
  in $\pi^-$ + Cu $\rightarrow$ $\pi^+$ + X interactions
  with $-3.0$~GeV/{\it c} beam momentum;
  the first error is statistical, the second systematic; 
 $p_{\rm T}$ in GeV/{\it c}, polar angle $\theta$ in degrees.}
 \label{pip.pimcu3}
 \begin{center}
 \begin{tabular}{|c||c|c|rcrcr||c|c|rcrcr|}
 \hline
   & \multicolumn{7}{c||}{$20<\theta<30$}
  & \multicolumn{7}{c|}{$30<\theta<40$} \\
 \hline
 $p_{\rm T}$ & $\langle p_{\rm T} \rangle$ & $\langle \theta \rangle$
  & \multicolumn{5}{c||}{${\rm d}^2 \sigma /{\rm d}p{\rm d}\Omega$}
  &$\langle p_{\rm T} \rangle$ & $\langle \theta \rangle$
  & \multicolumn{5}{c|}{${\rm d}^2 \sigma /{\rm d}p{\rm d}\Omega$} \\
 \hline
0.10--0.13 & 0.115 &  25.1 &  105.80 & $\!\!\pm\!\!$ &   5.24 & $\!\!\pm\!\!$ &   7.84 & 0.115 &  35.0 &  106.42 & $\!\!\pm\!\!$ &   5.05 & $\!\!\pm\!\!$ &   8.09 \\
0.13--0.16 & 0.145 &  24.8 &  140.10 & $\!\!\pm\!\!$ &   5.75 & $\!\!\pm\!\!$ &   8.95 & 0.145 &  35.0 &  123.12 & $\!\!\pm\!\!$ &   5.18 & $\!\!\pm\!\!$ &   7.51 \\
0.16--0.20 & 0.179 &  25.0 &  148.23 & $\!\!\pm\!\!$ &   4.96 & $\!\!\pm\!\!$ &   7.91 & 0.179 &  34.9 &  129.96 & $\!\!\pm\!\!$ &   4.42 & $\!\!\pm\!\!$ &   6.59 \\
0.20--0.24 & 0.219 &  25.1 &  157.18 & $\!\!\pm\!\!$ &   4.95 & $\!\!\pm\!\!$ &   6.94 & 0.219 &  34.7 &  141.06 & $\!\!\pm\!\!$ &   4.65 & $\!\!\pm\!\!$ &   6.28 \\
0.24--0.30 & 0.268 &  25.0 &  146.13 & $\!\!\pm\!\!$ &   3.88 & $\!\!\pm\!\!$ &   5.32 & 0.268 &  34.8 &  125.60 & $\!\!\pm\!\!$ &   3.56 & $\!\!\pm\!\!$ &   4.50 \\
0.30--0.36 & 0.326 &  25.2 &  119.20 & $\!\!\pm\!\!$ &   3.53 & $\!\!\pm\!\!$ &   4.07 & 0.325 &  34.8 &  113.31 & $\!\!\pm\!\!$ &   3.39 & $\!\!\pm\!\!$ &   3.62 \\
\hline
 \end{tabular}
 \end{center}
 \end{scriptsize}
 \end{table*}
  
 \begin{table*}[h]
 \begin{scriptsize}
 \caption{Double-differential inclusive
  cross-section ${\rm d}^2 \sigma /{\rm d}p{\rm d}\Omega$
  [mb/(GeV/{\it c} sr)] of the production of $\pi^-$'s
  in $\pi^-$ + Cu $\rightarrow$ $\pi^-$ + X interactions
  with $-3.0$~GeV/{\it c} beam momentum;
  the first error is statistical, the second systematic; 
 $p_{\rm T}$ in GeV/{\it c}, polar angle $\theta$ in degrees.}
 \label{pim.pimcu3}
 \begin{center}
 \begin{tabular}{|c||c|c|rcrcr||c|c|rcrcr|}
 \hline
   & \multicolumn{7}{c||}{$20<\theta<30$}
  & \multicolumn{7}{c|}{$30<\theta<40$} \\
 \hline
 $p_{\rm T}$ & $\langle p_{\rm T} \rangle$ & $\langle \theta \rangle$
  & \multicolumn{5}{c||}{${\rm d}^2 \sigma /{\rm d}p{\rm d}\Omega$}
  &$\langle p_{\rm T} \rangle$ & $\langle \theta \rangle$
  & \multicolumn{5}{c|}{${\rm d}^2 \sigma /{\rm d}p{\rm d}\Omega$} \\
 \hline
0.10--0.13 & 0.116 &  24.9 &  184.40 & $\!\!\pm\!\!$ &   7.12 & $\!\!\pm\!\!$ &  13.72 & 0.116 &  35.2 &  182.94 & $\!\!\pm\!\!$ &   7.09 & $\!\!\pm\!\!$ &  13.98 \\
0.13--0.16 & 0.146 &  24.9 &  210.24 & $\!\!\pm\!\!$ &   6.99 & $\!\!\pm\!\!$ &  12.67 & 0.146 &  35.1 &  185.66 & $\!\!\pm\!\!$ &   6.38 & $\!\!\pm\!\!$ &  11.24 \\
0.16--0.20 & 0.182 &  24.8 &  239.03 & $\!\!\pm\!\!$ &   6.33 & $\!\!\pm\!\!$ &  12.28 & 0.181 &  35.0 &  199.88 & $\!\!\pm\!\!$ &   5.56 & $\!\!\pm\!\!$ &  10.06 \\
0.20--0.24 & 0.221 &  25.0 &  234.53 & $\!\!\pm\!\!$ &   6.14 & $\!\!\pm\!\!$ &   9.83 & 0.221 &  34.7 &  205.78 & $\!\!\pm\!\!$ &   5.66 & $\!\!\pm\!\!$ &   8.70 \\
0.24--0.30 & 0.272 &  24.9 &  210.26 & $\!\!\pm\!\!$ &   4.74 & $\!\!\pm\!\!$ &   7.29 & 0.272 &  34.9 &  184.50 & $\!\!\pm\!\!$ &   4.36 & $\!\!\pm\!\!$ &   6.29 \\
0.30--0.36 & 0.332 &  25.1 &  177.48 & $\!\!\pm\!\!$ &   4.36 & $\!\!\pm\!\!$ &   5.44 & 0.332 &  34.8 &  164.46 & $\!\!\pm\!\!$ &   4.09 & $\!\!\pm\!\!$ &   4.84 \\
\hline
 \end{tabular}
 \end{center}
 \end{scriptsize}
 \end{table*}
\clearpage


  
 \begin{table*}[h]
 \begin{scriptsize}
 \caption{Double-differential inclusive
  cross-section ${\rm d}^2 \sigma /{\rm d}p{\rm d}\Omega$
  [mb/(GeV/{\it c} sr)] of the production of protons
  in p + Cu $\rightarrow$ p + X interactions
  with $+5.0$~GeV/{\it c} beam momentum;
  the first error is statistical, the second systematic; 
 $p_{\rm T}$ in GeV/{\it c}, polar angle $\theta$ in degrees.}
 \label{pro.procu5}
 \begin{center}
 \begin{tabular}{|c||c|c|rcrcr||c|c|rcrcr|}
 \hline
   & \multicolumn{7}{c||}{$20<\theta<30$}
  & \multicolumn{7}{c|}{$30<\theta<40$} \\
 \hline
 $p_{\rm T}$ & $\langle p_{\rm T} \rangle$ & $\langle \theta \rangle$
  & \multicolumn{5}{c||}{${\rm d}^2 \sigma /{\rm d}p{\rm d}\Omega$}
  &$\langle p_{\rm T} \rangle$ & $\langle \theta \rangle$
  & \multicolumn{5}{c|}{${\rm d}^2 \sigma /{\rm d}p{\rm d}\Omega$} \\
 \hline
0.20--0.24 & 0.220 &  25.1 &  490.74 & $\!\!\pm\!\!$ &  10.91 & $\!\!\pm\!\!$ &  26.52 &  &  &  \multicolumn{5}{c|}{ } \\
0.24--0.30 & 0.270 &  25.3 &  473.49 & $\!\!\pm\!\!$ &   8.32 & $\!\!\pm\!\!$ &  22.37 & 0.271 &  34.9 &  504.43 & $\!\!\pm\!\!$ &   8.78 & $\!\!\pm\!\!$ &  22.18 \\
0.30--0.36 & 0.330 &  25.2 &  402.84 & $\!\!\pm\!\!$ &   7.59 & $\!\!\pm\!\!$ &  16.31 & 0.331 &  35.0 &  448.88 & $\!\!\pm\!\!$ &   7.95 & $\!\!\pm\!\!$ &  16.70 \\
0.36--0.42 & 0.390 &  25.2 &  352.60 & $\!\!\pm\!\!$ &   7.19 & $\!\!\pm\!\!$ &  12.41 & 0.391 &  35.1 &  399.77 & $\!\!\pm\!\!$ &   7.52 & $\!\!\pm\!\!$ &  12.73 \\
0.42--0.50 & 0.461 &  25.2 &  285.95 & $\!\!\pm\!\!$ &   5.48 & $\!\!\pm\!\!$ &   8.92 & 0.461 &  35.1 &  312.22 & $\!\!\pm\!\!$ &   5.69 & $\!\!\pm\!\!$ &   8.69 \\
0.50--0.60 & 0.550 &  25.2 &  215.47 & $\!\!\pm\!\!$ &   4.24 & $\!\!\pm\!\!$ &   6.75 & 0.551 &  35.1 &  237.47 & $\!\!\pm\!\!$ &   4.45 & $\!\!\pm\!\!$ &   6.78 \\
\hline
 \end{tabular}
 \end{center}
 \end{scriptsize}
 \end{table*}
  
 \begin{table*}[h]
 \begin{scriptsize}
 \caption{Double-differential inclusive
  cross-section ${\rm d}^2 \sigma /{\rm d}p{\rm d}\Omega$
  [mb/(GeV/{\it c} sr)] of the production of $\pi^+$'s
  in p + Cu $\rightarrow$ $\pi^+$ + X interactions
  with $+5.0$~GeV/{\it c} beam momentum;
  the first error is statistical, the second systematic; 
 $p_{\rm T}$ in GeV/{\it c}, polar angle $\theta$ in degrees.}
 \label{pip.procu5}
 \begin{center}
 \begin{tabular}{|c||c|c|rcrcr||c|c|rcrcr|}
 \hline
   & \multicolumn{7}{c||}{$20<\theta<30$}
  & \multicolumn{7}{c|}{$30<\theta<40$} \\
 \hline
 $p_{\rm T}$ & $\langle p_{\rm T} \rangle$ & $\langle \theta \rangle$
  & \multicolumn{5}{c||}{${\rm d}^2 \sigma /{\rm d}p{\rm d}\Omega$}
  &$\langle p_{\rm T} \rangle$ & $\langle \theta \rangle$
  & \multicolumn{5}{c|}{${\rm d}^2 \sigma /{\rm d}p{\rm d}\Omega$} \\
 \hline
0.10--0.13 & 0.116 &  24.9 &  168.02 & $\!\!\pm\!\!$ &   8.06 & $\!\!\pm\!\!$ &  12.38 & 0.116 &  34.7 &  154.11 & $\!\!\pm\!\!$ &   7.72 & $\!\!\pm\!\!$ &  11.60 \\
0.13--0.16 & 0.145 &  24.8 &  213.31 & $\!\!\pm\!\!$ &   8.53 & $\!\!\pm\!\!$ &  13.01 & 0.145 &  34.9 &  155.03 & $\!\!\pm\!\!$ &   6.89 & $\!\!\pm\!\!$ &   9.31 \\
0.16--0.20 & 0.180 &  24.9 &  235.27 & $\!\!\pm\!\!$ &   7.38 & $\!\!\pm\!\!$ &  11.72 & 0.181 &  34.6 &  175.62 & $\!\!\pm\!\!$ &   6.24 & $\!\!\pm\!\!$ &   8.78 \\
0.20--0.24 & 0.220 &  24.9 &  236.54 & $\!\!\pm\!\!$ &   7.31 & $\!\!\pm\!\!$ &  10.27 & 0.221 &  34.6 &  182.94 & $\!\!\pm\!\!$ &   6.29 & $\!\!\pm\!\!$ &   7.88 \\
0.24--0.30 & 0.270 &  24.8 &  205.89 & $\!\!\pm\!\!$ &   5.59 & $\!\!\pm\!\!$ &   7.78 & 0.270 &  34.7 &  171.78 & $\!\!\pm\!\!$ &   5.00 & $\!\!\pm\!\!$ &   6.11 \\
0.30--0.36 & 0.331 &  25.0 &  158.63 & $\!\!\pm\!\!$ &   4.75 & $\!\!\pm\!\!$ &   5.09 & 0.330 &  34.8 &  132.77 & $\!\!\pm\!\!$ &   4.38 & $\!\!\pm\!\!$ &   4.19 \\
\hline
 \end{tabular}
 \end{center}
 \end{scriptsize}
 \end{table*}
  
 \begin{table*}[h]
 \begin{scriptsize}
 \caption{Double-differential inclusive
  cross-section ${\rm d}^2 \sigma /{\rm d}p{\rm d}\Omega$
  [mb/(GeV/{\it c} sr)] of the production of $\pi^-$'s
  in p + Cu $\rightarrow$ $\pi^-$ + X interactions
  with $+5.0$~GeV/{\it c} beam momentum;
  the first error is statistical, the second systematic; 
 $p_{\rm T}$ in GeV/{\it c}, polar angle $\theta$ in degrees.}
 \label{pim.procu5}
 \begin{center}
 \begin{tabular}{|c||c|c|rcrcr||c|c|rcrcr|}
 \hline
   & \multicolumn{7}{c||}{$20<\theta<30$}
  & \multicolumn{7}{c|}{$30<\theta<40$} \\
 \hline
 $p_{\rm T}$ & $\langle p_{\rm T} \rangle$ & $\langle \theta \rangle$
  & \multicolumn{5}{c||}{${\rm d}^2 \sigma /{\rm d}p{\rm d}\Omega$}
  &$\langle p_{\rm T} \rangle$ & $\langle \theta \rangle$
  & \multicolumn{5}{c|}{${\rm d}^2 \sigma /{\rm d}p{\rm d}\Omega$} \\
 \hline
0.10--0.13 & 0.115 &  25.2 &  156.73 & $\!\!\pm\!\!$ &   8.05 & $\!\!\pm\!\!$ &  13.14 & 0.115 &  35.1 &  135.97 & $\!\!\pm\!\!$ &   6.96 & $\!\!\pm\!\!$ &  10.65 \\
0.13--0.16 & 0.145 &  25.0 &  177.63 & $\!\!\pm\!\!$ &   7.80 & $\!\!\pm\!\!$ &  11.44 & 0.145 &  35.0 &  167.15 & $\!\!\pm\!\!$ &   7.25 & $\!\!\pm\!\!$ &  10.65 \\
0.16--0.20 & 0.180 &  25.0 &  172.82 & $\!\!\pm\!\!$ &   6.35 & $\!\!\pm\!\!$ &   8.92 & 0.179 &  34.8 &  138.91 & $\!\!\pm\!\!$ &   5.52 & $\!\!\pm\!\!$ &   7.28 \\
0.20--0.24 & 0.219 &  24.9 &  165.03 & $\!\!\pm\!\!$ &   6.18 & $\!\!\pm\!\!$ &   7.46 & 0.219 &  35.0 &  159.01 & $\!\!\pm\!\!$ &   5.96 & $\!\!\pm\!\!$ &   7.07 \\
0.24--0.30 & 0.268 &  24.9 &  132.03 & $\!\!\pm\!\!$ &   4.42 & $\!\!\pm\!\!$ &   4.76 & 0.268 &  35.0 &  119.79 & $\!\!\pm\!\!$ &   4.14 & $\!\!\pm\!\!$ &   4.33 \\
0.30--0.36 & 0.327 &  25.0 &   95.54 & $\!\!\pm\!\!$ &   3.77 & $\!\!\pm\!\!$ &   3.18 & 0.327 &  34.8 &  103.32 & $\!\!\pm\!\!$ &   3.90 & $\!\!\pm\!\!$ &   3.43 \\
\hline
 \end{tabular}
 \end{center}
 \end{scriptsize}
 \end{table*}
  
 \begin{table*}[h]
 \begin{scriptsize}
 \caption{Double-differential inclusive
  cross-section ${\rm d}^2 \sigma /{\rm d}p{\rm d}\Omega$
  [mb/(GeV/{\it c} sr)] of the production of protons
  in $\pi^+$ + Cu $\rightarrow$ p + X interactions
  with $+5.0$~GeV/{\it c} beam momentum;
  the first error is statistical, the second systematic; 
 $p_{\rm T}$ in GeV/{\it c}, polar angle $\theta$ in degrees.}
 \label{pro.pipcu5}
 \begin{center}
 \begin{tabular}{|c||c|c|rcrcr||c|c|rcrcr|}
 \hline
   & \multicolumn{7}{c||}{$20<\theta<30$}
  & \multicolumn{7}{c|}{$30<\theta<40$} \\
 \hline
 $p_{\rm T}$ & $\langle p_{\rm T} \rangle$ & $\langle \theta \rangle$
  & \multicolumn{5}{c||}{${\rm d}^2 \sigma /{\rm d}p{\rm d}\Omega$}
  &$\langle p_{\rm T} \rangle$ & $\langle \theta \rangle$
  & \multicolumn{5}{c|}{${\rm d}^2 \sigma /{\rm d}p{\rm d}\Omega$} \\
 \hline
0.20--0.24 & 0.221 &  25.3 &  487.58 & $\!\!\pm\!\!$ &   9.85 & $\!\!\pm\!\!$ &  26.98 &  &  &  \multicolumn{5}{c|}{ } \\
0.24--0.30 & 0.269 &  25.2 &  408.01 & $\!\!\pm\!\!$ &   6.90 & $\!\!\pm\!\!$ &  19.72 & 0.271 &  35.0 &  452.54 & $\!\!\pm\!\!$ &   7.48 & $\!\!\pm\!\!$ &  20.35 \\
0.30--0.36 & 0.330 &  25.2 &  349.24 & $\!\!\pm\!\!$ &   6.31 & $\!\!\pm\!\!$ &  14.63 & 0.330 &  35.1 &  402.65 & $\!\!\pm\!\!$ &   6.75 & $\!\!\pm\!\!$ &  15.50 \\
0.36--0.42 & 0.390 &  25.3 &  290.63 & $\!\!\pm\!\!$ &   5.83 & $\!\!\pm\!\!$ &  10.77 & 0.390 &  35.0 &  339.08 & $\!\!\pm\!\!$ &   6.19 & $\!\!\pm\!\!$ &  11.40 \\
0.42--0.50 & 0.460 &  25.2 &  239.68 & $\!\!\pm\!\!$ &   4.44 & $\!\!\pm\!\!$ &   7.99 & 0.460 &  35.1 &  274.23 & $\!\!\pm\!\!$ &   4.77 & $\!\!\pm\!\!$ &   8.19 \\
0.50--0.60 & 0.550 &  25.3 &  170.53 & $\!\!\pm\!\!$ &   3.31 & $\!\!\pm\!\!$ &   5.77 & 0.550 &  35.1 &  201.59 & $\!\!\pm\!\!$ &   3.65 & $\!\!\pm\!\!$ &   6.21 \\
\hline
 \end{tabular}
 \end{center}
 \end{scriptsize}
 \end{table*}
  
 \begin{table*}[h]
 \begin{scriptsize}
 \caption{Double-differential inclusive
  cross-section ${\rm d}^2 \sigma /{\rm d}p{\rm d}\Omega$
  [mb/(GeV/{\it c} sr)] of the production of $\pi^+$'s
  in $\pi^+$ + Cu $\rightarrow$ $\pi^+$ + X interactions
  with $+5.0$~GeV/{\it c} beam momentum;
  the first error is statistical, the second systematic; 
 $p_{\rm T}$ in GeV/{\it c}, polar angle $\theta$ in degrees.}
 \label{pip.pipcu5}
 \begin{center}
 \begin{tabular}{|c||c|c|rcrcr||c|c|rcrcr|}
 \hline
   & \multicolumn{7}{c||}{$20<\theta<30$}
  & \multicolumn{7}{c|}{$30<\theta<40$} \\
 \hline
 $p_{\rm T}$ & $\langle p_{\rm T} \rangle$ & $\langle \theta \rangle$
  & \multicolumn{5}{c||}{${\rm d}^2 \sigma /{\rm d}p{\rm d}\Omega$}
  &$\langle p_{\rm T} \rangle$ & $\langle \theta \rangle$
  & \multicolumn{5}{c|}{${\rm d}^2 \sigma /{\rm d}p{\rm d}\Omega$} \\
 \hline
0.10--0.13 & 0.116 &  24.7 &  183.45 & $\!\!\pm\!\!$ &   7.39 & $\!\!\pm\!\!$ &  13.86 & 0.116 &  35.1 &  200.03 & $\!\!\pm\!\!$ &   7.79 & $\!\!\pm\!\!$ &  15.35 \\
0.13--0.16 & 0.146 &  24.8 &  254.15 & $\!\!\pm\!\!$ &   8.26 & $\!\!\pm\!\!$ &  16.05 & 0.145 &  34.8 &  187.99 & $\!\!\pm\!\!$ &   6.75 & $\!\!\pm\!\!$ &  11.70 \\
0.16--0.20 & 0.181 &  24.7 &  278.40 & $\!\!\pm\!\!$ &   7.09 & $\!\!\pm\!\!$ &  14.21 & 0.181 &  34.7 &  236.11 & $\!\!\pm\!\!$ &   6.48 & $\!\!\pm\!\!$ &  12.20 \\
0.20--0.24 & 0.220 &  24.8 &  330.01 & $\!\!\pm\!\!$ &   7.68 & $\!\!\pm\!\!$ &  14.77 & 0.220 &  34.7 &  240.76 & $\!\!\pm\!\!$ &   6.43 & $\!\!\pm\!\!$ &  10.72 \\
0.24--0.30 & 0.271 &  24.9 &  301.66 & $\!\!\pm\!\!$ &   6.04 & $\!\!\pm\!\!$ &  11.68 & 0.270 &  34.8 &  238.51 & $\!\!\pm\!\!$ &   5.26 & $\!\!\pm\!\!$ &   8.73 \\
0.30--0.36 & 0.331 &  24.7 &  253.97 & $\!\!\pm\!\!$ &   5.38 & $\!\!\pm\!\!$ &   8.34 & 0.330 &  34.8 &  205.73 & $\!\!\pm\!\!$ &   4.89 & $\!\!\pm\!\!$ &   6.71 \\
\hline
 \end{tabular}
 \end{center}
 \end{scriptsize}
 \end{table*}
  
 \begin{table*}[h]
 \begin{scriptsize}
 \caption{Double-differential inclusive
  cross-section ${\rm d}^2 \sigma /{\rm d}p{\rm d}\Omega$
  [mb/(GeV/{\it c} sr)] of the production of $\pi^-$'s
  in $\pi^+$ + Cu $\rightarrow$ $\pi^-$ + X interactions
  with $+5.0$~GeV/{\it c} beam momentum;
  the first error is statistical, the second systematic; 
 $p_{\rm T}$ in GeV/{\it c}, polar angle $\theta$ in degrees.}
 \label{pim.pipcu5}
 \begin{center}
 \begin{tabular}{|c||c|c|rcrcr||c|c|rcrcr|}
 \hline
   & \multicolumn{7}{c||}{$20<\theta<30$}
  & \multicolumn{7}{c|}{$30<\theta<40$} \\
 \hline
 $p_{\rm T}$ & $\langle p_{\rm T} \rangle$ & $\langle \theta \rangle$
  & \multicolumn{5}{c||}{${\rm d}^2 \sigma /{\rm d}p{\rm d}\Omega$}
  &$\langle p_{\rm T} \rangle$ & $\langle \theta \rangle$
  & \multicolumn{5}{c|}{${\rm d}^2 \sigma /{\rm d}p{\rm d}\Omega$} \\
 \hline
0.10--0.13 & 0.115 &  25.1 &  200.20 & $\!\!\pm\!\!$ &   7.97 & $\!\!\pm\!\!$ &  16.47 & 0.115 &  35.0 &  178.01 & $\!\!\pm\!\!$ &   7.07 & $\!\!\pm\!\!$ &  14.29 \\
0.13--0.16 & 0.145 &  24.7 &  225.10 & $\!\!\pm\!\!$ &   7.74 & $\!\!\pm\!\!$ &  14.86 & 0.144 &  34.7 &  186.67 & $\!\!\pm\!\!$ &   6.84 & $\!\!\pm\!\!$ &  12.36 \\
0.16--0.20 & 0.180 &  24.9 &  227.21 & $\!\!\pm\!\!$ &   6.47 & $\!\!\pm\!\!$ &  12.14 & 0.179 &  34.9 &  180.59 & $\!\!\pm\!\!$ &   5.59 & $\!\!\pm\!\!$ &   9.85 \\
0.20--0.24 & 0.219 &  24.8 &  238.32 & $\!\!\pm\!\!$ &   6.60 & $\!\!\pm\!\!$ &  11.24 & 0.220 &  34.7 &  179.42 & $\!\!\pm\!\!$ &   5.62 & $\!\!\pm\!\!$ &   8.40 \\
0.24--0.30 & 0.268 &  24.9 &  203.11 & $\!\!\pm\!\!$ &   4.91 & $\!\!\pm\!\!$ &   7.61 & 0.269 &  34.8 &  164.35 & $\!\!\pm\!\!$ &   4.32 & $\!\!\pm\!\!$ &   6.22 \\
0.30--0.36 & 0.328 &  24.9 &  163.72 & $\!\!\pm\!\!$ &   4.39 & $\!\!\pm\!\!$ &   5.59 & 0.327 &  35.0 &  139.18 & $\!\!\pm\!\!$ &   4.05 & $\!\!\pm\!\!$ &   4.86 \\
\hline
 \end{tabular}
 \end{center}
 \end{scriptsize}
 \end{table*}
  
 \begin{table*}[h]
 \begin{scriptsize}
 \caption{Double-differential inclusive
  cross-section ${\rm d}^2 \sigma /{\rm d}p{\rm d}\Omega$
  [mb/(GeV/{\it c} sr)] of the production of protons
  in $\pi^-$ + Cu $\rightarrow$ p + X interactions
  with $-5.0$~GeV/{\it c} beam momentum;
  the first error is statistical, the second systematic; 
 $p_{\rm T}$ in GeV/{\it c}, polar angle $\theta$ in degrees.}
 \label{pro.pimcu5}
 \begin{center}
 \begin{tabular}{|c||c|c|rcrcr||c|c|rcrcr|}
 \hline
   & \multicolumn{7}{c||}{$20<\theta<30$}
  & \multicolumn{7}{c|}{$30<\theta<40$} \\
 \hline
 $p_{\rm T}$ & $\langle p_{\rm T} \rangle$ & $\langle \theta \rangle$
  & \multicolumn{5}{c||}{${\rm d}^2 \sigma /{\rm d}p{\rm d}\Omega$}
  &$\langle p_{\rm T} \rangle$ & $\langle \theta \rangle$
  & \multicolumn{5}{c|}{${\rm d}^2 \sigma /{\rm d}p{\rm d}\Omega$} \\
 \hline
0.20--0.24 & 0.219 &  25.2 &  408.33 & $\!\!\pm\!\!$ &   9.83 & $\!\!\pm\!\!$ &  22.13 &  &  &  \multicolumn{5}{c|}{ } \\
0.24--0.30 & 0.267 &  25.2 &  365.49 & $\!\!\pm\!\!$ &   7.25 & $\!\!\pm\!\!$ &  17.35 & 0.269 &  34.9 &  412.34 & $\!\!\pm\!\!$ &   7.74 & $\!\!\pm\!\!$ &  18.21 \\
0.30--0.36 & 0.326 &  25.2 &  305.38 & $\!\!\pm\!\!$ &   6.58 & $\!\!\pm\!\!$ &  12.42 & 0.326 &  35.2 &  356.95 & $\!\!\pm\!\!$ &   7.03 & $\!\!\pm\!\!$ &  13.37 \\
0.36--0.42 & 0.385 &  25.2 &  248.79 & $\!\!\pm\!\!$ &   5.85 & $\!\!\pm\!\!$ &   8.81 & 0.385 &  35.0 &  292.51 & $\!\!\pm\!\!$ &   6.31 & $\!\!\pm\!\!$ &   9.40 \\
0.42--0.50 & 0.452 &  25.4 &  199.83 & $\!\!\pm\!\!$ &   4.54 & $\!\!\pm\!\!$ &   6.29 & 0.451 &  35.2 &  232.23 & $\!\!\pm\!\!$ &   4.84 & $\!\!\pm\!\!$ &   6.54 \\
0.50--0.60 & 0.536 &  25.3 &  142.76 & $\!\!\pm\!\!$ &   3.40 & $\!\!\pm\!\!$ &   4.53 & 0.537 &  35.0 &  180.82 & $\!\!\pm\!\!$ &   3.86 & $\!\!\pm\!\!$ &   5.23 \\
\hline
 \end{tabular}
 \end{center}
 \end{scriptsize}
 \end{table*}
  
 \begin{table*}[h]
 \begin{scriptsize}
 \caption{Double-differential inclusive
  cross-section ${\rm d}^2 \sigma /{\rm d}p{\rm d}\Omega$
  [mb/(GeV/{\it c} sr)] of the production of $\pi^+$'s
  in $\pi^-$ + Cu $\rightarrow$ $\pi^+$ + X interactions
  with $-5.0$~GeV/{\it c} beam momentum;
  the first error is statistical, the second systematic; 
 $p_{\rm T}$ in GeV/{\it c}, polar angle $\theta$ in degrees.}
 \label{pip.pimcu5}
 \begin{center}
 \begin{tabular}{|c||c|c|rcrcr||c|c|rcrcr|}
 \hline
   & \multicolumn{7}{c||}{$20<\theta<30$}
  & \multicolumn{7}{c|}{$30<\theta<40$} \\
 \hline
 $p_{\rm T}$ & $\langle p_{\rm T} \rangle$ & $\langle \theta \rangle$
  & \multicolumn{5}{c||}{${\rm d}^2 \sigma /{\rm d}p{\rm d}\Omega$}
  &$\langle p_{\rm T} \rangle$ & $\langle \theta \rangle$
  & \multicolumn{5}{c|}{${\rm d}^2 \sigma /{\rm d}p{\rm d}\Omega$} \\
 \hline
0.10--0.13 & 0.115 &  24.8 &  188.30 & $\!\!\pm\!\!$ &   8.56 & $\!\!\pm\!\!$ &  14.69 & 0.115 &  35.0 &  151.51 & $\!\!\pm\!\!$ &   7.33 & $\!\!\pm\!\!$ &  11.23 \\
0.13--0.16 & 0.145 &  25.0 &  220.45 & $\!\!\pm\!\!$ &   8.50 & $\!\!\pm\!\!$ &  13.48 & 0.144 &  34.8 &  166.26 & $\!\!\pm\!\!$ &   6.96 & $\!\!\pm\!\!$ &   9.81 \\
0.16--0.20 & 0.179 &  24.9 &  237.40 & $\!\!\pm\!\!$ &   7.26 & $\!\!\pm\!\!$ &  11.84 & 0.178 &  34.7 &  185.43 & $\!\!\pm\!\!$ &   6.30 & $\!\!\pm\!\!$ &   9.14 \\
0.20--0.24 & 0.219 &  24.8 &  241.44 & $\!\!\pm\!\!$ &   7.17 & $\!\!\pm\!\!$ &  10.29 & 0.218 &  34.8 &  196.53 & $\!\!\pm\!\!$ &   6.53 & $\!\!\pm\!\!$ &   8.53 \\
0.24--0.30 & 0.267 &  24.8 &  227.58 & $\!\!\pm\!\!$ &   5.69 & $\!\!\pm\!\!$ &   8.14 & 0.267 &  34.7 &  174.77 & $\!\!\pm\!\!$ &   4.98 & $\!\!\pm\!\!$ &   6.16 \\
0.30--0.36 & 0.326 &  24.8 &  178.95 & $\!\!\pm\!\!$ &   5.00 & $\!\!\pm\!\!$ &   5.56 & 0.325 &  34.8 &  159.38 & $\!\!\pm\!\!$ &   4.71 & $\!\!\pm\!\!$ &   4.84 \\
\hline
 \end{tabular}
 \end{center}
 \end{scriptsize}
 \end{table*}
  
 \begin{table*}[h]
 \begin{scriptsize}
 \caption{Double-differential inclusive
  cross-section ${\rm d}^2 \sigma /{\rm d}p{\rm d}\Omega$
  [mb/(GeV/{\it c} sr)] of the production of $\pi^-$'s
  in $\pi^-$ + Cu $\rightarrow$ $\pi^-$ + X interactions
  with $-5.0$~GeV/{\it c} beam momentum;
  the first error is statistical, the second systematic; 
 $p_{\rm T}$ in GeV/{\it c}, polar angle $\theta$ in degrees.}
 \label{pim.pimcu5}
 \begin{center}
 \begin{tabular}{|c||c|c|rcrcr||c|c|rcrcr|}
 \hline
   & \multicolumn{7}{c||}{$20<\theta<30$}
  & \multicolumn{7}{c|}{$30<\theta<40$} \\
 \hline
 $p_{\rm T}$ & $\langle p_{\rm T} \rangle$ & $\langle \theta \rangle$
  & \multicolumn{5}{c||}{${\rm d}^2 \sigma /{\rm d}p{\rm d}\Omega$}
  &$\langle p_{\rm T} \rangle$ & $\langle \theta \rangle$
  & \multicolumn{5}{c|}{${\rm d}^2 \sigma /{\rm d}p{\rm d}\Omega$} \\
 \hline
0.10--0.13 & 0.116 &  24.9 &  281.13 & $\!\!\pm\!\!$ &  10.60 & $\!\!\pm\!\!$ &  21.53 & 0.116 &  34.8 &  266.41 & $\!\!\pm\!\!$ &  10.06 & $\!\!\pm\!\!$ &  19.95 \\
0.13--0.16 & 0.147 &  24.8 &  317.36 & $\!\!\pm\!\!$ &  10.11 & $\!\!\pm\!\!$ &  18.81 & 0.146 &  34.9 &  245.92 & $\!\!\pm\!\!$ &   8.80 & $\!\!\pm\!\!$ &  14.63 \\
0.16--0.20 & 0.181 &  24.7 &  358.09 & $\!\!\pm\!\!$ &   9.21 & $\!\!\pm\!\!$ &  18.51 & 0.181 &  34.7 &  278.72 & $\!\!\pm\!\!$ &   7.89 & $\!\!\pm\!\!$ &  13.83 \\
0.20--0.24 & 0.221 &  24.8 &  347.49 & $\!\!\pm\!\!$ &   8.74 & $\!\!\pm\!\!$ &  14.29 & 0.221 &  34.7 &  269.49 & $\!\!\pm\!\!$ &   7.68 & $\!\!\pm\!\!$ &  11.28 \\
0.24--0.30 & 0.271 &  25.0 &  322.00 & $\!\!\pm\!\!$ &   6.93 & $\!\!\pm\!\!$ &  11.17 & 0.272 &  34.7 &  252.74 & $\!\!\pm\!\!$ &   6.05 & $\!\!\pm\!\!$ &   8.48 \\
0.30--0.36 & 0.332 &  24.9 &  279.91 & $\!\!\pm\!\!$ &   6.37 & $\!\!\pm\!\!$ &   8.00 & 0.331 &  34.8 &  219.62 & $\!\!\pm\!\!$ &   5.63 & $\!\!\pm\!\!$ &   6.27 \\
\hline
 \end{tabular}
 \end{center}
 \end{scriptsize}
 \end{table*}
\clearpage


  
 \begin{table*}[h]
 \begin{scriptsize}
 \caption{Double-differential inclusive
  cross-section ${\rm d}^2 \sigma /{\rm d}p{\rm d}\Omega$
  [mb/(GeV/{\it c} sr)] of the production of protons
  in p + Cu $\rightarrow$ p + X interactions
  with $+8.0$~GeV/{\it c} beam momentum;
  the first error is statistical, the second systematic; 
 $p_{\rm T}$ in GeV/{\it c}, polar angle $\theta$ in degrees.}
 \label{pro.procu8}
 \begin{center}
 \begin{tabular}{|c||c|c|rcrcr||c|c|rcrcr|}
 \hline
   & \multicolumn{7}{c||}{$20<\theta<30$}
  & \multicolumn{7}{c|}{$30<\theta<40$} \\
 \hline
 $p_{\rm T}$ & $\langle p_{\rm T} \rangle$ & $\langle \theta \rangle$
  & \multicolumn{5}{c||}{${\rm d}^2 \sigma /{\rm d}p{\rm d}\Omega$}
  &$\langle p_{\rm T} \rangle$ & $\langle \theta \rangle$
  & \multicolumn{5}{c|}{${\rm d}^2 \sigma /{\rm d}p{\rm d}\Omega$} \\
 \hline
0.20--0.24 & 0.221 &  25.0 &  498.06 & $\!\!\pm\!\!$ &  11.13 & $\!\!\pm\!\!$ &  27.02 &  &  &  \multicolumn{5}{c|}{ } \\
0.24--0.30 & 0.269 &  25.1 &  465.98 & $\!\!\pm\!\!$ &   8.62 & $\!\!\pm\!\!$ &  22.48 & 0.270 &  35.0 &  536.91 & $\!\!\pm\!\!$ &   9.51 & $\!\!\pm\!\!$ &  23.67 \\
0.30--0.36 & 0.329 &  25.1 &  395.20 & $\!\!\pm\!\!$ &   8.00 & $\!\!\pm\!\!$ &  17.68 & 0.329 &  35.0 &  462.94 & $\!\!\pm\!\!$ &   8.27 & $\!\!\pm\!\!$ &  17.47 \\
0.36--0.42 & 0.389 &  25.2 &  324.77 & $\!\!\pm\!\!$ &   7.24 & $\!\!\pm\!\!$ &  14.03 & 0.389 &  35.2 &  390.95 & $\!\!\pm\!\!$ &   7.91 & $\!\!\pm\!\!$ &  14.33 \\
0.42--0.50 & 0.458 &  25.1 &  288.49 & $\!\!\pm\!\!$ &   5.98 & $\!\!\pm\!\!$ &  11.38 & 0.458 &  35.0 &  301.76 & $\!\!\pm\!\!$ &   6.15 & $\!\!\pm\!\!$ &  12.28 \\
0.50--0.60 & 0.548 &  25.2 &  203.61 & $\!\!\pm\!\!$ &   4.44 & $\!\!\pm\!\!$ &   9.01 & 0.547 &  35.0 &  229.10 & $\!\!\pm\!\!$ &   4.80 & $\!\!\pm\!\!$ &   9.91 \\
\hline
 \end{tabular}
 \end{center}
 \end{scriptsize}
 \end{table*}
  
 \begin{table*}[h]
 \begin{scriptsize}
 \caption{Double-differential inclusive
  cross-section ${\rm d}^2 \sigma /{\rm d}p{\rm d}\Omega$
  [mb/(GeV/{\it c} sr)] of the production of $\pi^+$'s
  in p + Cu $\rightarrow$ $\pi^+$ + X interactions
  with $+8.0$~GeV/{\it c} beam momentum;
  the first error is statistical, the second systematic; 
 $p_{\rm T}$ in GeV/{\it c}, polar angle $\theta$ in degrees.}
 \label{pip.procu8}
 \begin{center}
 \begin{tabular}{|c||c|c|rcrcr||c|c|rcrcr|}
 \hline
   & \multicolumn{7}{c||}{$20<\theta<30$}
  & \multicolumn{7}{c|}{$30<\theta<40$} \\
 \hline
 $p_{\rm T}$ & $\langle p_{\rm T} \rangle$ & $\langle \theta \rangle$
  & \multicolumn{5}{c||}{${\rm d}^2 \sigma /{\rm d}p{\rm d}\Omega$}
  &$\langle p_{\rm T} \rangle$ & $\langle \theta \rangle$
  & \multicolumn{5}{c|}{${\rm d}^2 \sigma /{\rm d}p{\rm d}\Omega$} \\
 \hline
0.10--0.13 & 0.115 &  25.0 &  237.06 & $\!\!\pm\!\!$ &  10.27 & $\!\!\pm\!\!$ &  18.35 & 0.115 &  35.0 &  209.12 & $\!\!\pm\!\!$ &   9.05 & $\!\!\pm\!\!$ &  15.45 \\
0.13--0.16 & 0.146 &  24.8 &  266.29 & $\!\!\pm\!\!$ &   9.61 & $\!\!\pm\!\!$ &  15.72 & 0.145 &  34.9 &  227.97 & $\!\!\pm\!\!$ &   8.87 & $\!\!\pm\!\!$ &  13.37 \\
0.16--0.20 & 0.180 &  24.8 &  296.43 & $\!\!\pm\!\!$ &   8.54 & $\!\!\pm\!\!$ &  14.97 & 0.181 &  34.8 &  246.20 & $\!\!\pm\!\!$ &   7.76 & $\!\!\pm\!\!$ &  12.23 \\
0.20--0.24 & 0.220 &  24.8 &  303.16 & $\!\!\pm\!\!$ &   8.45 & $\!\!\pm\!\!$ &  13.09 & 0.219 &  34.8 &  253.77 & $\!\!\pm\!\!$ &   7.72 & $\!\!\pm\!\!$ &  10.90 \\
0.24--0.30 & 0.270 &  24.8 &  287.57 & $\!\!\pm\!\!$ &   6.70 & $\!\!\pm\!\!$ &  10.51 & 0.270 &  34.8 &  234.25 & $\!\!\pm\!\!$ &   6.05 & $\!\!\pm\!\!$ &   8.35 \\
0.30--0.36 & 0.328 &  24.7 &  231.39 & $\!\!\pm\!\!$ &   5.88 & $\!\!\pm\!\!$ &   7.31 & 0.329 &  34.9 &  192.12 & $\!\!\pm\!\!$ &   5.48 & $\!\!\pm\!\!$ &   5.99 \\
\hline
 \end{tabular}
 \end{center}
 \end{scriptsize}
 \end{table*}
  
 \begin{table*}[h]
 \begin{scriptsize}
 \caption{Double-differential inclusive
  cross-section ${\rm d}^2 \sigma /{\rm d}p{\rm d}\Omega$
  [mb/(GeV/{\it c} sr)] of the production of $\pi^-$'s
  in p + Cu $\rightarrow$ $\pi^-$ + X interactions
  with $+8.0$~GeV/{\it c} beam momentum;
  the first error is statistical, the second systematic; 
 $p_{\rm T}$ in GeV/{\it c}, polar angle $\theta$ in degrees.}
 \label{pim.procu8}
 \begin{center}
 \begin{tabular}{|c||c|c|rcrcr||c|c|rcrcr|}
 \hline
   & \multicolumn{7}{c||}{$20<\theta<30$}
  & \multicolumn{7}{c|}{$30<\theta<40$} \\
 \hline
 $p_{\rm T}$ & $\langle p_{\rm T} \rangle$ & $\langle \theta \rangle$
  & \multicolumn{5}{c||}{${\rm d}^2 \sigma /{\rm d}p{\rm d}\Omega$}
  &$\langle p_{\rm T} \rangle$ & $\langle \theta \rangle$
  & \multicolumn{5}{c|}{${\rm d}^2 \sigma /{\rm d}p{\rm d}\Omega$} \\
 \hline
0.10--0.13 & 0.115 &  24.9 &  257.38 & $\!\!\pm\!\!$ &  10.30 & $\!\!\pm\!\!$ &  19.92 & 0.116 &  34.9 &  220.14 & $\!\!\pm\!\!$ &   9.23 & $\!\!\pm\!\!$ &  16.98 \\
0.13--0.16 & 0.145 &  25.0 &  277.00 & $\!\!\pm\!\!$ &   9.65 & $\!\!\pm\!\!$ &  16.60 & 0.145 &  34.8 &  240.77 & $\!\!\pm\!\!$ &   8.92 & $\!\!\pm\!\!$ &  14.54 \\
0.16--0.20 & 0.180 &  24.9 &  305.58 & $\!\!\pm\!\!$ &   8.54 & $\!\!\pm\!\!$ &  15.60 & 0.180 &  34.7 &  242.05 & $\!\!\pm\!\!$ &   7.50 & $\!\!\pm\!\!$ &  12.15 \\
0.20--0.24 & 0.220 &  24.8 &  281.02 & $\!\!\pm\!\!$ &   8.06 & $\!\!\pm\!\!$ &  12.19 & 0.220 &  34.8 &  225.27 & $\!\!\pm\!\!$ &   7.08 & $\!\!\pm\!\!$ &   9.61 \\
0.24--0.30 & 0.269 &  25.0 &  241.69 & $\!\!\pm\!\!$ &   6.04 & $\!\!\pm\!\!$ &   8.36 & 0.269 &  34.9 &  199.75 & $\!\!\pm\!\!$ &   5.47 & $\!\!\pm\!\!$ &   6.93 \\
0.30--0.36 & 0.329 &  24.9 &  181.16 & $\!\!\pm\!\!$ &   5.20 & $\!\!\pm\!\!$ &   5.44 & 0.328 &  34.9 &  165.46 & $\!\!\pm\!\!$ &   5.01 & $\!\!\pm\!\!$ &   5.00 \\
\hline
 \end{tabular}
 \end{center}
 \end{scriptsize}
 \end{table*}
  
 \begin{table*}[h]
 \begin{scriptsize}
 \caption{Double-differential inclusive
  cross-section ${\rm d}^2 \sigma /{\rm d}p{\rm d}\Omega$
  [mb/(GeV/{\it c} sr)] of the production of protons
  in $\pi^+$ + Cu $\rightarrow$ p + X interactions
  with $+8.0$~GeV/{\it c} beam momentum;
  the first error is statistical, the second systematic; 
 $p_{\rm T}$ in GeV/{\it c}, polar angle $\theta$ in degrees.}
 \label{pro.pipcu8}
 \begin{center}
 \begin{tabular}{|c||c|c|rcrcr||c|c|rcrcr|}
 \hline
   & \multicolumn{7}{c||}{$20<\theta<30$}
  & \multicolumn{7}{c|}{$30<\theta<40$} \\
 \hline
 $p_{\rm T}$ & $\langle p_{\rm T} \rangle$ & $\langle \theta \rangle$
  & \multicolumn{5}{c||}{${\rm d}^2 \sigma /{\rm d}p{\rm d}\Omega$}
  &$\langle p_{\rm T} \rangle$ & $\langle \theta \rangle$
  & \multicolumn{5}{c|}{${\rm d}^2 \sigma /{\rm d}p{\rm d}\Omega$} \\
 \hline
0.20--0.24 & 0.220 &  25.1 &  416.01 & $\!\!\pm\!\!$ &  14.70 & $\!\!\pm\!\!$ &  22.76 &  &  &  \multicolumn{5}{c|}{ } \\
0.24--0.30 & 0.269 &  25.0 &  380.56 & $\!\!\pm\!\!$ &  11.26 & $\!\!\pm\!\!$ &  18.56 & 0.271 &  35.0 &  414.40 & $\!\!\pm\!\!$ &  11.65 & $\!\!\pm\!\!$ &  18.50 \\
0.30--0.36 & 0.330 &  25.3 &  325.06 & $\!\!\pm\!\!$ &  10.52 & $\!\!\pm\!\!$ &  14.76 & 0.329 &  35.1 &  376.78 & $\!\!\pm\!\!$ &  10.81 & $\!\!\pm\!\!$ &  14.48 \\
0.36--0.42 & 0.389 &  25.2 &  248.43 & $\!\!\pm\!\!$ &   9.09 & $\!\!\pm\!\!$ &  10.94 & 0.389 &  35.2 &  295.13 & $\!\!\pm\!\!$ &   9.95 & $\!\!\pm\!\!$ &  11.08 \\
0.42--0.50 & 0.459 &  25.0 &  209.70 & $\!\!\pm\!\!$ &   7.29 & $\!\!\pm\!\!$ &   8.39 & 0.458 &  35.1 &  237.82 & $\!\!\pm\!\!$ &   7.89 & $\!\!\pm\!\!$ &   9.90 \\
0.50--0.60 & 0.547 &  25.0 &  159.53 & $\!\!\pm\!\!$ &   5.59 & $\!\!\pm\!\!$ &   7.19 & 0.546 &  35.0 &  180.66 & $\!\!\pm\!\!$ &   6.13 & $\!\!\pm\!\!$ &   7.97 \\
\hline
 \end{tabular}
 \end{center}
 \end{scriptsize}
 \end{table*}
  
 \begin{table*}[h]
 \begin{scriptsize}
 \caption{Double-differential inclusive
  cross-section ${\rm d}^2 \sigma /{\rm d}p{\rm d}\Omega$
  [mb/(GeV/{\it c} sr)] of the production of $\pi^+$'s
  in $\pi^+$ + Cu $\rightarrow$ $\pi^+$ + X interactions
  with $+8.0$~GeV/{\it c} beam momentum;
  the first error is statistical, the second systematic; 
 $p_{\rm T}$ in GeV/{\it c}, polar angle $\theta$ in degrees.}
 \label{pip.pipcu8}
 \begin{center}
 \begin{tabular}{|c||c|c|rcrcr||c|c|rcrcr|}
 \hline
   & \multicolumn{7}{c||}{$20<\theta<30$}
  & \multicolumn{7}{c|}{$30<\theta<40$} \\
 \hline
 $p_{\rm T}$ & $\langle p_{\rm T} \rangle$ & $\langle \theta \rangle$
  & \multicolumn{5}{c||}{${\rm d}^2 \sigma /{\rm d}p{\rm d}\Omega$}
  &$\langle p_{\rm T} \rangle$ & $\langle \theta \rangle$
  & \multicolumn{5}{c|}{${\rm d}^2 \sigma /{\rm d}p{\rm d}\Omega$} \\
 \hline
0.10--0.13 & 0.116 &  24.9 &  243.74 & $\!\!\pm\!\!$ &  15.29 & $\!\!\pm\!\!$ &  20.22 & 0.116 &  34.8 &  220.68 & $\!\!\pm\!\!$ &  13.78 & $\!\!\pm\!\!$ &  16.83 \\
0.13--0.16 & 0.146 &  24.8 &  266.57 & $\!\!\pm\!\!$ &  14.01 & $\!\!\pm\!\!$ &  16.00 & 0.146 &  35.2 &  243.01 & $\!\!\pm\!\!$ &  13.29 & $\!\!\pm\!\!$ &  14.38 \\
0.16--0.20 & 0.180 &  24.9 &  342.04 & $\!\!\pm\!\!$ &  13.29 & $\!\!\pm\!\!$ &  17.37 & 0.181 &  34.9 &  272.33 & $\!\!\pm\!\!$ &  11.79 & $\!\!\pm\!\!$ &  13.42 \\
0.20--0.24 & 0.220 &  24.9 &  383.02 & $\!\!\pm\!\!$ &  13.70 & $\!\!\pm\!\!$ &  16.52 & 0.220 &  34.7 &  263.25 & $\!\!\pm\!\!$ &  11.39 & $\!\!\pm\!\!$ &  11.22 \\
0.24--0.30 & 0.269 &  24.7 &  331.76 & $\!\!\pm\!\!$ &  10.40 & $\!\!\pm\!\!$ &  12.20 & 0.270 &  34.8 &  277.36 & $\!\!\pm\!\!$ &   9.56 & $\!\!\pm\!\!$ &   9.88 \\
0.30--0.36 & 0.330 &  24.9 &  299.54 & $\!\!\pm\!\!$ &   9.78 & $\!\!\pm\!\!$ &   9.45 & 0.329 &  34.7 &  240.59 & $\!\!\pm\!\!$ &   8.88 & $\!\!\pm\!\!$ &   7.49 \\
\hline
 \end{tabular}
 \end{center}
 \end{scriptsize}
 \end{table*}
  
 \begin{table*}[h]
 \begin{scriptsize}
 \caption{Double-differential inclusive
  cross-section ${\rm d}^2 \sigma /{\rm d}p{\rm d}\Omega$
  [mb/(GeV/{\it c} sr)] of the production of $\pi^-$'s
  in $\pi^+$ + Cu $\rightarrow$ $\pi^-$ + X interactions
  with $+8.0$~GeV/{\it c} beam momentum;
  the first error is statistical, the second systematic; 
 $p_{\rm T}$ in GeV/{\it c}, polar angle $\theta$ in degrees.}
 \label{pim.pipcu8}
 \begin{center}
 \begin{tabular}{|c||c|c|rcrcr||c|c|rcrcr|}
 \hline
   & \multicolumn{7}{c||}{$20<\theta<30$}
  & \multicolumn{7}{c|}{$30<\theta<40$} \\
 \hline
 $p_{\rm T}$ & $\langle p_{\rm T} \rangle$ & $\langle \theta \rangle$
  & \multicolumn{5}{c||}{${\rm d}^2 \sigma /{\rm d}p{\rm d}\Omega$}
  &$\langle p_{\rm T} \rangle$ & $\langle \theta \rangle$
  & \multicolumn{5}{c|}{${\rm d}^2 \sigma /{\rm d}p{\rm d}\Omega$} \\
 \hline
0.10--0.13 & 0.116 &  24.9 &  288.86 & $\!\!\pm\!\!$ &  15.82 & $\!\!\pm\!\!$ &  22.16 & 0.115 &  34.7 &  239.12 & $\!\!\pm\!\!$ &  13.90 & $\!\!\pm\!\!$ &  18.39 \\
0.13--0.16 & 0.146 &  24.5 &  302.36 & $\!\!\pm\!\!$ &  14.51 & $\!\!\pm\!\!$ &  18.16 & 0.144 &  35.0 &  225.33 & $\!\!\pm\!\!$ &  12.54 & $\!\!\pm\!\!$ &  13.75 \\
0.16--0.20 & 0.181 &  24.8 &  334.13 & $\!\!\pm\!\!$ &  12.89 & $\!\!\pm\!\!$ &  17.25 & 0.180 &  34.9 &  235.80 & $\!\!\pm\!\!$ &  10.67 & $\!\!\pm\!\!$ &  11.87 \\
0.20--0.24 & 0.221 &  24.9 &  297.00 & $\!\!\pm\!\!$ &  11.98 & $\!\!\pm\!\!$ &  12.99 & 0.220 &  34.7 &  230.91 & $\!\!\pm\!\!$ &  10.38 & $\!\!\pm\!\!$ &   9.85 \\
0.24--0.30 & 0.269 &  24.9 &  266.63 & $\!\!\pm\!\!$ &   9.12 & $\!\!\pm\!\!$ &   9.28 & 0.269 &  34.9 &  205.35 & $\!\!\pm\!\!$ &   7.97 & $\!\!\pm\!\!$ &   7.14 \\
0.30--0.36 & 0.329 &  24.7 &  212.58 & $\!\!\pm\!\!$ &   8.11 & $\!\!\pm\!\!$ &   6.37 & 0.329 &  34.9 &  184.82 & $\!\!\pm\!\!$ &   7.65 & $\!\!\pm\!\!$ &   5.57 \\
\hline
 \end{tabular}
 \end{center}
 \end{scriptsize}
 \end{table*}
  
 \begin{table*}[h]
 \begin{scriptsize}
 \caption{Double-differential inclusive
  cross-section ${\rm d}^2 \sigma /{\rm d}p{\rm d}\Omega$
  [mb/(GeV/{\it c} sr)] of the production of protons
  in $\pi^-$ + Cu $\rightarrow$ p + X interactions
  with $-8.0$~GeV/{\it c} beam momentum;
  the first error is statistical, the second systematic; 
 $p_{\rm T}$ in GeV/{\it c}, polar angle $\theta$ in degrees.}
 \label{pro.pimcu8}
 \begin{center}
 \begin{tabular}{|c||c|c|rcrcr||c|c|rcrcr|}
 \hline
   & \multicolumn{7}{c||}{$20<\theta<30$}
  & \multicolumn{7}{c|}{$30<\theta<40$} \\
 \hline
 $p_{\rm T}$ & $\langle p_{\rm T} \rangle$ & $\langle \theta \rangle$
  & \multicolumn{5}{c||}{${\rm d}^2 \sigma /{\rm d}p{\rm d}\Omega$}
  &$\langle p_{\rm T} \rangle$ & $\langle \theta \rangle$
  & \multicolumn{5}{c|}{${\rm d}^2 \sigma /{\rm d}p{\rm d}\Omega$} \\
 \hline
0.20--0.24 & 0.220 &  25.3 &  339.98 & $\!\!\pm\!\!$ &   6.99 & $\!\!\pm\!\!$ &  18.36 &  &  &  \multicolumn{5}{c|}{ } \\
0.24--0.30 & 0.270 &  25.2 &  328.63 & $\!\!\pm\!\!$ &   5.48 & $\!\!\pm\!\!$ &  15.84 & 0.267 &  35.3 &  415.67 & $\!\!\pm\!\!$ &  53.20 & $\!\!\pm\!\!$ &  18.36 \\
0.30--0.36 & 0.329 &  25.3 &  267.73 & $\!\!\pm\!\!$ &   4.97 & $\!\!\pm\!\!$ &  12.12 & 0.329 &  35.1 &  321.00 & $\!\!\pm\!\!$ &   5.22 & $\!\!\pm\!\!$ &  12.20 \\
0.36--0.42 & 0.390 &  25.2 &  225.01 & $\!\!\pm\!\!$ &   4.61 & $\!\!\pm\!\!$ &  10.03 & 0.389 &  35.1 &  259.09 & $\!\!\pm\!\!$ &   4.84 & $\!\!\pm\!\!$ &   9.69 \\
0.42--0.50 & 0.458 &  25.2 &  175.83 & $\!\!\pm\!\!$ &   3.49 & $\!\!\pm\!\!$ &   7.54 & 0.458 &  35.1 &  198.95 & $\!\!\pm\!\!$ &   3.77 & $\!\!\pm\!\!$ &   8.29 \\
0.50--0.60 & 0.547 &  25.2 &  132.81 & $\!\!\pm\!\!$ &   2.69 & $\!\!\pm\!\!$ &   5.90 & 0.547 &  35.0 &  149.55 & $\!\!\pm\!\!$ &   2.94 & $\!\!\pm\!\!$ &   6.95 \\
\hline
 \end{tabular}
 \end{center}
 \end{scriptsize}
 \end{table*}
  
 \begin{table*}[h]
 \begin{scriptsize}
 \caption{Double-differential inclusive
  cross-section ${\rm d}^2 \sigma /{\rm d}p{\rm d}\Omega$
  [mb/(GeV/{\it c} sr)] of the production of $\pi^+$'s
  in $\pi^-$ + Cu $\rightarrow$ $\pi^+$ + X interactions
  with $-8.0$~GeV/{\it c} beam momentum;
  the first error is statistical, the second systematic; 
 $p_{\rm T}$ in GeV/{\it c}, polar angle $\theta$ in degrees.}
 \label{pip.pimcu8}
 \begin{center}
 \begin{tabular}{|c||c|c|rcrcr||c|c|rcrcr|}
 \hline
   & \multicolumn{7}{c||}{$20<\theta<30$}
  & \multicolumn{7}{c|}{$30<\theta<40$} \\
 \hline
 $p_{\rm T}$ & $\langle p_{\rm T} \rangle$ & $\langle \theta \rangle$
  & \multicolumn{5}{c||}{${\rm d}^2 \sigma /{\rm d}p{\rm d}\Omega$}
  &$\langle p_{\rm T} \rangle$ & $\langle \theta \rangle$
  & \multicolumn{5}{c|}{${\rm d}^2 \sigma /{\rm d}p{\rm d}\Omega$} \\
 \hline
0.10--0.13 & 0.116 &  25.1 &  215.19 & $\!\!\pm\!\!$ &   7.07 & $\!\!\pm\!\!$ &  15.99 & 0.116 &  34.9 &  199.77 & $\!\!\pm\!\!$ &   6.65 & $\!\!\pm\!\!$ &  14.64 \\
0.13--0.16 & 0.145 &  25.0 &  250.03 & $\!\!\pm\!\!$ &   6.92 & $\!\!\pm\!\!$ &  14.48 & 0.145 &  35.1 &  212.91 & $\!\!\pm\!\!$ &   6.34 & $\!\!\pm\!\!$ &  12.11 \\
0.16--0.20 & 0.180 &  24.8 &  275.54 & $\!\!\pm\!\!$ &   5.98 & $\!\!\pm\!\!$ &  13.30 & 0.180 &  34.9 &  214.22 & $\!\!\pm\!\!$ &   5.27 & $\!\!\pm\!\!$ &  10.23 \\
0.20--0.24 & 0.220 &  24.8 &  294.36 & $\!\!\pm\!\!$ &   6.16 & $\!\!\pm\!\!$ &  12.80 & 0.220 &  34.8 &  228.87 & $\!\!\pm\!\!$ &   5.47 & $\!\!\pm\!\!$ &   9.65 \\
0.24--0.30 & 0.269 &  24.8 &  270.93 & $\!\!\pm\!\!$ &   4.77 & $\!\!\pm\!\!$ &   9.52 & 0.269 &  34.8 &  215.56 & $\!\!\pm\!\!$ &   4.30 & $\!\!\pm\!\!$ &   7.46 \\
0.30--0.36 & 0.329 &  24.8 &  228.62 & $\!\!\pm\!\!$ &   4.33 & $\!\!\pm\!\!$ &   6.84 & 0.329 &  34.8 &  181.03 & $\!\!\pm\!\!$ &   3.89 & $\!\!\pm\!\!$ &   5.25 \\
\hline
 \end{tabular}
 \end{center}
 \end{scriptsize}
 \end{table*}
\input{table.pim.pimcu8.tex}
\clearpage


\input{table.pro.procu12.tex}
\input{table.pip.procu12.tex}
\input{table.pim.procu12.tex}
\input{table.pro.pipcu12.tex}
\input{table.pip.pipcu12.tex}
\input{table.pim.pipcu12.tex}
\input{table.pro.pimcu12.tex}
\input{table.pip.pimcu12.tex}
\input{table.pim.pimcu12.tex}
\clearpage


\input{table.pro.procu15.tex}
\input{table.pip.procu15.tex}
\input{table.pim.procu15.tex}
\input{table.pro.pipcu15.tex}
\input{table.pip.pipcu15.tex}
\input{table.pim.pipcu15.tex}
\input{table.pro.pimcu15.tex}
\input{table.pip.pimcu15.tex}
\input{table.pim.pimcu15.tex}
\clearpage

\end{document}